\documentclass[10pt]{article}
\usepackage[T1]{fontenc}
\usepackage[latin9]{inputenc}
\usepackage{geometry}
\usepackage{natbib}
\usepackage{color}
\usepackage[usenames,dvipsnames]{xcolor}
\usepackage{amsmath}
\usepackage{amssymb}
\usepackage{graphicx}
\usepackage{esint}
\usepackage{soul}
\usepackage{framed}

\makeatletter

\usepackage{setspace}
\usepackage{amsthm}
\usepackage{graphics}
\usepackage{subfigure}
\usepackage[affil-it,auth-sc]{authblk}
\usepackage{booktabs}
\usepackage{array}

\newcolumntype{L}[1]{>{\raggedright\let\newline\\\arraybackslash\hspace{0pt}}m{#1}}
\newcolumntype{C}[1]{>{\centering\let\newline\\\arraybackslash\hspace{0pt}}m{#1}}
\newcolumntype{R}[1]{>{\raggedleft\let\newline\\\arraybackslash\hspace{0pt}}m{#1}}

\newcommand{\beq}{\begin{equation}}
\newcommand{\eeq}{\end{equation}}

\newcommand{\beqar}{\begin{eqnarray}}
\newcommand{\eeqar}{\end{eqnarray}}
\newcommand{\bit}{\begin{itemize}}
\newcommand{\eit}{\end{itemize}}
\newcommand{\benum}{\begin{enumerate}}
\newcommand{\eenum}{\end{enumerate}}
\newcommand{\barr}{\begin{array}}
\newcommand{\earr}{\end{array}}

\newcommand\eq[1]{(\ref{#1})}
\newcommand{\bfm}[1]{\mbox{\boldmath ${#1}$}}        

\newcommand{\jump}[2]{[\mbox{\hspace{-#1em}}[#2]\mbox{\hspace{-#1em}}]}

\newcommand{\bjump}[2]{\left[\mbox{\hspace{-#1em}}\left[#2\right]\mbox{\hspace{-#1em}}\right]}

\def\XXint#1#2#3{{\setbox0=\hbox{$#1{#2#3}{\int}$}
   \vcenter{\hbox{$#2#3$}}\kern-.5\wd0}}

\def\b0{\mbox{\boldmath $0$}}

\def\bb{\mbox{\boldmath $b$}}

\def\be{\mbox{\boldmath $e$}}

\def\bj{\mbox{\boldmath $j$}}

\def\bn{\mbox{\boldmath $n$}}

\def\bq{\mbox{\boldmath $q$}}

\def\bu{\mbox{\boldmath $u$}}
\def\bv{\mbox{\boldmath $v$}}

\def\bx{\mbox{\boldmath $x$}}

\def\bD{\mbox{\boldmath $D$}}
\def\bE{\mbox{\boldmath $E$}}

\def\bH{\mbox{\boldmath $H$}}

\def\bJ{\mbox{\boldmath $J$}}
\def\bK{\mbox{\boldmath $K$}}

\def\bQ{\mbox{\boldmath $Q$}}

\def\bU{\mbox{\boldmath $U$}}

\def\f0{\ensuremath{\mathbb{O}}}

\newcommand{\Gve}{\varepsilon}

\newcommand{\BGa}{\bfm\alpha}
\newcommand{\BGb}{\bfm\beta}

\newcommand{\BGve}{\bfm\varepsilon}

\newcommand{\BGs}{\bfm\sigma}

\newcommand{\BGx}{\bfm\xi}

\newcommand{\BGz}{\bfm\zeta}

\newcommand{\mA}{\ensuremath{\mathcal{A}}}

\newcommand{\mL}{\ensuremath{\mathcal{L}}}

\newcommand{\mQ}{\ensuremath{\mathcal{Q}}}

\graphicspath{{./}{./figures/}}

\title{Multiscale asymptotic homogenization analysis \\ of thermo-diffusive composite materials}

\author[1]{A. Bacigalupo\footnote{Corresponding author. Tel.: +39 0583 4326613, email address: andrea.bacigalupo@imtlucca.it}}
\author[2]{L. Morini}
\author[2]{A. Piccolroaz}

\affil[1]{IMT Institute for Advanced Studies, Lucca, Italy}
\affil[2]{Department of Civil, Environmental and Mechanical Engineering, University of Trento, Italy}

\makeatother
\begin{document}


\maketitle 
\begin{abstract}
In this paper an asymptotic homogenization method for the analysis of 
composite materials with periodic microstructure in presence of thermodiffusion is 
described. Appropriate down-scaling relations correlating the microscopic fields to the macroscopic displacements, 
temperature and chemical potential are introduced. The effects of the material inhomogeneities
are described by perturbation functions derived from the solution of recursive cell problems. Exact expressions for the overall
elastic and thermodiffusive constants of the equivalent first order thermodiffusive continuum are derived. The proposed approach is
applied to the case of a two-dimensional bi-phase orthotropic layered material, 
where the effective elastic and thermodiffusive properties can be determined analytically. 
Considering this illustrative example and assuming periodic body forces, heat and mass sources acting on the medium, the solution performed by 
the first order homogenization approach is compared with the numerical results obtained by the heterogeneous model. \\

\end{abstract}

\section{Introduction}
Composite materials are extensively used in industrial practice. Indeed, many advanced engineering applications,
such as aerospace, aircraft, green building, biomedical, energetics and electronics require the design and the use of heterogeneous multiphase materials. Due to
the microstructural effects as well as the interaction between their constituents, these materials may present several favorable physical properties, as for example high stiffness, improved
strength and toughness, enhanced thermal conductivity, mass diffusivity or electrical permittivity. 

Recently, multiphase composite materials have been largely used in the design and fabrication of battery devices, 
in particular of lithium-ion batteries and solid oxide fuel cells \citep{Nakajo1, Dev1, Ellis1}. Since high operational temperatures can be 
reached and intense particle fluxes are needed for maintaining the electrical current, the components of
such battery devices are subject to severe thermomechanical stresses as well as stresses induced by the particle diffusion, which
can cause damage and crack formation, compromising the performance of the devices in terms of power generation and
energy conversion efficiency \citep{Atkinson1, Delette1}. Modelling the mechanical and thermodiffusive
properties of the components of such battery devices represent a crucial issue in order to predict these phenomena and then to ensure the
successful manufacture and the reliability of the systems.

The macroscopic behavior of thermodiffusive composite materials used for realizing lithium-ion batteries and solid oxide fuel cells 
is influenced by multiphysics phenomena occurring at scale-lengths characteristic of the microscopic constituents, 
which is small compared to the macroscopic dimension (i.e. structural size) \citep{Richard1, Bove1, Haji1}. Consequently, multiscale techniques represent an appropriate and powerful tool
for modelling the effects of the microstructures on the macroscopic mechanical and thermodiffusive properties of these materials. In particular, for composites with periodic
microstructures, homogenization techniques represent an useful and advantageous method for providing a rigorous and synthetic description of the effects of the microscopic 
phases on the overall properties of the materials. The application of these approaches makes possible to avoid the challenging numerical computations 
required by computational modelling of heterogeneous media. 

Several homogenization techniques have been proposed for studying overall static and dynamic elastic properties of composite materials with periodic microstructures, such as
the asymptotic (see for example \cite{Bensou1, BakhPan1, Gambin1, Allaire1, Boutin1, Meguid1, Boutin2, Andrianov1, Tran1}), the variational-asymptotic 
methods (see for example \cite{SmiCher1,PeerFleck1, Smysh1, Bacigalupo2, Bacigalupo3}) and the computational approaches 
(see for example \cite{Forest1, Forest3, Kouznetsova1, Kouznetsova2, Kaczmarc1, Forest2, Bacigalupo6, Bacigalupo5,Bacigalupo1, DeBellis1, DeBellis2, Bacca1, Bacca2, Bacca3}). These techniques
associate to the considered heterogeneous material at the micro-scale, described by a standard Cauchy continuum, an equivalent
homogenous medium at the macro-scale. The behavior of the equivalent macroscopic material can be described by means of a first order continuum or alternatively a non-local
medium. Multiscale asymptotic and computational homogenization procedures have been also proposed for the analysis of heterogeneous media in presence of multiphysics
phenomena, such as thermomechanical \citep{Schrefler1, Schrefler3, Aboudi1} and thermo-magneto-electro-elastic  \citep{Camacho1} deformations. Recently, these methods have been
applied for studying the influence of the microstructural effects on the macroscopical mechanical behavior and operative performances of lithium-ion batteries \citep{salvadori1} and
solid oxide fuel cells \citep{Bacigalupo7}. The overall properties of periodic multilayered structures characterizing such energy devices can be efficiently described by means of 
homogenization methods developed for periodic composite materials. Nevertheless, to the author's knowledge, a rigorous asymptotic procedure accounting for the effects of the microstructures on both macroscopic 
elastic and thermodiffusive properties of composite materials as well as on the coupling between these properties is still unknown in literature.

In this paper, an original asymptotic homogenization method for modelling the static elastic, thermal and diffusive properties of periodic thermodiffusive
composite materials is proposed. The rigorous approach developed in \cite{BakhPan1, SmiCher1, Bacigalupo2} and \citet{Bacigalupo3} is extended in order to account for
the effects of the microstructures on the macroscopic temperature and chemical potential of the materials and on the stresses induced by these fields. 
The displacements, temperature and chemical potential at the micro- and macro-scale are related
through an asymptotic expansion of the microscopic fields in terms of characteristic size $\varepsilon$ of the microstructure.
This expansion depends both on the macroscopic strains, temperature and chemical potential gradients and on unknown perturbation functions
accounting for the effects of the heterogeneities. Perturbation functions representing the effects of the 
material microstructures on the displacement, temperature, chemical potential and on the coupling effects between these fields are introduced.
These perturbation functions, depending only on the properties of the microstructure, are obtained through the solution of non-homogeneous
problems on the cell with periodic boundary conditions. 

Similarly to the procedure proposed in \citet{SmiCher1} and  \citet{Bacigalupo2}, averaged field equations of infinite order are obtained, and their formal solution is performed by representing the macroscopic displacements, 
temperature and chemical potential in terms of power series. Field equation for the homogenized first order thermodiffusive continuum are derived,
and exact expressions for the overall elastic and thermodiffusive constants of this equivalent medium are obtained. 
The proposed formulation is applied to
the case of a two-dimensional bi-phase orthotropic layered material. The effective elastic and thermodiffusive constants corresponding
to this example are determined analytically using the general expressions derived by the homogenization procedure. 
The solution performed by the proposed approach is compared with the numerical results obtained by the heterogeneous model assuming 
periodic body forces, heat and mass sources acting on the considered bi-phase layered composite.

The article is organized as follows: in Section 2 the geometry of the considered thermodiffusive composite material 
with periodic microstructure is illustrated, and the corresponding constitutive relations and balance equations are introduced. The developed
multiscale asymptotic homogenization technique is described in Section 3, based on down-scaling relations correlating the microscopic fields 
to the macroscopic displacements, temperature and chemical potential. The unknown perturbation functions describing the effects
of the material heterogeneities are defined as solutions of the corresponding non-homogeneous cell problems. In the same Section, 
averaged field equations of infinite order are obtained, and a solution scheme based on asymptotic expansion of the macroscopic displacements,
temperature and chemical potential field is reported. Field equations and explicit expressions for the overall elastic and thermodiffusive
constants of the equivalent first order homogeneous continuum are derived in Section 4. As just anticipated, the proposed approach
is applied for studying overall properties of two-dimensional bi-phase orthotropic layered materials in Section 5. Finally, a critical discussion
about the obtained results is reported together with conclusions and future perspectives in Section 6.

\section{Governing equations of periodic multiphase materials in presence
of thermodiffusion}

\begin{figure}
\centering

\includegraphics[scale=0.8]{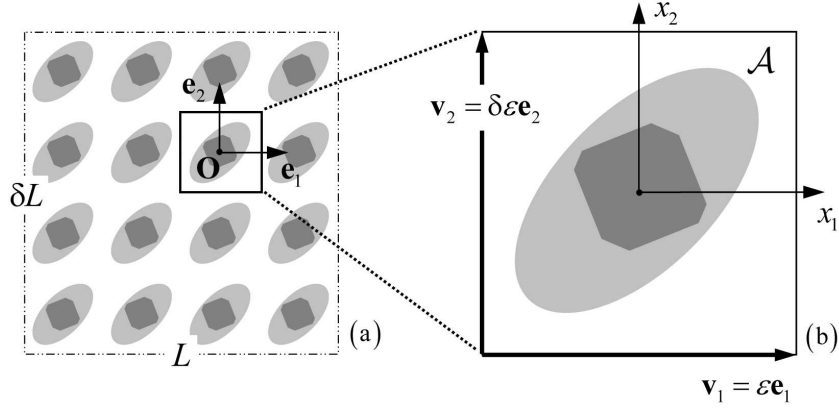}

\protect\protect\protect\caption{(a) Heterogeneous material \textendash{} Periodic domain $L$ ; (b)
Periodic cell $\mA$ and periodicity vectors. }

\label{fig01} 
\end{figure}

Let us consider an heterogeneous composite material having periodic
micro-structure and subject to stresses induced by temperature changes,
mass diffusion and body forces. The two-dimensional geometry shown
in Fig.~\ref{fig01} is assumed for the system. Considering small
strains approximation, the constituent elements of the medium are
modelled as a linear thermodiffusive elastic Cauchy continua. The
material point is identified by position vector $\bx=x_{1}\be_{1}+x_{2}\be_{2}$
referred to a system of coordinates with origin at point $O$ and
orthogonal base $\left\{ \be_{1},\be_{2}\right\} $. The periodic
cell $\mA=[0,\varepsilon]\times[0,\delta\varepsilon]$ with characteristic
size $\varepsilon$ is illustrated in Fig.~\ref{fig01}b. The entire
periodic medium can be obtained spanning the cell $\mA$ by the two
orthogonal vectors $\bv_{1}=d_{1}\be_{1}=\varepsilon\be_{1},\bv_{2}=d_{2}\be_{2}=\delta\varepsilon\be_{2}$.

According to the periodicity of the material, $\mA$ is the elementary
cell period of the elasticity tensor $\mathbb{C}^{(m,\varepsilon)}(\bx)$:
\begin{equation}
\mathbb{C}^{(m,\varepsilon)}(\bx+\bv_{i})=\mathbb{C}^{(m,\varepsilon)}(\bx),\ \ i=1,2,\ \ \forall\bx\in\mA, \label{C}
\end{equation}
where the superscript $m$ stands for {\em microscopic} field.
Similarly, the heat conduction tensor $\bK^{(m,\varepsilon)}(\bx)$
and the thermal dilatation tensor $\BGa^{(m,\varepsilon)}(\bx)$ are
defined as follows 
\begin{equation}
\bK^{(m,\varepsilon)}(\bx+\bv_{i})=\bK^{(m,\varepsilon)}(\bx),\ \ \BGa^{(m,\varepsilon)}(\bx+\bv_{i})=\BGa^{(m,\varepsilon)}(\bx),\ \ i=1,2,\ \ \forall\bx\in\mA,\label{Kalpha}
\end{equation}
and then the mass diffusion tensor $\bD^{(m,\varepsilon)}(\bx)$ and
diffusive expansion tensor $\BGb^{(m,\varepsilon)}(\bx)$ become 
\begin{equation}
\bD^{(m,\varepsilon)}(\bx+\bv_{i})=\bD^{(m,\varepsilon)}(\bx),\ \ \BGb^{(m,\varepsilon)}(\bx+\bv_{i})=\BGb^{(m,\varepsilon)}(\bx),\ \ i=1,2,\ \ \forall\bx\in\mA.\label{Dbeta}
\end{equation}
The tensors (\ref{C}), (\ref{Kalpha}) and (\ref{Dbeta}) are commonly referred to 
as $\mA-$periodic functions.

The system is subject to body forces $\bb(\bx)$, heat source $r(\bx)$
and mass source $s(\bx)$ which are assumed to be $\mL-$periodic
with period $\mL=[0,L]\times[0,\delta L]$ and to have vanishing mean
values on $\mL$. Since $L$ is a large multiple of $\varepsilon$, 
then $\mL$ can be assumed to be a representative portion of the overall
body. This means that the body forces, heat sources and mass sources
are characterized by a period much greater than the microstructural
size $\varepsilon$.

Following the procedure reported in \citet{Bacigalupo2}, a non-dimensional
unit cell $\mQ=[0,1]\times[0,\delta]$ that reproduces the periodic
microstructure by rescaling with the small parameter $\varepsilon$
is introduced. Two distinct scales are represented by the macroscopic
(slow) variables $\bx\in\mA$ and the microscopic (fast) variable
$\BGx=\bx/\varepsilon\in\mQ$ (see for example \cite{BakhPan1, SmiCher1} and \cite{Bacigalupo2}). The constitutive tensors (\ref{C}),
(\ref{Kalpha}) and (\ref{Dbeta}) are functions of the microscopic
variable, whereas the body forces, heat sources and mass sources depend
by the slow macroscopic variable. Consequently, the mapping of both
the elasticity and thermodiffusive tensors may be defined on $\mQ$
as follows: $\mathbb{C}^{(m,\varepsilon)}(\bx)=\mathbb{C}^{m}(\BGx=\bx/\varepsilon),\ \bK^{(m,\varepsilon)}(\bx)=\bK^{m}(\BGx=\bx/\varepsilon),\ \BGa^{(m,\varepsilon)}(\bx)=\BGa^{m}(\BGx=\bx/\varepsilon),\bD^{(m,\varepsilon)}(\bx)=\bD^{m}(\BGx=\bx/\varepsilon),\ \BGb^{(m,\varepsilon)}(\bx)=\BGb^{m}(\BGx=\bx/\varepsilon)$,
respectively.

The relevant micro-fields are the micro-displacement $\bu(\bx)$,
the microscopic temperature $\theta(\bx)=T(\bx)-T_{0}$ ($T_0$ stands for the temperature of the natural state) and the microscopic
chemical potential $\eta(\bx)$. The micro-stress $\BGs(\bx)$,
the microscopic heat and mass fluxes $\bq(\bx)$ and $\bj(\bx)$ are
defined by the following constitutive relations: 
\begin{equation}
\BGs(\bx)=\mathbb{C}^{m}\left(\frac{\bx}{\varepsilon}\right)\BGve(\bx)-\BGa^{m}\left(\frac{\bx}{\varepsilon}\right)\theta(\bx)-\BGb^{m}\left(\frac{\bx}{\varepsilon}\right)\eta(\bx),\label{microstrain}
\end{equation}
\begin{equation}
\bq(\bx)=-\bK^{m}\left(\frac{\bx}{\varepsilon}\right)\nabla\theta(\bx),\quad\bj(\bx)=-\bD^{m}\left(\frac{\bx}{\varepsilon}\right)\nabla\eta(\bx),\label{microfluxes}
\end{equation}
where $\BGve(\bx)=\mbox{sym}\nabla\bu(\bx)$ is the micro-strain tensor
which is assumed to be zero at the fundamental state of the system.

Note that, in eqs.~(\ref{microfluxes}) describing the heat and mass fluxes, we confine ourselves to the essential effects and neglect coupling terms, 
which is an assumption generally accepted in the quasi-static theory of thermodiffusion, see for instance \citet{Now1}.

The micro-stresses (\ref{microstrain}) and the microscopic fluxes
(\ref{microfluxes}) satisfy the local balance equations on the domain
$\mA$ 
\begin{equation}
\nabla\cdot\BGs(\bx)+\bb(\bx)=\bfm{0}, \quad 
\nabla\cdot\bq(\bx)-r(\bx)=0, \quad 
\nabla\cdot\bj(\bx)-s(\bx)=0. \label{bal1}
\end{equation}
Substituting expressions (\ref{microstrain})-(\ref{microfluxes})
in equations (\ref{bal1}) and remembering the symmetry
of the elasticity tensor, the resulting set of partial
differential equations is written in the form 
\begin{equation}
\nabla\cdot\left(\mathbb{C}^{m}\left(\frac{\bx}{\varepsilon}\right)\nabla\bu(\bx)\right)-\nabla\cdot\left(\BGa^{m}\left(\frac{\bx}{\varepsilon}\right)\theta(\bx)\right)-\nabla\cdot\left(\BGb^{m}\left(\frac{\bx}{\varepsilon}\right)\eta(\bx)\right)+\bb(\bx)=\bfm{0}\label{field1}
\end{equation}
\begin{equation}
\nabla\cdot\left(\bK^{m}\left(\frac{\bx}{\varepsilon}\right)\nabla\theta(\bx)\right)+r(\bx)=0, \quad 
\nabla\cdot\left(\bD^{m}\left(\frac{\bx}{\varepsilon}\right)\nabla\eta(\bx)\right)+s(\bx)=0. \label{field2}
\end{equation}
%


Moreover, at the interface $\Sigma$ between two different phase of the material, the microscopic fields satisfy the following interface conditions:
\begin{equation}
\left.\jump{0.1}{\bu(\bx)}\right|_{\bx\in\Sigma}=0, \qquad 
\left.\bjump{0.3}{\left(\mathbb{C}^{m}\left(\frac{\bx}{\varepsilon}\right)\nabla\bu(\bx) -\BGa^{m}\left(\frac{\bx}{\varepsilon}\right)\theta(\bx)-\BGb^{m}\left(\frac{\bx}{\varepsilon}\right)\eta(\bx)\right)\bn}\right|_{\bx\in\Sigma}=0,
\label{intc1}
\end{equation}
\begin{equation}
\left.\jump{0.1}{\theta(\bx)}\right|_{\bx\in\Sigma}=0, \qquad 
\left.\bjump{0.3}{\bK^{m}\left(\frac{\bx}{\varepsilon}\right)\nabla\theta(\bx)\cdot\bn}\right|_{\bx\in\Sigma}=0,
\label{intc2}
\end{equation}
\begin{equation}
\left.\jump{0.1}{\eta(\bx)}\right|_{\bx\in\Sigma}=0, \qquad 
\left.\bjump{0.3}{\bD^{m}\left(\frac{\bx}{\varepsilon}\right)\nabla\eta(\bx)\cdot\bn}\right|_{\bx\in\Sigma}=0,
\label{intc3}
\end{equation}
where the notation $\jump{0.1}{f}=f^{i}(\Sigma)-f^{j}(\Sigma)$ denotes the difference between the values of a function $f$ at the interface $\Sigma$ separating the
phase $i$ from the phase $j$.

The micro-displacement, microscopic temperature and chemical potential
may be seen in the form $\bu(\bx,\BGx=\bx/\varepsilon),\ \theta(\bx,\BGx=\bx/\varepsilon),\eta(\bx,\BGx=\bx/\varepsilon)$
as functions of both the slow and the fast variable.

It is important to note that since $\bu(\bx,\BGx)$, $\theta(\bx,\BGx)$ and $\theta(\bx,\BGx)$ are assumed to be $\mQ-$periodic 
smoothing functions with respect to the variable $\bx$, the interface conditions \eq{intc1}-\eq{intc3} can be expressed directly in function of the fast variable $\BGx$ \citep{BakhPan1}.

The solution
of microscopic field equations (\ref{field1}), (\ref{field2}) is computationally very expensive and
provides too detailed results to be of practical use, so that it is
convenient to replace the heterogeneous model with an equivalent homogeneous
one to obtain equations whose coefficients are not rapidly oscillating
while their solutions are close to those of the original equations.

Further in the paper, assuming that the size of the microstructure
$\varepsilon$ is sufficiently small with respect to the structural
size $L$, an equivalent classical first order thermodiffusive continuum
is considered. The overall elastic moduli, thermal and diffusion expansion
tensors, thermal and diffusive conduction tensors of a homogeneous
continuum equivalent to periodic heterogeneous material reported in
Fig.~\ref{fig01} are derived by means of asymptotic homogenization
techniques  based on the generalization of down-scaling
relations. The overall elastic and thermodiffusive properties of the
homogeneous continuum are expressed in terms of geometrical, mechanical,
thermal and diffusive properties of the microstructure by means of
an asymptotic expansion for the microscopic fields. The asymptotic
expansion is performed in terms of the parameter $\varepsilon$ that
keeps the dependence on the slow variable $\bx$ separate from the
fast one $\BGx=\bx/\varepsilon$ such that two distinct scales are
represented.

In the equivalent homogenized continuum, the macro-displacement $\bU(\bx)$
of component $U_{i}$, the macroscopic temperature $\varTheta(\bx)$
and chemical potential $\varUpsilon(\bx)$ are defined at a point
$\bx$ in the reference $(\be_{i},\ i=1,2)$. The displacement gradient
is given by $\nabla\bU(\bx)=\frac{\partial U_{i}}{\partial x_{j}}\be_{i}\otimes\be_{j}=H_{ij}\be_{i}\otimes\be_{j}=\bH(\bx)$,and
then the macroscopic strain is $\bE(\bx)=\mbox{sym}\nabla\bU(\bx)$.
The macro-stress $\bfm{\varSigma}(\bx)$ associate to $\bE(\bx)$
are defined as $\bfm{\varSigma}(\bx)=\varSigma_{ij}\be_{i}\otimes\be_{j}$
with $\varSigma_{ij}=\varSigma_{ji}$, and the macroscopic heat and
mass fluxes are respectively: $\bQ(\bx)=Q_{i}\be_{i}$ and $\bJ(\bx)=J_{i}\be_{i}$.

\section{Multiscale analysis and asymptotic solution of the heterogeneous
problem}
\label{multiscale}

\subsection{Down-scaling and up-scaling relations}
\label{down}
Following the approaches developed in \cite{BakhPan1, SmiCher1, Bacigalupo3} and \cite{Bacigalupo2} 
for purely elastic problems in periodic heterogeneous media,
the microscopic displacement, temperature
and chemical potential fields are represented through an asymptotic
expansion with respect to the parameter $\varepsilon$, whose terms depend
on macroscopic fields and perturbation functions: 
\begin{align}
u_{k}\left(\bx,\BGx=\frac{\bx}{\varepsilon}\right) & =\left[U_{k}(\bx)+\sum_{l=1}^{+\infty}\varepsilon^{l}\sum_{|q|=l}N_{kpq}^{(l)}(\BGx)\frac{\partial^{l}U_{p}(\bx)}{\partial x_{q}}+\right.\nonumber \\
 & \;\;\;\;+\left.\sum_{l=1}^{+\infty}\varepsilon^{l}\sum_{|q|=l-1}\left(\tilde{N}_{kq-1}^{(l)}(\BGx)\frac{\partial^{l-1}\Theta(\bx)}{\partial x_{q}}+\hat{N}_{kq-1}^{(l)}(\BGx)\frac{\partial^{l-1}\Upsilon(\bx)}{\partial x_{q}}\right)\right]_{\BGx=\bx/\varepsilon}\nonumber \\
 & =U_{k}(\bx)+\varepsilon\left(N_{kpq_{1}}^{(1)}(\BGx)\frac{\partial U_{p}(\bx)}{\partial x_{q_{1}}}+\tilde{N}_{k}^{(1)}(\BGx)\Theta(\bx)+\hat{N}_{k}^{(1)}(\BGx)\Upsilon(\bx)\right)_{\BGx=\bx/\varepsilon}+\nonumber \\
 & \;\;\;\;+\varepsilon^{2}\left(N_{kpq_{1}q_{2}}^{(2)}(\BGx)\frac{\partial^{2}U_{p}(\bx)}{\partial x_{q_{1}}\partial x_{q_{2}}}+\tilde{N}_{kq_{1}}^{(2)}(\BGx)\frac{\partial\Theta(\bx)}{\partial x_{q_{1}}}+\hat{N}_{kq_{1}}^{(2)}(\BGx)\frac{\partial\Upsilon(\bx)}{\partial x_{q_{1}}}\right)_{\BGx=\bx/\varepsilon}+\cdots,\nonumber \\
\label{Uasym}
\end{align}
\begin{align}
\theta\left(\bx,\BGx=\frac{\bx}{\varepsilon}\right) & =\left[\Theta(\bx)+\sum_{l=1}^{+\infty}\varepsilon^{l}\sum_{|q|=l}\left(M_{q}^{(l)}(\BGx)\frac{\partial^{l}\Theta(\bx)}{\partial x_{q}}\right)\right]_{\BGx=\bx/\varepsilon}\nonumber \\
 & =\Theta(\bx)+\varepsilon\left(M_{q_{1}}^{(1)}(\BGx)\frac{\partial\Theta(\bx)}{\partial x_{q_{1}}}\right)_{\BGx=\bx/\varepsilon}+\varepsilon^{2}\left(M_{q_{1}q_{2}}^{(2)}(\BGx)\frac{\partial^{2}\Theta(\bx)}{\partial x_{q_{1}}\partial x_{q_{2}}}\right)_{\BGx=\bx/\varepsilon}+\cdots,\nonumber \\
\label{Tasym}
\end{align}
\begin{align}
\eta\left(\bx,\BGx=\frac{\bx}{\varepsilon}\right) & =\left[\Upsilon(\bx)+\sum_{l=1}^{+\infty}\varepsilon^{l}\sum_{|q|=l}\left(W_{q}^{(l)}(\BGx)\frac{\partial^{l}\Upsilon(\bx)}{\partial x_{q}}\right)\right]_{\BGx=\bx/\varepsilon}\nonumber \\
 & =\Upsilon(\bx)+\varepsilon\left(W_{q_{1}}^{(1)}(\BGx)\frac{\partial\Upsilon(\bx)}{\partial x_{q_{1}}}\right)_{\BGx=\bx/\varepsilon}+\varepsilon^{2}\left(W_{q_{1}q_{2}}^{(2)}(\BGx)\frac{\partial^{2}\Upsilon(\bx)}{\partial x_{q_{1}}\partial x_{q_{2}}}\right)_{\BGx=\bx/\varepsilon}+\cdots.\nonumber \\
\label{Easym}
\end{align}

In equations (\ref{Uasym}), (\ref{Tasym}) and (\ref{Easym}) (commonly
known as \emph{down-scaling} relations), $q=q_{1},\cdots, q_{l}$ is
a multi-index and $\partial^{l}(\cdot)/\partial x_{q}=\partial^{l}(\cdot)/\partial x_{q_{1}}\cdots\partial x_{q_{l}}$. Due to their dependence on the slow space variable
$\bx$, the macroscopic fields $U_{k},\Theta$ and $\Upsilon$ are
$\mL-$periodic functions. $N_{kpq}^{(l)},M_{q}^{(l)}$ and $W_{q}^{(l)}$
are the mechanical, thermal and diffusive fluctuation functions,
respectively, whereas $\tilde{N}_{kq}^{(l)}$ and $\hat{N}_{kq}^{(l)}$
denote the additional fluctuation functions corresponding to the
contribution of the thermodiffusion to local displacement. All these
perturbation functions depend on the fast space variable $\BGx=\bx/\varepsilon$,
and moreover, as it will be shown in Section \ref{cellproblems}, they are $\mQ-$periodic. Similarly to the procedure reported
in \citet{SmiCher1} and \citet{Bacigalupo2}, the mean value of the fluctuation functions is assumed
to vanish on the unit cell $\mQ$, this means that the following normalization
conditions are satisfied:

\[
\left\langle N_{kpq}^{(l)}\right\rangle =\frac{1}{\delta}\int_{\mQ}N_{kpq}^{(l)}(\BGx)d\BGx=0,\quad\left\langle \tilde{N}_{kq}^{(l)}\right\rangle =\frac{1}{\delta}\int_{\mQ}\tilde{N}_{kq}^{(l)}(\BGx)d\BGx=0,\quad\left\langle \hat{N}_{kq}^{(l)}\right\rangle =\frac{1}{\delta}\int_{\mQ}\hat{N}_{kq}^{(l)}(\BGx)d\BGx=0,
\]
\begin{equation}
\left\langle M_{q}^{(l)}\right\rangle =\frac{1}{\delta}\int_{\mQ}M_{q}^{(l)}(\BGx)d\BGx=0,\quad\left\langle W_{q}^{(l)}\right\rangle =\frac{1}{\delta}\int_{\mQ}W_{q}^{(l)}(\BGx)d\BGx=0.\label{renorm-1}
\end{equation}

Introducing a new variable $\BGz\in\mQ$ and a vector $\varepsilon\BGz\in\mA$,
which represents the translations of the medium with respect to the
$\mL-$periodic body forces $\bb(\bx)$, heat sources $r(\bx)$ and
mass sources $s(\bx)$ \citep{Bacigalupo3}, it can be shown that any $\mQ-$periodic function
$g(\BGx+\BGz)$ satisfies the following invariance property:
\begin{equation}
\left\langle g(\BGx+\BGz)\right\rangle =\frac{1}{\delta}\int_{\mQ}g(\BGx+\BGz)d\BGz=\frac{1}{\delta}\int_{\mQ}g(\BGx+\BGz)d\BGx.\label{inv}
\end{equation}
According to the invariance property (\ref{inv}) and to the normalization
conditions (\ref{renorm-1}), the macroscopic fields can be defined
as the mean values of the microscopic quantities (\ref{Uasym}), (\ref{Tasym})
and (\ref{Easym}) evaluated on the unit cell $\mQ$: 
\begin{equation}
U_{k}(\bx)\doteq\left\langle u_{k}\left(\bx,\frac{\bx}{\varepsilon}+\BGz\right)\right\rangle ,\quad\Theta(\bx)\doteq\left\langle \theta\left(\bx,\frac{\bx}{\varepsilon}+\BGz\right)\right\rangle ,\quad\Upsilon(\bx)\doteq\left\langle \eta\left(\bx,\frac{\bx}{\varepsilon}+\BGz\right)\right\rangle ,\label{upscaling}
\end{equation}
Expressions (\ref{upscaling}) are commonly known as \emph{up-scaling}
relations.
More details regarding the structure of the down-scaling relations \eq{Uasym}, \eq{Tasym} and \eq{Easym} are provided in Appendix \ref{appdown}.

\subsection{First-order asymptotic analysis and derivation of the corresponding \emph{first-order cell problems}}
\label{cellproblems}

In order to derive exact expressions for the
fluctuation functions affecting the behavior of the microscopic
fields $u_{k},\theta,\eta$, the \emph{down-scaling} relations (\ref{Uasym}),
(\ref{Tasym}) and (\ref{Easym}) are substituted into the microscopic
field equations (\ref{field1}), (\ref{field2}).
Remembering the property $\frac{\partial}{\partial x_{j}}f(\boldsymbol{x},\boldsymbol{\xi}=\frac{\boldsymbol{x}}{\varepsilon})=\left(\frac{\partial f}{\partial x_{j}}+\frac{1}{\varepsilon}\frac{\partial f}{\partial\xi_{j}}\right)_{\boldsymbol{\xi}=\boldsymbol{x}/\varepsilon}=\left(\frac{\partial f}{\partial x_{j}}+\frac{f_{,j}}{\varepsilon}\right)_{\boldsymbol{\xi}=\boldsymbol{x}/\varepsilon}$,
equation (\ref{field1}) become to the first order approximation
\begin{align}
\varepsilon^{-1} & \biggl\{\left[\left(C_{ijkl}^{\Gve}N_{kpq_{1},l}^{(1)}\right)_{,j}+C_{ijpq_{1},j}^{\Gve}\right]H_{pq_{1}}(\bx)+\left[\left(C_{ijkl}^{\Gve}\tilde{N}_{k,l}^{(1)}\right)_{,j}-\alpha_{ij,j}^{\Gve}\right]\Theta(\boldsymbol{x})\nonumber \\
 & +\left[\left(C_{ijkl}^{\Gve}\hat{N}_{k,l}^{(1)}\right)_{,j}-\beta_{ij,j}^{\Gve}\right]\Upsilon(\boldsymbol{x})\biggr\} +\cdots\cdots+b_{i}(\bx)=0,\quad\quad i=1,2,\label{eq:asymech}
\end{align}
where $H_{pq_{1}}=\partial U_{p}/\partial x_{q_{1}}$ are the components
of the macroscopic displacement gradient tensor previously defined.
Equations (\ref{field2}) assume the following form
\begin{align}
\varepsilon^{-1} & \biggl[\left(K_{ij}^{\Gve}M_{q_{1},j}^{(1)}\right)_{,i}+K_{iq_{1},i}^{\Gve}\biggr]\frac{\partial\Theta}{\partial x_{q_{1}}} +\cdots\cdots+r(\bx)=0,\label{eq:asytemp}
\end{align}
\begin{align}
\varepsilon^{-1} & \biggl[\left(D_{ik}^{\Gve}W_{q_{1},j}^{(1)}\right)_{,i}+D_{iq_{1},i}^{\Gve}\biggr]\frac{\partial\Upsilon}{\partial x_{q_{1}}} +\cdots\cdots+s(\bx)=0.\label{eq:asydiff}
\end{align}
In order to transform the field equation (\ref{eq:asymech}), (\ref{eq:asytemp})
and (\ref{eq:asydiff}) in a PDEs system with constant coefficients,
in which the unknowns are the macroscopic quantities $U_{k}(\mathbf{x})$, $\Theta(\mathbf{x})$
and $\varUpsilon(\mathbf{x})$, the fluctuation functions have to
satisfy non-homogeneous equations (\emph{first-order cell problems}) reported
below. 

At the order $\varepsilon^{-1}$ from the equation (\ref{eq:asymech}) we derive: 
\begin{equation}
\left(C_{ijkl}^{\Gve}N_{kpq_{1},l}^{(1)}\right)_{,j} +C_{ijpq_{1},j}^{\Gve}=n_{ipq_{1}}^{(1)}, \quad
\left(C_{ijkl}^{\Gve}\tilde{N}_{k,l}^{(1)}\right)_{,j} -\alpha_{ij,j}^{\Gve}=\tilde{n}_{i}^{(1)}, \quad
\left(C_{ijkl}^{\Gve}\hat{N}_{k,l}^{(1)}\right)_{,j} -\beta_{ij,j}^{\Gve}=\hat{n}_{i}^{(1)}, \label{cellmech-1}
\end{equation}
whereas from thermodiffusion equations (\ref{eq:asytemp}) and (\ref{eq:asydiff})
we obtain: 
\begin{equation}
\left(K_{ij}^{\Gve}M_{q_{1},j}^{(1)}\right)_{,i} +K_{iq_{1},i}^{\Gve}=m_{q_{1}}^{(1)}, \quad
\left(D_{ij}^{\Gve}W_{q_{1},j}^{(1)}\right)_{,i} +D_{iq_{1},i}^{\Gve}=w_{q_{1}}^{(1)}, \label{cellthermodiff-1}
\end{equation}
where: 
\[
n_{ipq_{1}}^{(1)}=\langle C_{ijpq_{1},j}^{\Gve}\rangle=0,\quad\tilde{n}_{i}^{(1)}=-\langle\alpha_{ij,j}^{\Gve}\rangle=0,\quad\hat{n}_{i}^{(1)}=-\langle\beta_{ij,j}^{\Gve}\rangle=0,
\]
\begin{equation}
m_{q_{1}}^{(1)}=\langle K_{iq_{1},i}^{\Gve}\rangle=0,\quad w_{q_{1}}^{(1)}=\langle D_{iq_{1},i}^{\Gve}\rangle=0.\label{nmw0}
\end{equation}
The properties (\ref{nmw0}) are consequence of the $\mQ-$periodicity
of the components $C_{ijpq_{1}}^{\Gve},\alpha_{ij}^{\Gve},\beta_{ij}^{\Gve},K_{iq_{1}}^{\Gve}$
and $D_{iq_{1}}^{\Gve}$. 
Note that in equations (\ref{eq:asymech})--(\ref{nmw0}) the derivatives should be understood in the generalized sense.

The perturbation functions characterizing the \emph{down-scaling}
relations (\ref{Uasym}), (\ref{Tasym}), and (\ref{Easym})
are obtained by the solution of the previously defined cells problems,
derived by imposing the normalization conditions (\ref{renorm-1}).

\section{Homogenized thermodiffusive Cauchy continuum: field equations and
overall properties}
\label{seccauchy}
The field equations of the first order homogeneous continuum can be obtained
by the zero order terms (equations (\ref{inf_asymech}) and (\ref{inf_diff0})) of the sequence of PDEs derived applying the
asymptotic analysis to the averaged field equation, see Appendix \ref{apphigh}. This implies that
the macroscopic displacement, temperature and chemical potential are
approximated as follows: 
\begin{equation}
U_{p}(\mathbf{x})\approx U_{p}^{(0)}(\mathbf{x}),\quad\Theta(\mathbf{x})\approx\Theta^{(0)}(\mathbf{x}),\quad\Upsilon(\mathbf{x})\approx\Upsilon^{(0)}(\mathbf{x}).\label{approx}
\end{equation}
Alternatively, the field equations of the equivalent Cauchy continuum
can be derived considering only the terms of order $\varepsilon^{0}$
in the equations (\ref{mech_inf}), (\ref{thermo_inf}) and (\ref{diff_inf}).

The field equations of an homogeneous first order continuum in presence
of thermodiffusion are given by 
\begin{equation}
C_{iq_{1}pq_{2}}\frac{\partial^{2}U_{p}}{\partial x_{q_{1}}\partial x_{q_{2}}}-\alpha_{iq_{1}}\frac{\partial\Theta}{\partial x_{q_{1}}}-\beta_{iq_{1}}\frac{\partial\Upsilon}{\partial x_{q_{1}}}+b_{i}=0,\label{ovmech}
\end{equation}
\begin{equation}
K_{q_{1}q_{2}}\frac{\partial^{2}\Theta}{\partial x_{q_{1}}\partial x_{q_{2}}}+r=0,\quad
D_{q_{1}q_{2}}\frac{\partial^{2}\Upsilon}{\partial x_{q_{1}}\partial x_{q_{2}}}+s=0,\label{ovdiff}
\end{equation}
where $C_{iq_{1}pq_{2}}$ are the components of the overall elastic
tensor, $\alpha_{iq_{1}}$ and $\beta_{iq_{1}}$ are respectively
the components of the overall thermal dilatation and diffusive expansion
tensors, $K_{q_{1}q_{2}}$ denotes the components of the overall heat
conduction tensor and $D_{q_{1}q_{2}}$ represents the components
of the overall mass diffusion tensor. Remembering the approximation
(\ref{approx}), the macroscopic field equations (\ref{ovmech})--(\ref{ovdiff}) can be compared to the zero
order terms of the averaged field equation (\ref{inf_asymech}) and (\ref{inf_diff0}) for determining the overall properties of the
thermodiffusive Cauchy continuum. In order to relate the coefficients
$n_{ipq_{1}q_{2}}^{(2)}$, $\tilde{n}_{iq_{1}}^{(2)}$, $\hat{n}_{iq_{1}}^{(2)}$,
$m_{q_{1}q_{2}}^{(2)}$, $w_{q_{1}q_{2}}^{(2)}$ contained in the
equations (\ref{inf_asymech}) and (\ref{inf_diff0})
to the overall elastic and thermodiffusive constants of the media
$C_{iq_{1}pq_{2}}$, $\alpha_{iq_{1}}$, $\beta_{iq_{1}}$, $K_{q_{1}q_{2}}$,
$D_{q_{1}q_{2}}$, the symmetries of the tensors of components $n_{ipq_{1}q_{2}}^{(2)}$,
$\tilde{n}_{iq_{1}}^{(2)}$, $\hat{n}_{iq_{1}}^{(2)}$, $m_{q_{1}q_{2}}^{(2)}$,
$w_{q_{1}q_{2}}^{(2)}$, and the ellipticity of the field equations
(\ref{inf_asymech}) and (\ref{inf_diff0})
are required. A demonstration of these properties is reported in Appendix
\ref{appover}. As a consequence of these properties, it can be observed that:
$n_{ipq_{1}q_{2}}^{(2)}=\frac{1}{2}(C_{iq_{1}pq_{2}}+C_{iq_{2}pq_{1}})$,
$\tilde{n}_{iq_{1}}^{(2)}=\alpha_{iq_{1}}$, $\hat{n}_{iq_{1}}^{(2)}=\beta_{iq_{1}}$,
$m_{q_{1}q_{2}}^{(2)}=K_{q_{1}q_{2}}$ and $w_{q_{1}q_{2}}^{(2)}=D_{q_{1}q_{2}}$.
In particular, comparing the field equation (\ref{ovmech}) to 
(\ref{inf_asymech}), and remembering the relationship between $n_{ipq_{1}q_{2}}^{(2)}$
and $C_{iq_{1}pq_{2}}$, it is easy to note that due to the repetition
of the indexes $q_{1}$ and $q_{2}$: $C_{iq_{1}pq_{2}}\frac{\partial^{2}U_{p}}{\partial x_{q_{1}} \partial x_{q_{2}}}=n_{ipq_{1}q_{2}}^{(2)}\frac{\partial^{2}U_{p}}
{\partial x_{q_{1}} \partial x_{q_{2}}}=\frac{1}{2}(C_{iq_{1}pq_{2}}+C_{iq_{2}pq_{1}})\frac{\partial^{2}U_{p}}
{\partial x_{q_{1}} \partial x_{q_{2}}}$.

The overall elastic and thermodiffusive tensors, obtained in terms
of fluctuation functions, and the components of microscopic elastic
and thermodiffusive tensors, take the form (see Appendix \ref{apphigh} for details):
\[
C_{iq_{1}pq_{2}}=\frac{1}{4}\left\langle C_{rjkl}^{\varepsilon}\left(N_{riq_{1},j}^{(1)}+\delta_{ri}\delta_{jq_{1}}+N_{rq_{1}i,j}^{(1)}+\delta_{rq_{1}}\delta_{ij}\right)\left(N_{kpq_{2},l}^{(1)}+\delta_{kp}\delta_{q_{2},l}+N_{kq_{2}p,l}^{(1)}+\delta_{kq_{2}}\delta_{lp}\right)\right\rangle 
\]
\[
\alpha_{iq_{1}}=\left\langle C_{iq_{1}kl}^{\Gve}\tilde{N}_{k,l}^{(1)}-\alpha_{iq_{1}}^{\Gve}\right\rangle , \quad 
\beta_{iq_{1}}=\left\langle C_{iq_{1}kl}^{\Gve}\hat{N}_{k,l}^{(1)}-\beta_{iq_{1}}^{\Gve}\right\rangle ,
\]
\begin{equation}
K_{q_{1}q_{2}}=\left\langle K_{ij}^{\varepsilon}(M_{q_{1},j}^{(1)}+\delta_{jq_{1}})(M_{q_{2},i}^{(1)}+\delta_{iq_{2}})\right\rangle , \quad
D_{q_{1}q_{2}}=\left\langle D_{ij}^{\varepsilon}(W_{q_{1},j}^{(1)}+\delta_{jq_{1}})(W_{q_{2},i}^{(1)}+\delta_{iq_{2}})\right\rangle .\label{overall}
\end{equation}
The components $C_{iq_{1}pq_{2}}$, $K_{q_{1}q_{2}}$ and $D_{q_{1}q_{2}}$
of the overall constitutive tensors of the material coincide with
those derived by asymptotic homogenization techniques applied to uncoupled
static elastic \citep{BakhPan1,SmiCher1, Bacigalupo2} and heat conduction problems \citep{Schrefler3} in media
with periodic microstructures. The components $\alpha_{iq_{1}}$ and
$\beta_{iq_{1}}$ of the coupling thermodiffusive tensors have been
obtained by means of a consistent generalization of the down-scaling
relations (\ref{Uasym}) (\ref{Tasym}) and (\ref{Easym}). These
expressions relate the microscopic displacement field to the macroscopic
displacements, temperature, chemical potential and their higher order
gradients.

\section{Illustrative example: homogenization of bi-phase orthotropic layered
materials in presence of thermodiffusion}

The general results obtained are now applied to the case of a bi-phase
layered material in presence of thermodiffusion. Exact analytical
expressions for the overall elastic and thermodiffusive constants
are derived. Considering a two-dimensional infinite thermodiffusive
medium subject to periodic body forces, heat and/or mass sources,
the solution obtained applying the proposed homogenized model is compared
with the results provided by the analysis of the corresponding heterogeneous
problem.

\begin{figure}[!htcb]
\centering

\includegraphics[scale=0.8]{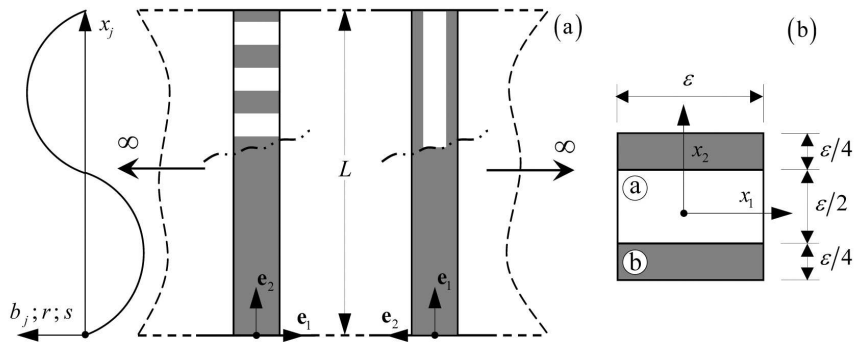}

\protect\protect\protect\caption{(a) Heterogeneous and homogenized models with $\mL$ -periodic body
force $b_{j}$, heat sources $r(x_{j})$ and mass sources $s(x_{j})$;
(b) Periodic cell and constituents: bi-phase layered material. }

\label{fig02} 
\end{figure}

\subsection{Perturbation functions and overall constitutive constants: exact analytical expressions}
\label{ovortho}
Let us consider a layered body obtained as an unbounded $d_{2}-$periodic
arrangement of two different layers having thickness $a$ and $b$,
where $d_{2}=\varepsilon=a+b$ and $\zeta=a/b$ are defined. The phases
are assumed homogeneous and orthotropic, with an orthotropic axis
coincident with the layering direction $\mathbf{e}_{1}$, the geometry of the system is shown in Fig.~\ref{fig02}. The orthotropic
symmetry is assumed for both the elastic and thermodiffusive tensors.
The micro-fluctuation functions $N_{riq_{1}}^{(1)}$, $\tilde{N}_{k}^{(1)}$,
$\hat{N}_{k}^{(1)}$, $M_{q_{1}}^{(1)}$ and $W_{q_{1}}^{(1)}$ are
analytically obtained through the solution of the cell problems formulated
in Section \ref{cellproblems} (see equations (\ref{cellmech-1})
and (\ref{cellthermodiff-1}) and conditions (\ref{nmw0})). Due to
the particular properties of symmetry of the microstructure, these
functions depend only on the fast variable $\xi_{2}$. This variable
is perpendicular to the layering direction $\mathbf{e}_{1}$ (see
Fig.~\ref{fig02}). The non-vanishing micro-fluctuation functions $N_{riq_{1}}^{(1)}$,
$\tilde{N}_{k}^{(1)}$ and $\hat{N}_{k}^{(1)}$, obtained by solving
the cell problem of order $\varepsilon^{-1}$ (\ref{cellmech-1})
are: 
\begin{align}
 & N_{211}^{(1)\_a}=\frac{C_{1122}^{b}-C_{1122}^{a}}{C_{2222}^{a}-\zeta C_{2222}^{b}}\xi_{2}^{a};\quad N_{211}^{(1)\_b}=\zeta\frac{C_{1122}^{a}-C_{1122}^{b}}{C_{2222}^{a}-\zeta C_{2222}^{b}}\xi_{2}^{b};\nonumber \\
 & N_{222}^{(1)\_a}=\frac{C_{2222}^{b}-C_{2222}^{a}}{C_{2222}^{a}-\zeta C_{2222}^{b}}\xi_{2}^{a};\quad N_{211}^{(1)\_b}=\zeta\frac{C_{1122}^{a}-C_{1122}^{b}}{C_{2222}^{a}-\zeta C_{2222}^{b}}\xi_{2}^{b};\nonumber \\
 & N_{112}^{(1)\_a}=N_{121}^{1\_a}=\frac{C_{1212}^{b}-C_{1212}^{a}}{C_{1212}^{a}-\zeta C_{1212}^{b}}\xi_{2}^{a};\quad N_{112}^{(1)\_b}=N_{121}^{1\_b}=\zeta\frac{C_{1212}^{a}-C_{1212}^{b}}{C_{1212}^{a}-\zeta C_{1212}^{b}}\xi_{2}^{b};\label{pert_mech}
\end{align}
\begin{equation}
\tilde{N}_{2}^{(1)\_a}=-\frac{\alpha_{22}^{b}-\alpha_{22}^{a}}{C_{2222}^{a}+\zeta C_{2222}^{b}}\xi_{2}^{a};\quad\tilde{N}_{2}^{(1)\_b}=\zeta\frac{\alpha_{22}^{b}-\alpha_{22}^{a}}{C_{2222}^{a}+\zeta C_{2222}^{b}}\xi_{2}^{b};\label{pert_couple1}
\end{equation}
\begin{equation}
\hat{N}_{2}^{(1)\_a}=-\frac{\beta_{22}^{b}-\beta_{22}^{a}}{C_{2222}^{a}+\zeta C_{2222}^{b}}\xi_{2}^{a};\quad\hat{N}_{2}^{(1)\_b}=\zeta\frac{\beta_{22}^{b}-\beta_{22}^{a}}{C_{2222}^{a}+\zeta C_{2222}^{b}}\xi_{2}^{b};\label{pert_couple2}
\end{equation}
where $\xi_{2}^{a}\in\left[-\frac{\zeta}{2(\zeta+1)},\frac{\zeta}{2(\zeta+1)}\right]$
and $\xi_{2}^{b}\in\left[-\frac{1}{2(\zeta+1)},\frac{1}{2(\zeta+1)}\right]$
are non-dimensional vertical coordinates centered in each layer. The
non-vanishing fluctuation functions associate with the thermodiffusion
equations, derived by the solution of the cell problems of order $\varepsilon^{-1}$
(\ref{cellthermodiff-1}) are: 
\begin{equation}
M_{2}^{(1)\_a}=-\frac{K_{22}^{a}-K_{22}^{b}}{K_{22}^{a}-\zeta K_{22}^{b}}\xi_{2}^{a};\quad M_{2}^{(1)\_b}=\zeta\frac{K_{22}^{a}-K_{22}^{b}}{K_{22}^{a}-\zeta K_{22}^{b}}\xi_{2}^{b};\label{pert_thermo}
\end{equation}
\begin{equation}
W_{2}^{(1)\_a}=-\frac{D_{22}^{a}-D_{22}^{b}}{D_{22}^{a}-\zeta D_{22}^{b}}\xi_{2}^{a};\quad W_{2}^{(1)\_b}=\zeta\frac{D_{22}^{a}-D_{22}^{b}}{D_{22}^{a}-\zeta D_{22}^{b}}\xi_{2}^{b}.\label{pert_diff}
\end{equation}
Note that the superscripts $^{a,b}$ denote that the elastic and thermodiffusive
constants are referred respectively to the phases $a$ and $b$.

In order to derive the overall elastic and thermodiffusive constants
corresponding to a first order equivalent continuum, the fluctuation
functions (\ref{pert_mech}), (\ref{pert_couple1}), (\ref{pert_couple2}),
(\ref{pert_thermo}) and (\ref{pert_diff}) are used into expressions
(\ref{overall}). The components of the overall elastic tensor $C_{iq_{1}pq_{2}}$
take the form: 
\[
C_{1111}=\frac{\zeta^{2}C_{1111}^{a}C_{2222}^{b}+\zeta(C_{1111}^{b}C_{2222}^{b}-(C_{1122}^{a})^{2}+2C_{1122}^{a}C_{1122}^{b}-(C_{1122}^{b})^{2}+C_{1111}^{a}C_{2222}^{a})+C_{1111}^{b}C_{2222}^{a}}{(\zeta+1)(C_{2222}^{a}+\zeta C_{2222}^{b})};
\]
\begin{equation}
C_{2222}=\frac{(\zeta+1)C_{2222}^{a}C_{2222}^{b}}{C_{2222}^{a}+\zeta C_{2222}^{b}}, \quad
C_{1212}=\frac{(\zeta+1)C_{1212}^{a}C_{1212}^{b}}{C_{1212}^{a}+\zeta C_{1212}^{b}}, \quad
C_{1122}=\frac{C_{1122}^{b}C_{2222}^{a}+\zeta C_{1122}^{a}C_{2222}^{b}}{C_{2222}^{a}+\zeta C_{2222}^{b}}.\label{omC}
\end{equation}
The non-vanishing components of the thermal dilatation tensor $\alpha_{iq_{1}}$
and diffusive expansion tensor $\beta_{iq_{1}}$ are respectively
given by 
\[
\alpha_{11}=-\frac{\zeta(C_{1122}^{b}\alpha_{22}^{b}-C_{1122}^{b}\alpha_{22}^{a}-C_{2222}^{b}\alpha_{11}^{b}-C_{1122}^{a}\alpha_{22}^{b}+C_{1122}^{a}\alpha_{22}^{a}-C_{2222}^{a}\alpha_{11}^{a})-\zeta^{2}C_{2222}^{b}\alpha_{11}^{a}-C_{2222}^{a}\alpha_{11}^{b}}{(\zeta+1)(C_{2222}^{a}+\zeta C_{2222}^{b})};
\]
\begin{equation}
\alpha_{22}=\frac{\zeta C_{2222}^{b}\alpha_{22}^{a}+\alpha_{22}^{b}C_{2222}^{a}}{C_{2222}^{a}+\zeta C_{2222}^{b}};\label{omalpha}
\end{equation}
\[
\beta_{11}=-\frac{\zeta(C_{1122}^{b}\beta_{22}^{b}-C_{1122}^{b}\beta_{22}^{a}-C_{2222}^{b}\beta_{11}^{b}-C_{1122}^{a}\beta_{22}^{b}+C_{1122}^{a}\beta_{22}^{a}-C_{2222}^{a}\beta_{11}^{a})-\zeta^{2}C_{2222}^{b}\beta_{11}^{a}-C_{2222}^{a}\beta_{11}^{b}}{(\zeta+1)(C_{2222}^{a}+\zeta C_{2222}^{b})};
\]
\begin{equation}
\beta_{22}=\frac{\zeta C_{2222}^{b}\beta_{22}^{a}+\beta_{22}^{b}C_{2222}^{a}}{C_{2222}^{a}+\zeta C_{2222}^{b}}.\label{ombeta}
\end{equation}
The non-vanishing components of the heat conduction tensor $K_{iq_{1}}$
and mass diffusion tensor $D_{iq_{1}}$ take the form 
\begin{equation}
K_{11}=\frac{K_{11}^{b}+\zeta K_{11}^{a}}{\zeta+1}, \quad
K_{22}=\frac{(\zeta+1)K_{22}^{a}K_{22}^{b}}{K_{22}^{a}+\zeta K_{22}^{b}};\label{omK}
\end{equation}
\begin{equation}
D_{11}=\frac{D_{11}^{b}+\zeta D_{11}^{a}}{\zeta+1}, \quad
D_{22}=\frac{(\zeta+1)D_{22}^{a}D_{22}^{b}}{D_{22}^{a}+\zeta D_{22}^{b}}.\label{omD}
\end{equation}
Considering the case of isotropic phases, the components of the elasticity
tensor become $C_{1111}^{\varsigma}=C_{2222}^{\varsigma}=\frac{\tilde{E}_{\varsigma}}{1-\tilde{\nu}_{\varsigma}^{2}}$,
$C_{1122}^{\varsigma}=\frac{\tilde{\nu}_{\varsigma}\tilde{E}_{\varsigma}}{1-\tilde{\nu}_{\varsigma}^{2}}$,
$C_{1212}^{\varsigma}=\frac{\tilde{E}_{\varsigma}}{2(1+\tilde{\nu}_{\varsigma})}$,
(with $\varsigma=a,b$), where for plane-strain: $\tilde{E}_{\varsigma}=\frac{E_{\varsigma}}{1-\nu_{\varsigma}^{2}}$,
$\tilde{\nu}_{\varsigma}=\frac{\nu_{\varsigma}}{1-\nu_{\varsigma}}$,
whereas for plane-stress: $\tilde{E}_{\varsigma}=E_{\varsigma}$,
$\tilde{\nu}_{\varsigma}=\nu_{\varsigma}$, being $E_{\varsigma}$
the Young's modulus and $\nu_{\varsigma}$ the Poisson's ratio, respectively.
The components of the thermal dilatation and diffusive expansion tensors
take respectively the forms: $\alpha_{11}^{\varsigma}=\alpha_{22}^{\varsigma}=\alpha^{\varsigma}$,
$\beta_{11}^{\varsigma}=\beta_{22}^{\varsigma}=\beta^{\varsigma}$
(note that the coefficients $\alpha^{\varsigma}$ and $\beta^{\varsigma}$
can be expressed in terms of the linear isotropic thermal and diffusive
expansion coefficients and the elastic moduli \citep{Now1, NowB}.
The components of the heat conduction and mass diffusion tensors finally become $K_{11}^{\varsigma}=K_{22}^{\varsigma}=K^{\varsigma}$
and $D_{11}^{\varsigma}=D_{22}^{\varsigma}=D^{\varsigma}$. The overall elastic and thermodiffusive constants for the case of isotropic phases
are reported in Appendix \ref{appover}. 

By an asymptotic expansion of the constants (\ref{omC}), (\ref{omalpha}), (\ref{ombeta}), (\ref{omK}) and (\ref{omD}) in terms
of the concentration of the two constituents phases (not reported here for conciseness), it can be easily shown that, 
if the concentration of the phase $a$ vanishes, the overall elastic and thermodiffusive constants of the bi-phase layered material  
tend to the values corresponding to phase $b$. Conversely, if the concentration of the phase $a$ tends to one, the same expressions tend to 
the elastic and thermodiffusive constants of the phase $a$.

\begin{figure}[!htcb]
\centering

\includegraphics[width=60mm]{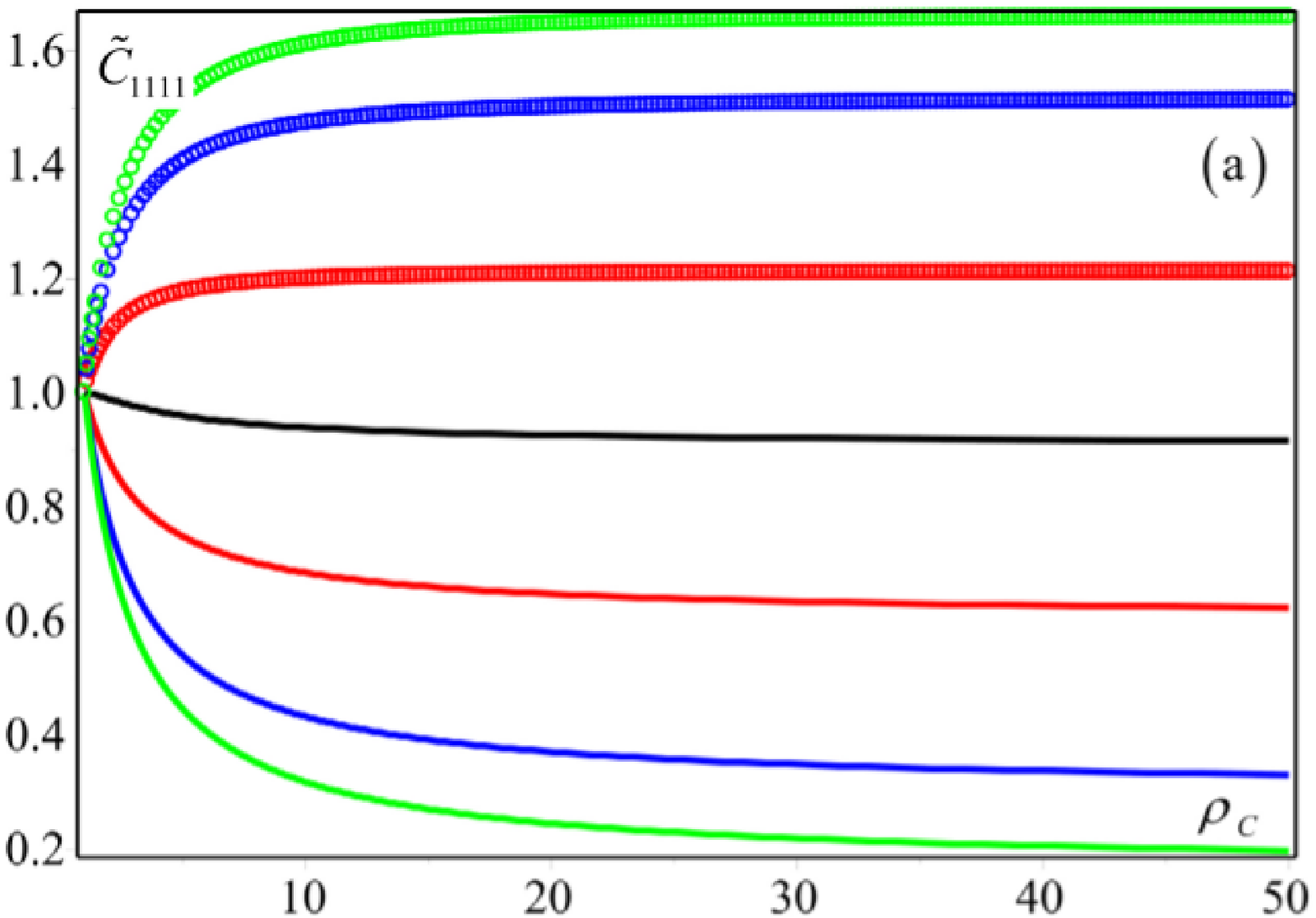}\hspace{5mm}\includegraphics[width=60mm]{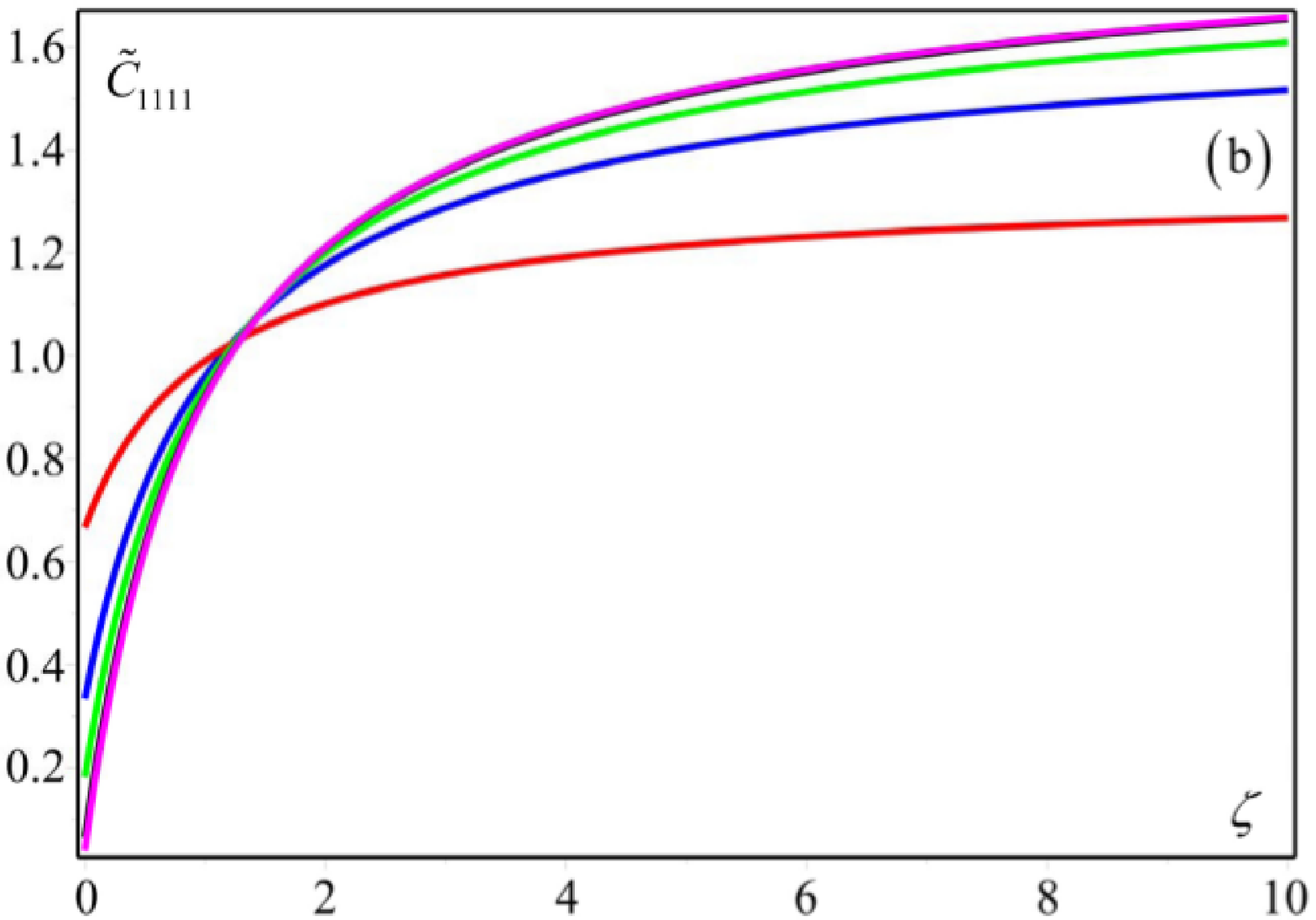}

\protect\protect\caption{(a) Dimensionless constant $\tilde{C}_{1111}$ vs. the ratio $\rho_{C}$
for $\tilde{\nu}_{a}=\tilde{\nu}_{b}=0.3$ and for different values of the geometric
ratio $\zeta$: $\zeta=1/10$ green line, $\zeta=1/5$ blue line,
$\zeta=1/2$ red line, $\zeta=1/1$ black line, $\zeta=2$ red points,
$\zeta=5$ blue points, $\zeta=10$ green points. (b) Dimensionless
constant $\tilde{C}_{1111}$ vs. the geometric ratio $\zeta$ for
$\tilde{\nu}_{a}=\tilde{\nu}_{b}=0.3$ and for different values of the ratio $\rho_{C}$:
$\rho_{C}=2$ red line, $\rho_{C}=5$ blue line, $\rho_{C}=10$ green
line, $\rho_{C}=30$ black line, $\rho_{C}=50$ violet line. }

\label{C1111} 
\end{figure}

\begin{figure}[!htcb]
\centering

\includegraphics[width=60mm]{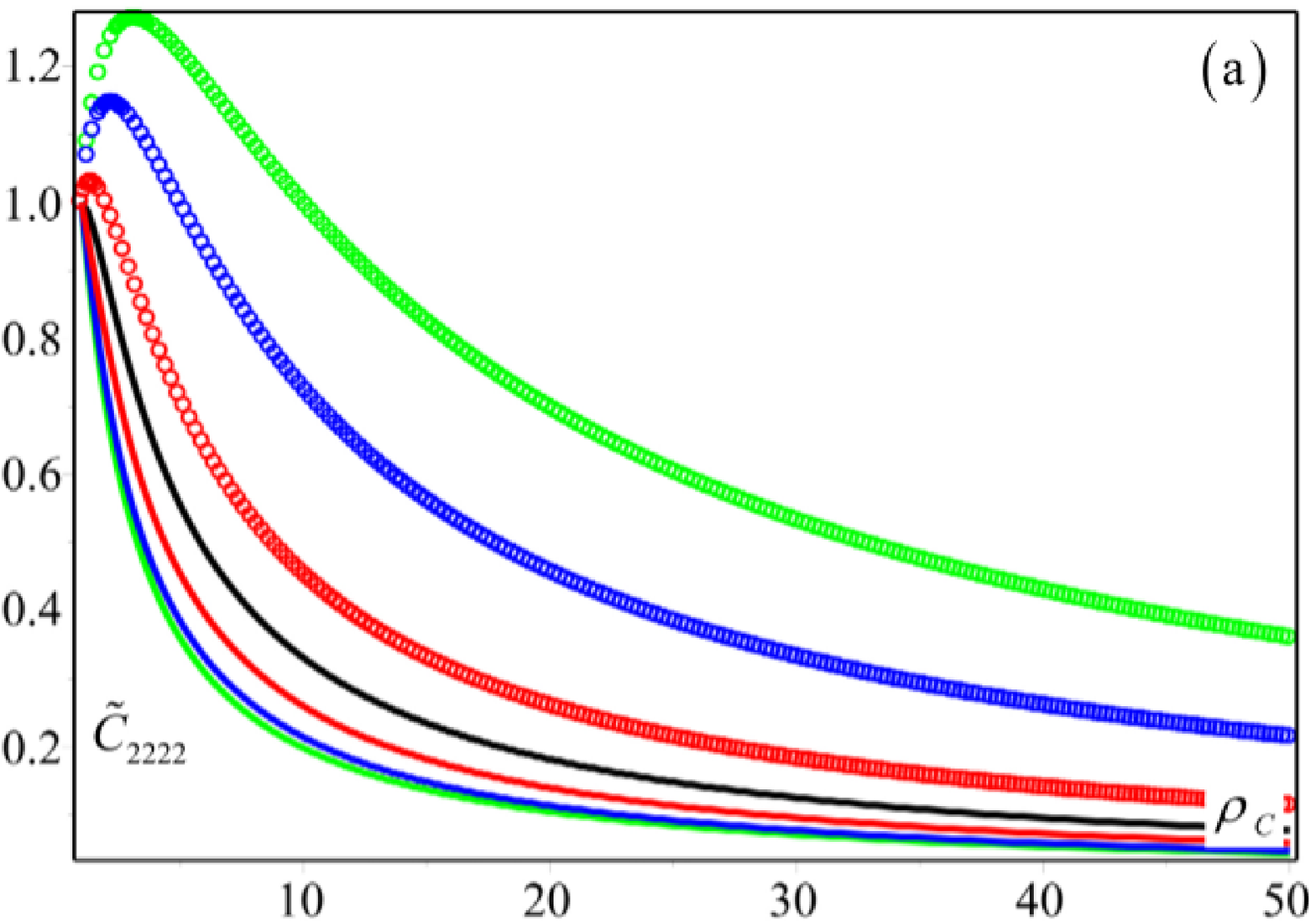}\hspace{5mm}\includegraphics[width=60mm]{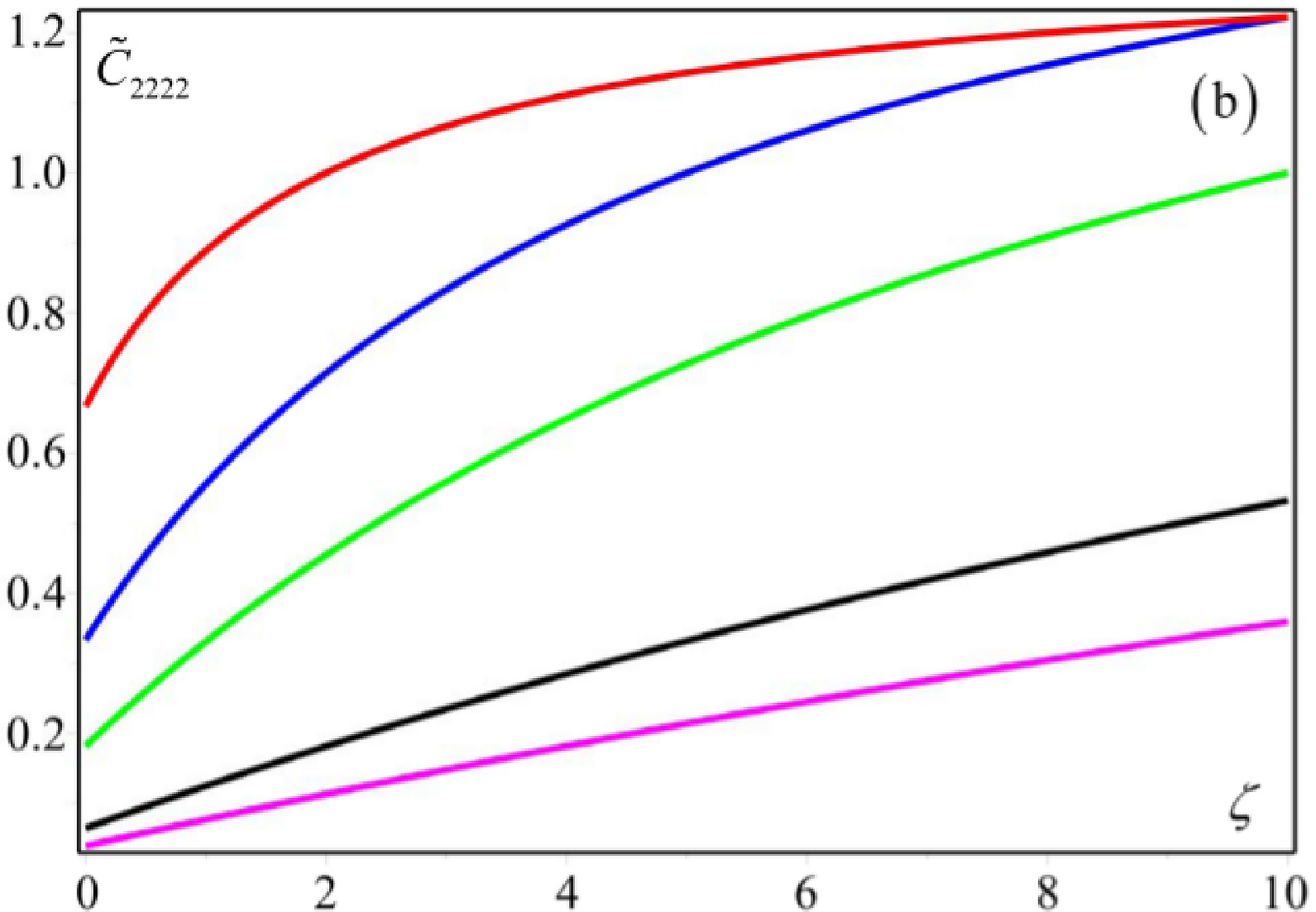}

\protect\protect\caption{(a) Dimensionless constant $\tilde{C}_{2222}$ vs. the ratio $\rho_{C}$
for $\tilde{\nu}_{a}=\tilde{\nu}_{b}=0.3$ and for different values of the geometric
ratio $\zeta$: $\zeta=1/10$ green line, $\zeta=1/5$ blue line,
$\zeta=1/2$ red line, $\zeta=1/1$ black line, $\zeta=2$ red points,
$\zeta=5$ blue points, $\zeta=10$ green points. (b) Dimensionless
constant $\tilde{C}_{2222}$ vs. the geometric ratio $\zeta$ for
$\tilde{\nu}_{a}=\tilde{\nu}_{b}=0.3$ and for different values of the ratio $\rho_{C}$:
$\rho_{C}=2$ red line, $\rho_{C}=5$ blue line, $\rho_{C}=10$ green
line, $\rho_{C}=30$ black line, $\rho_{C}=50$ violet line. }

\label{C2222} 
\end{figure}

In order to simplify the required computations, for the illustrative examples both the phases are assumed to be isotropic,
and then the overall elastic and thermodiffusive constants reported in Appendix \ref{appover} are used. These constants can be represented in the 
non-dimensional form:
\[
\tilde{C}_{iq_{1}pq_{2}}(\rho_{C},\zeta,\tilde{\nu}_{a},\tilde{\nu}_{b})=\frac{C_{iq_{1}pq_{2}}}{\hat{C}_{iq_{1}pq_{2}}}, \quad  
\tilde{\alpha}_{iq_{1}}(\rho_{C},\rho_{\alpha},\zeta,\tilde{\nu}_{a},\tilde{\nu}_{b})=\frac{\alpha_{iq_{1}}}{\hat{\alpha}_{iq_{1}}}, \quad 
\tilde{\beta}_{iq_{1}}(\rho_{C},\rho_{\beta},\zeta,\tilde{\nu}_{a},\tilde{\nu}_{b})=\frac{\beta_{iq_{1}}}{\hat{\beta}_{iq_{1}}},
\]
\begin{equation}
 \tilde{K}_{iq_{1}}(\rho_{K}, \zeta)=\frac{K_{iq_{1}}}{\hat{K}_{iq_{1}}}, \quad \tilde{D}_{iq_{1}}(\rho_{D}, \zeta)=\frac{D_{iq_{1}}}{\hat{D}_{iq_{1}}},
\label{constND}
 \end{equation}
where $\rho_{C}=\tilde{E}_{a}/\tilde{E}_{b}, \rho_{\alpha}=\tilde{\alpha}^{a}/\tilde{\alpha}^{b}, \rho_{\beta}=\tilde{\beta}^{a}/\tilde{\beta}^{b}, 
\rho_{K}=\tilde{K}^{a}/\tilde{K}^{b}, \rho_{D}=\tilde{D}^{a}/\tilde{D}^{b}$, and $\hat{C}_{iq_{1}pq_{2}}=(C_{iq_{1}pq_{2}}^{a}+C_{iq_{1}pq_{2}}^{b})/2,
\hat{\alpha}_{iq_{1}}=(\alpha_{iq_{1}}^{a}+\alpha_{iq_{1}}^{b})/2, \hat{\beta}_{iq_{1}}=(\beta_{iq_{1}}^{a}+\beta_{iq_{1}}^{b})/2, 
\hat{K}_{iq_{1}}=(K_{iq_{1}}^{a}+K_{iq_{1}}^{b})/2, \hat{D}_{iq_{1}}=(D_{iq_{1}}^{a}+D_{iq_{1}}^{b})/2$. It is important to note that if the Poisson's coefficients of the two phases 
are identical (i.e. $\nu_{a}=\nu_{b}$), the non-dimensional overall elastic and thermodiffusive constants (\ref{constND}) possess the following property:
\[
 \tilde{C}_{iq_{1}pq_{2}}(\rho_{C},\zeta,\tilde{\nu})=\tilde{C}_{iq_{1}pq_{2}}(\rho_{C}^{-1},\zeta^{-1},\tilde{\nu}), \quad
 \tilde{\alpha}_{iq_{1}}(\rho_{C},\rho_{\alpha},\zeta,\tilde{\nu})=\tilde{\alpha}_{iq_{1}}(\rho_{C}^{-1},\rho_{\alpha}^{-1},\zeta^{-1},\tilde{\nu}), 
\]
\[
 \tilde{\beta}_{iq_{1}}(\rho_{C},\rho_{\alpha},\zeta,\tilde{\nu})=\tilde{\beta}_{iq_{1}}(\rho_{C}^{-1},\rho_{\alpha}^{-1},\zeta^{-1},\tilde{\nu}), \quad
 \tilde{K}_{iq_{1}}(\rho_{K}, \zeta)=\tilde{K}_{iq_{1}}(\rho_{K}^{-1}, \zeta^{-1}),
\]
\begin{equation}
\tilde{D}_{iq_{1}}(\rho_{K}, \zeta)=\tilde{D}_{iq_{1}}(\rho_{K}^{-1}, \zeta^{-1}).
\end{equation}
The variation of the normalized components of the overall elasticity tensor $\tilde{C}_{1111}$ and $\tilde{C}_{2222}$ with the ratio $\rho_{C}$ is reported in Figs. \ref{C1111}$/(a)$ and \ref{C2222}$/(a)$, respectively. The same value of the
Poisson's coefficient $\tilde{\nu}=0.3$ has been assumed for both the phases, and  several values of the non-dimensional geometrical parameter $\zeta$ has been considered for the computations.
It can be observed that for $\rho_{C}=1$, corresponding to the case of two isotropic phases having identical elastic properties, 
the non-dimensional components of the overall elastic tensor assume the value $\tilde{C}_{iq_{1}pq_{2}}=1$ (i. e. $C_{iq_{1}pq_{2}}=C_{iq_{1}pq_{2}}^{a}=C_{iq_{1}pq_{2}}^{b}$).
In Figs. \ref{C1111}$/(b)$ and \ref{C2222}$/(b)$ the components $\tilde{C}_{1111}$ and $\tilde{C}_{2222}$ are plotted as functions of $\zeta$ for different values of $\rho_{C}$ considering the fixed Poisson's coefficient $\tilde{\nu}=0.3$ identical for both the phases.
For $\zeta\rightarrow 0$, the thickness of the phase $a$ vanishes. Consequently, the values of the overall elastic constants tends to those of the phase $b$: 
$C_{iq_{1}pq_{2}}=C_{iq_{1}pq_{2}}^{b}$, and the limit values assumed by the normalized components of the elastic tensor reported in the figures are $\tilde{C}_{iq_{1}pq_{2}}=2/(1+\rho_{C})$.
Conversely, for $\zeta\rightarrow+\infty$ the thickness of the phase $b$ tends to zero, and then $C_{iq_{1}pq_{2}}=C_{iq_{1}pq_{2}}^{a}$ and the non-dimensional constants plotted in Figs. 
\ref{C1111}$/(b)$-\ref{C2222}$/(b)$ assume the limit values $\tilde{C}_{iq_{1}pq_{2}}=2\rho_{C}/(1+\rho_{C}).$
Results (not reported here) show that the normalized components $\tilde{C}_{1212}$ and $\tilde{C}_{1122}$ have a behaviour very similar to that of the component $\tilde{C}_{2222}$.

\newpage

\begin{figure}[!htcb]
\centering

\includegraphics[width=60mm]{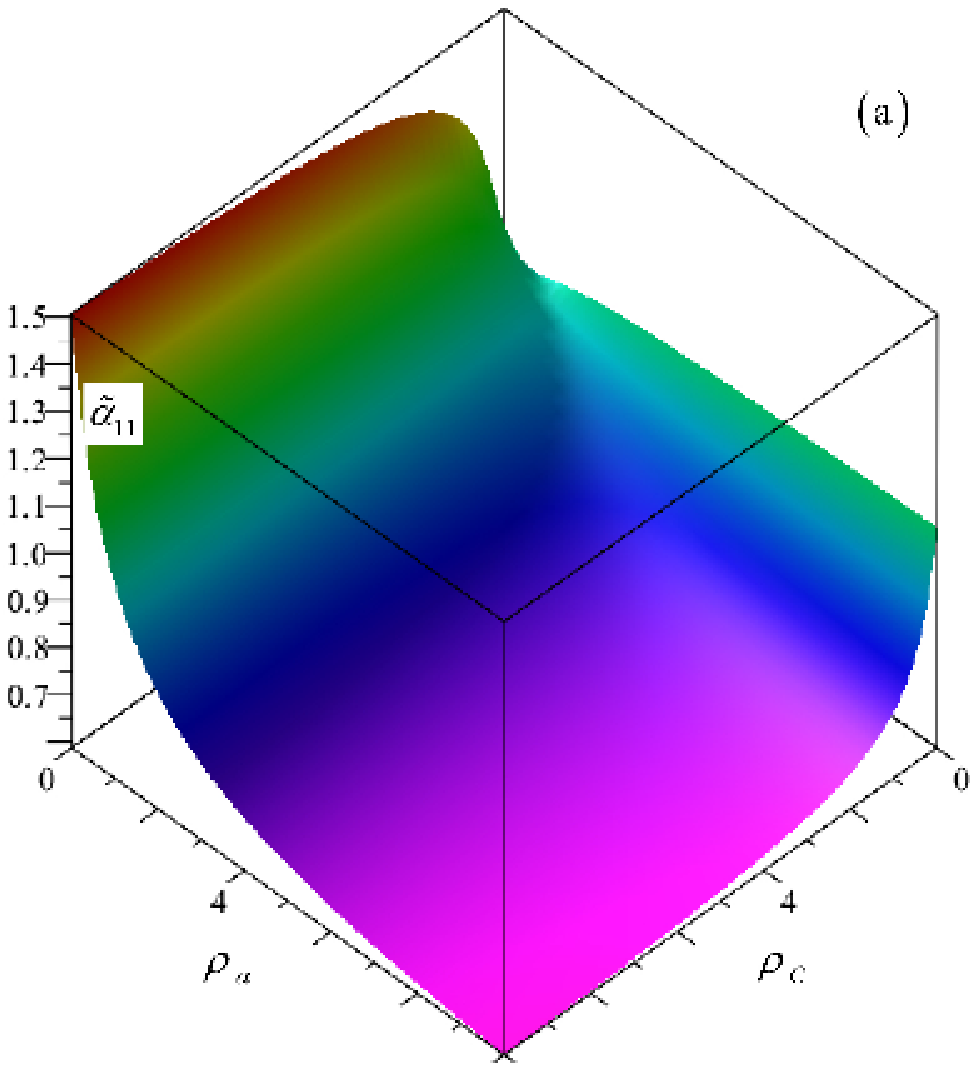}\hspace{5mm}\includegraphics[width=60mm]{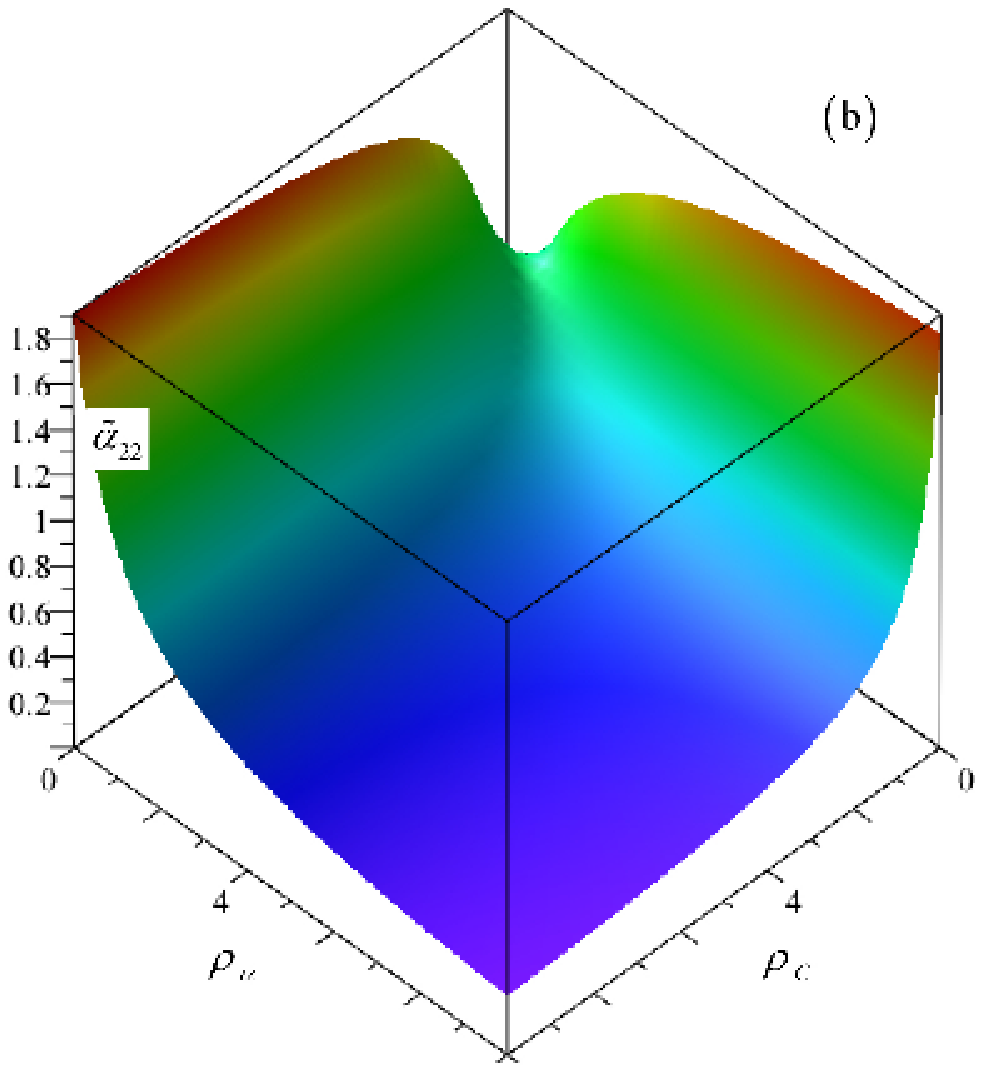}

\protect\protect\caption{Dimensionless component $\tilde{\alpha}_{11}$ (a), $\tilde{\alpha}_{22}$
(b) vs. the ratios $\rho_{C}$ and $\rho_{\alpha}$ for $\tilde{\nu}_{a}=\tilde{\nu}_{b}=0.3$
and $\zeta=1/2$. }

\label{alpha3d} 
\end{figure}

\begin{figure}[!htcb]
\centering

\includegraphics[width=60mm]{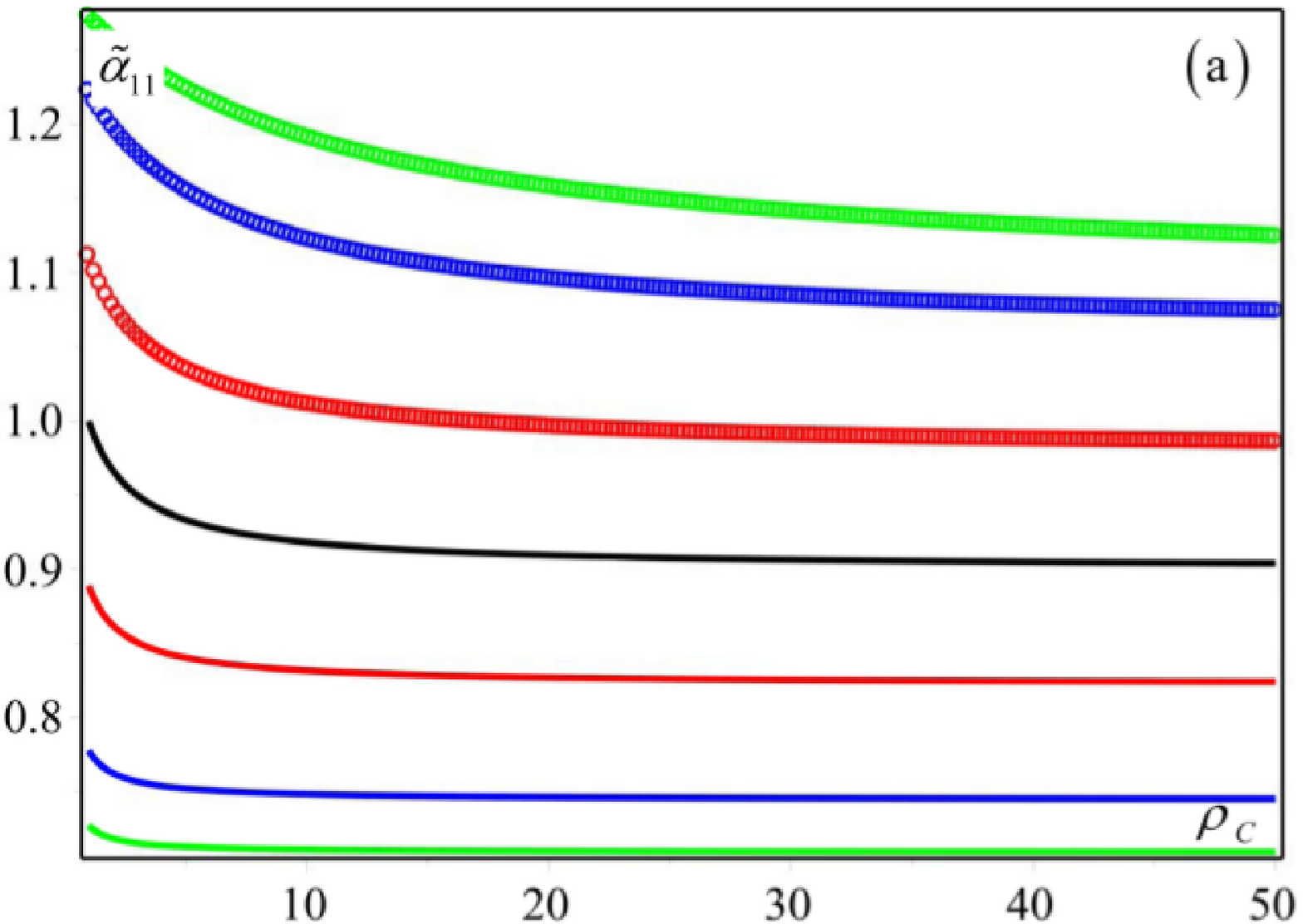}\hspace{5mm}\includegraphics[width=60mm]{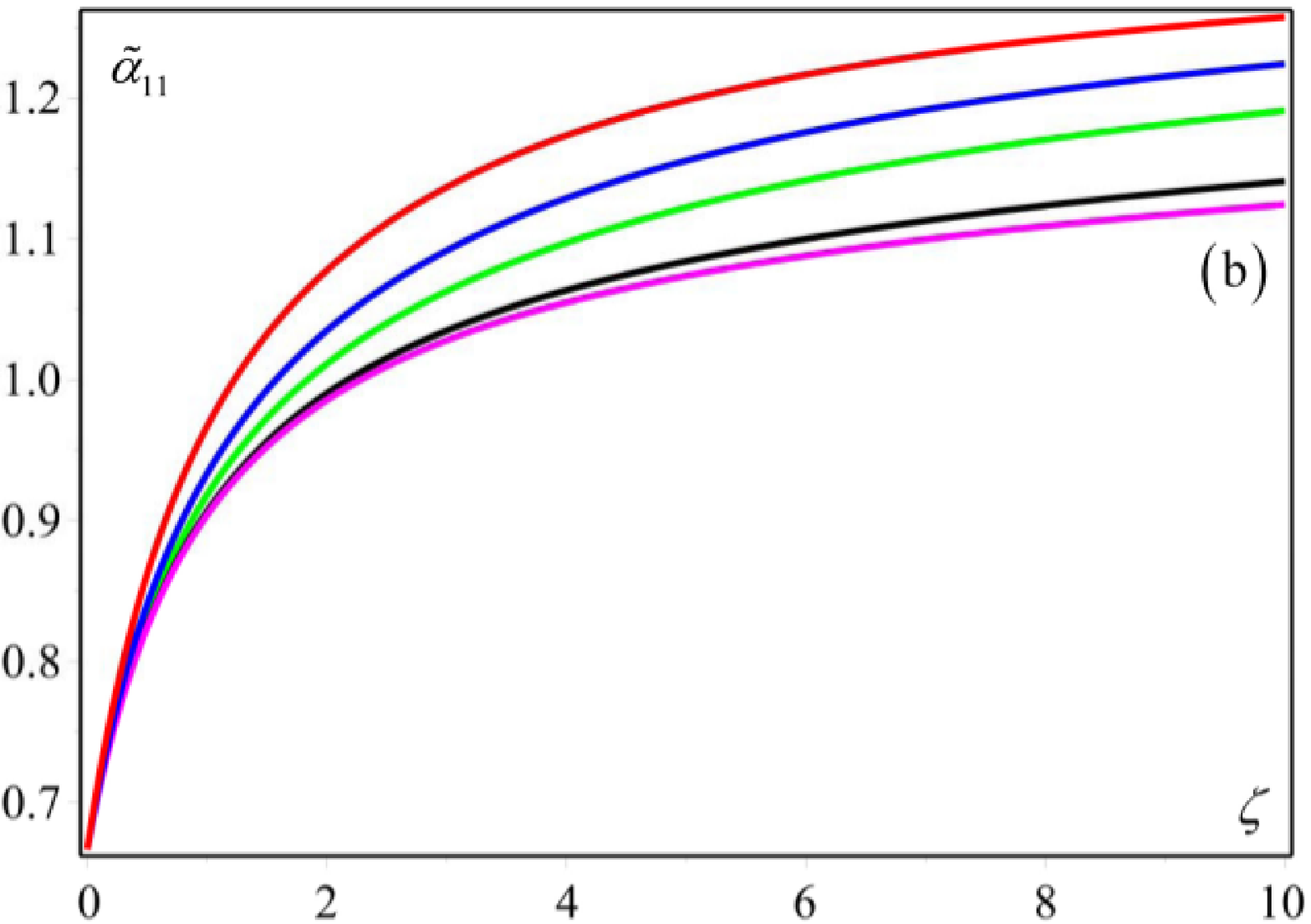}

\protect\protect\caption{Dimensionless component $\tilde{\alpha}_{11}$ vs. the ratio $\rho_{C}$ for
$\tilde{\nu}_{a}=\tilde{\nu}_{b}=0.3$, $\rho_{\alpha}=2$ and for different values
of the geometric ratio $\zeta$: $\zeta=1/10$ green line, $\zeta=1/5$
blue line, $\zeta=1/2$ red line, $\zeta=1/1$ black line, $\zeta=2$
red points, $\zeta=5$ blue points, $\zeta=10$ green points. (b)
Dimensionless component $\alpha_{11}$ vs. the ratio $\rho_{C}$ for
$\tilde{\nu}_{a}=\tilde{\nu}_{b}=0.3$, $\rho_{\alpha}=2$ and for different values
of the geometric ratio $\rho_{C}$: $\rho_{C}=2$ red line, $\rho_{C}=5$
blue line, $\rho_{C}=10$ green line, $\rho_{C}=30$ black line, $\rho_{C}=50$
violet line. }

\label{alpha11} 
\end{figure}

\begin{figure}[!htcb]
\centering

\includegraphics[width=60mm]{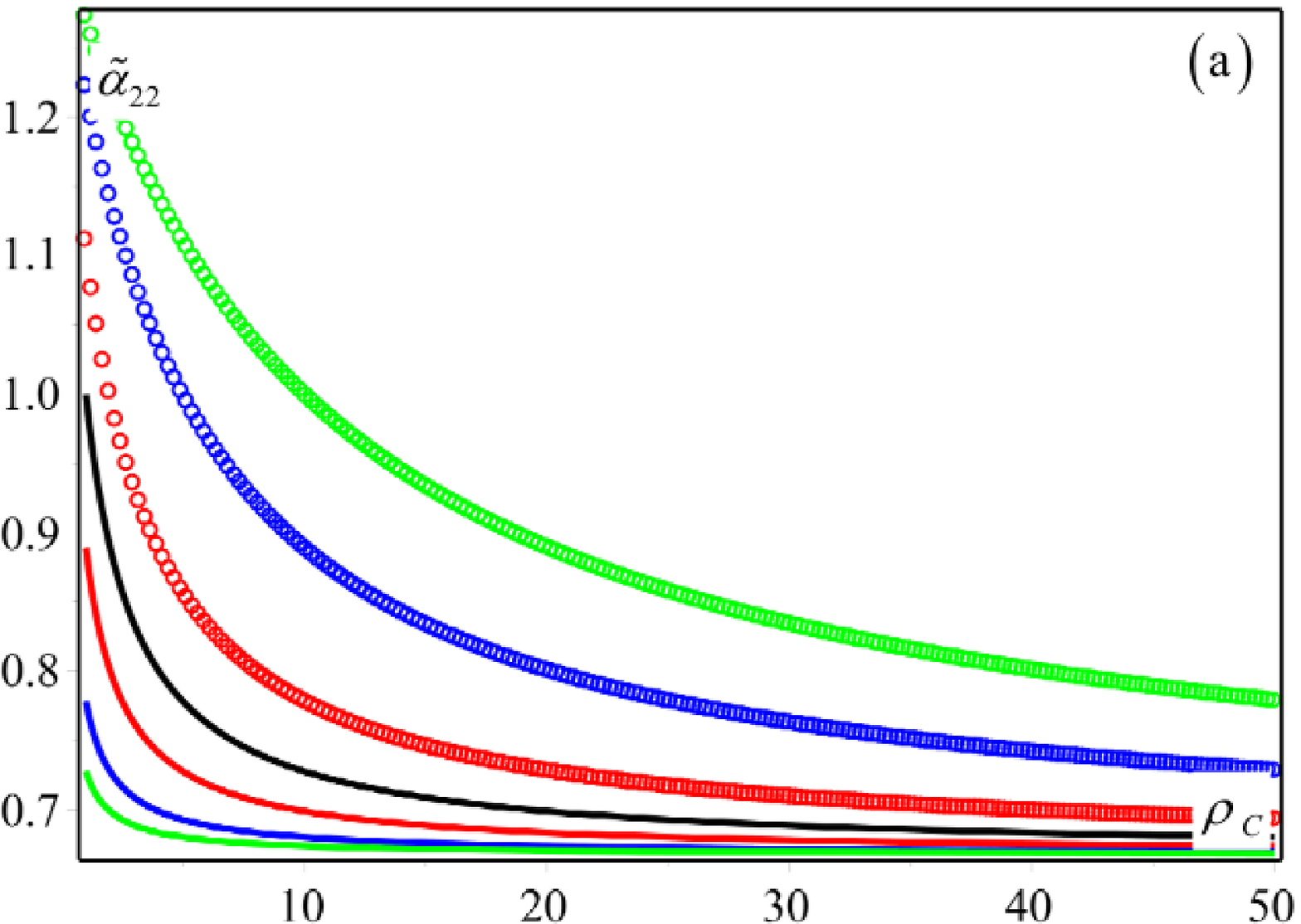}\hspace{5mm}\includegraphics[width=60mm]{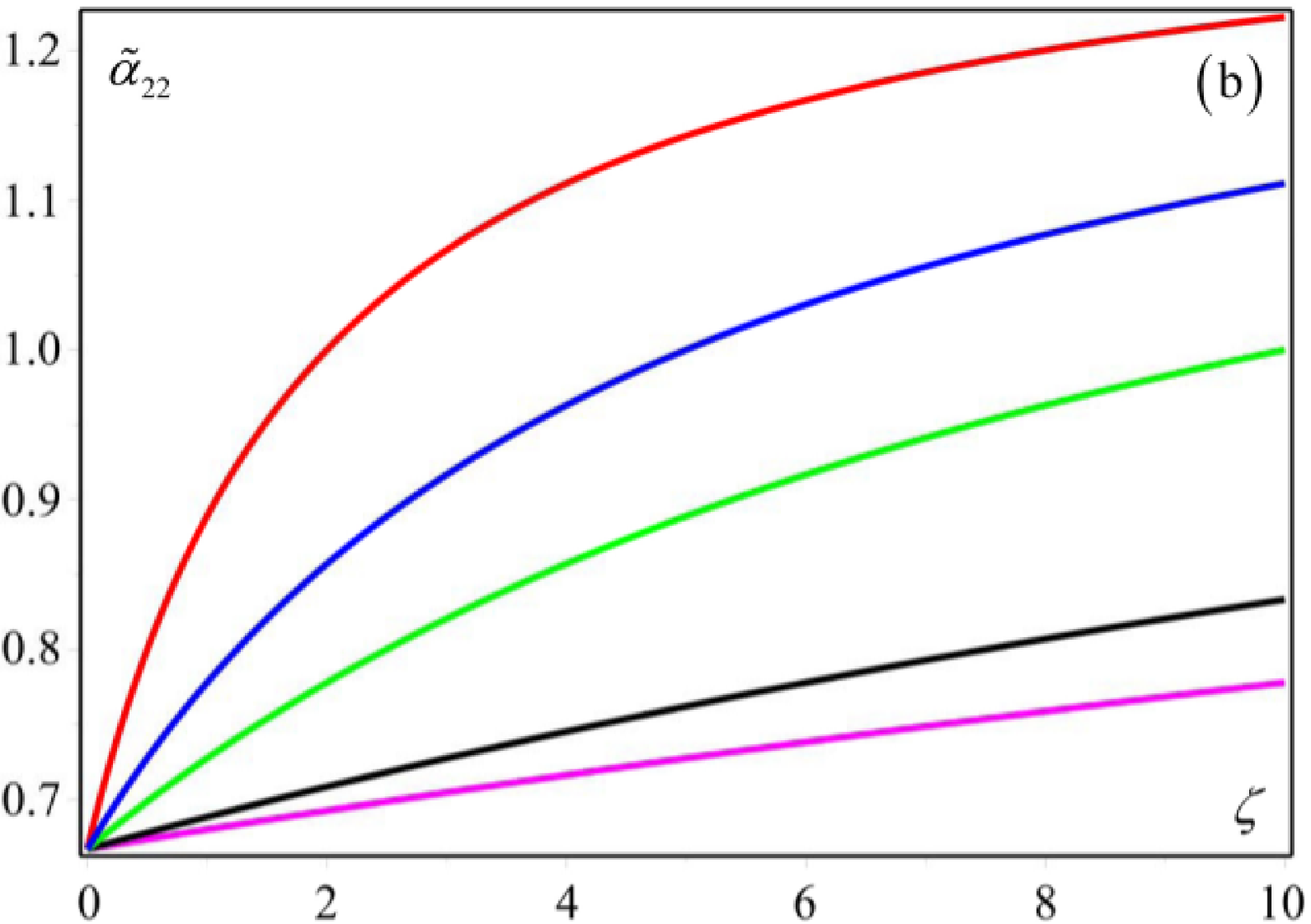}

\protect\protect\caption{Dimensionless component $\tilde{\alpha}_{22}$ vs. the ratio $\rho_{C}$ for
$\tilde{\nu}_{a}=\tilde{\nu}_{b}=0.3$, $\rho_{\alpha}=2$ and for different values
of the geometric ratio $\zeta$: $\zeta=1/10$ green line, $\zeta=1/5$
blue line, $\zeta=1/2$ red line, $\zeta=1/1$ black line, $\zeta=2$
red points, $\zeta=5$ blue points, $\zeta=10$ green points. (b)
Dimensionless component $\alpha_{22}$ vs. the ratio $\rho_{C}$ for
$\tilde{\nu}_{a}=\tilde{\nu}_{b}=0.3$, $\rho_{\alpha}=2$ and for different values
of the geometric ratio $\rho_{C}$: $\rho_{C}=2$ red line, $\rho_{C}=5$
blue line, $\rho_{C}=10$ green line, $\rho_{C}=30$ black line, $\rho_{C}=50$
violet line. }

\label{alpha22} 
\end{figure}

The three-dimensional plots reported in Fig.~\ref{alpha3d} show the variation of the normalized components of the overall 
thermal dilatation tensor $\tilde{\alpha}_{11}$ and $\tilde{\alpha}_{22}$ as functions of $\rho_{c}$ and $\rho_{\alpha}$, assuming $\tilde{\nu}=0.3$ for both the phases
and $\zeta=1$. In Figs. \ref{alpha11}$/(a)$ and \ref{alpha22}$/(a)$ the variation of $\tilde{\alpha}_{11}$ and $\tilde{\alpha}_{22}$ with the non-dimensional ratio $\rho_{C}$ is reported 
for several values of $\zeta$ assuming $\tilde{\nu}_{a}=\tilde{\nu}_{b}=\tilde{\nu}=0.3$ and $\rho_{\alpha}=2$. For $\rho_{C}=1$, corresponding to the case of two isotropic
phases with identical elastic constants but different thermal dilatation properties, the normalized components of the overall thermal dilatation tensor tend to the values 
$\tilde{\alpha}_{iq_{1}}=2(\zeta\rho_{\alpha}+1)/[(\rho_{\alpha}+1)(\zeta+1)]$ (i.e. $\alpha_{iq_{1}}=(\alpha_{iq_{1}}^{a}\zeta+\alpha_{iq_{1}}^{b})/(\zeta+1)$). In the case where $\rho_{C}=1$ and also $\rho_{\alpha}=1$, both the elastic and thermal dilatation tensors
of the two phases are identical, and then $\tilde{\alpha}_{iq_{1}}=1$. In Figs. \ref{alpha11}$/(b)$ and \ref{alpha22}$/(b)$ the same constants  $\hat{\alpha}_{11}$ and $\hat{\alpha}_{22}$
are plotted as functions of $\zeta$ for $\tilde{\nu}=0.3$, $\rho_{\alpha}=2$ and several different values of $\rho_{C}$. For $\zeta\rightarrow 0$,  
the thickness for the phase $a$ vanishes, and the elements of the overall thermal dilatation tensor tends to those of the phase $b$ (i.e. $\alpha_{iq_{1}}=\alpha_{iq_{1}}^{b}$). As it can be observed in the figures, in this 
case the normalized constants tend to a limit value which is the same for any value of $\rho_{C}$ (i.e. $\tilde{\alpha}_{iq_{1}}=2/(1+\rho_{\alpha})$). This value can be easily derived by using expressions for $\alpha_{11}$ and 
$\alpha_{22}$ reported in Appendix \ref{appover}. Conversely, for $\zeta\rightarrow+\infty$, the thickness of the layer $b$ tends to zero, the effective thermal dilatation constants tend to those
of the phase $a$ (i.e. $\alpha_{iq_{1}}=\alpha_{iq_{1}}^{a}$) and the normalized components  $\tilde{\alpha}_{11}$ and $\tilde{\alpha}_{22}$ reported in Figs. \ref{alpha11}$/(b)$ and \ref{alpha22}$/(b)$ assume the values 
$\tilde{\alpha}_{iq_{1}}=2\rho_{\alpha}/(1+\rho_{\alpha})$. The properties of the normalized elements of the overall diffusive expansion tensor $\tilde{\beta}_{11}$ and  $\tilde{\beta}_{22}$ 
are similar to those of $\tilde{\alpha}_{11}$ and  $\tilde{\alpha}_{22}$, and can be easily studied substituting the non-dimensional ratio $\rho_{\alpha}$ with $\rho_{\beta}$.

\begin{figure}[!htcb]
\centering

\includegraphics[width=60mm]{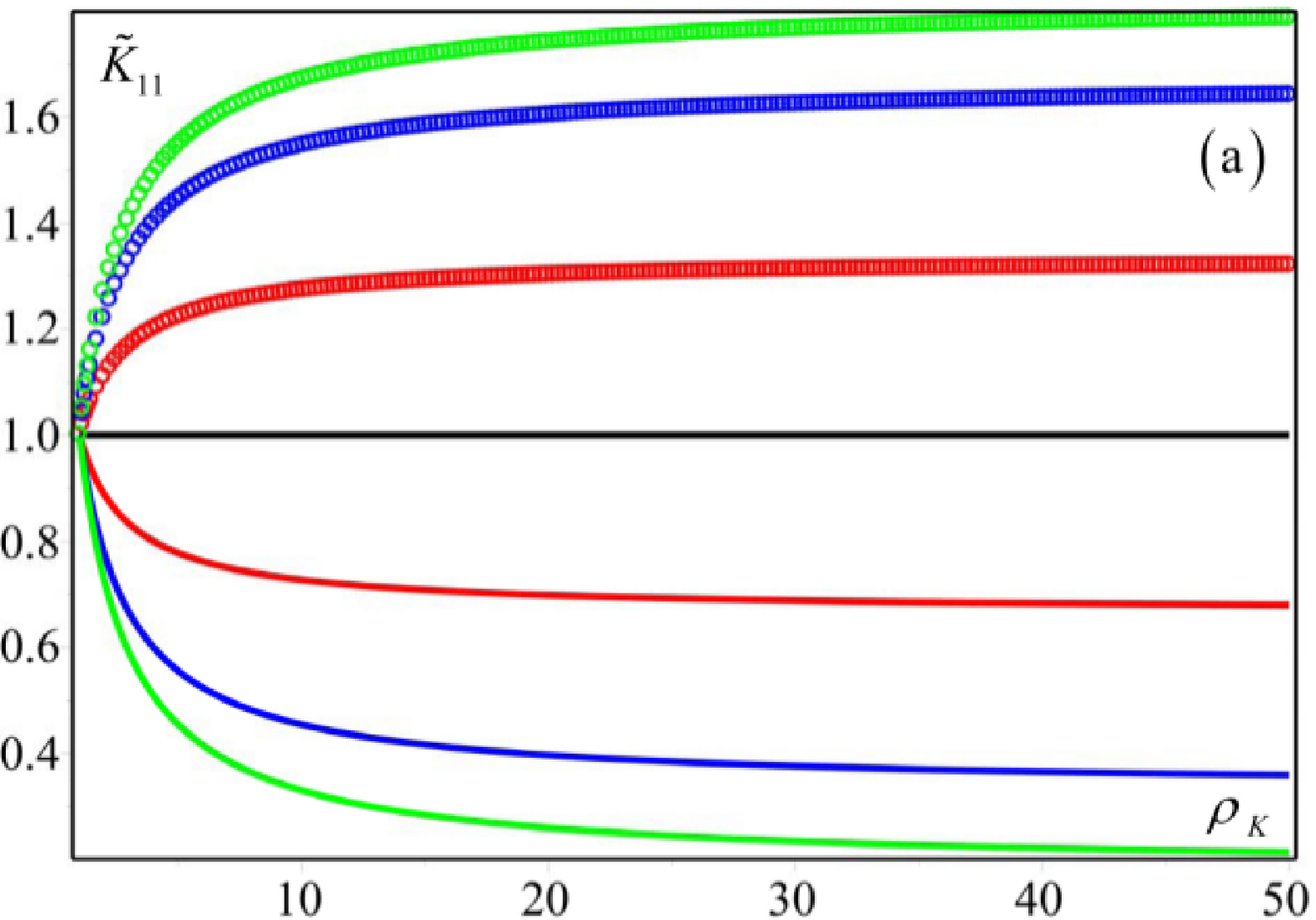}\hspace{5mm}\includegraphics[width=60mm]{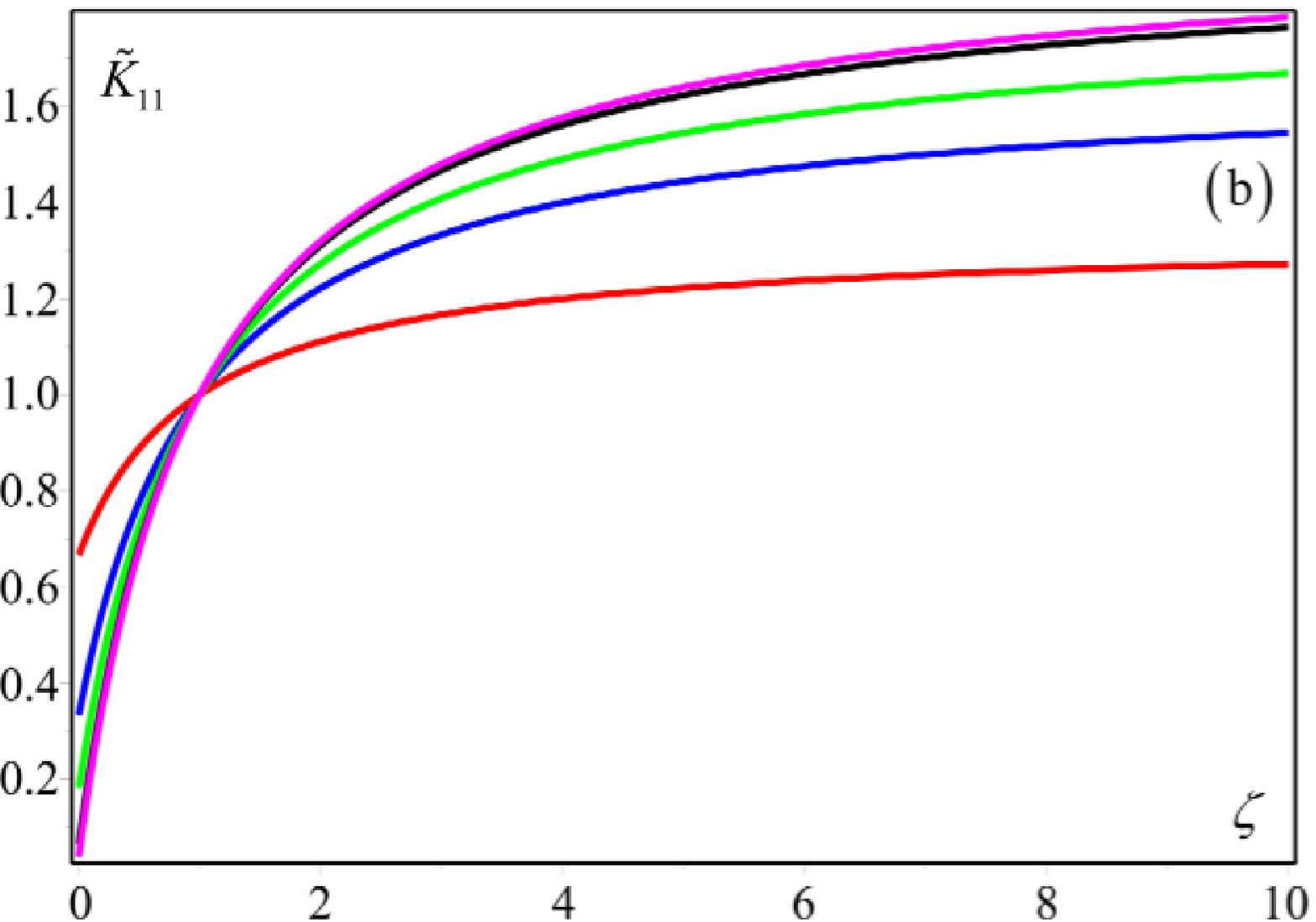}

\protect\protect\caption{Dimensionless component $\tilde{K}_{11}$ vs. the ratio $\rho_{K}$ for different
values of the geometric ratio $\zeta$: $\zeta=1/10$ green line,
$\zeta=1/5$ blue line, $\zeta=1/2$ red line, $\zeta=1/1$ black
line, $\zeta=2$ red points, $\zeta=5$ blue points, $\zeta=10$ green
points. (b) Dimensionless component $\tilde{K}_{11}$ vs. the ratio $\rho_{K}$
for different values of the ratio $\rho_{K}$: $\rho_{K}=2$ red line,
$\rho_{K}=5$ blue line, $\rho_{K}=10$ green line, $\rho_{K}=30$
black line, $\rho_{K}=50$ violet line. }

\label{K11} 
\end{figure}

\begin{figure}[!htcb]
\centering

\includegraphics[width=60mm]{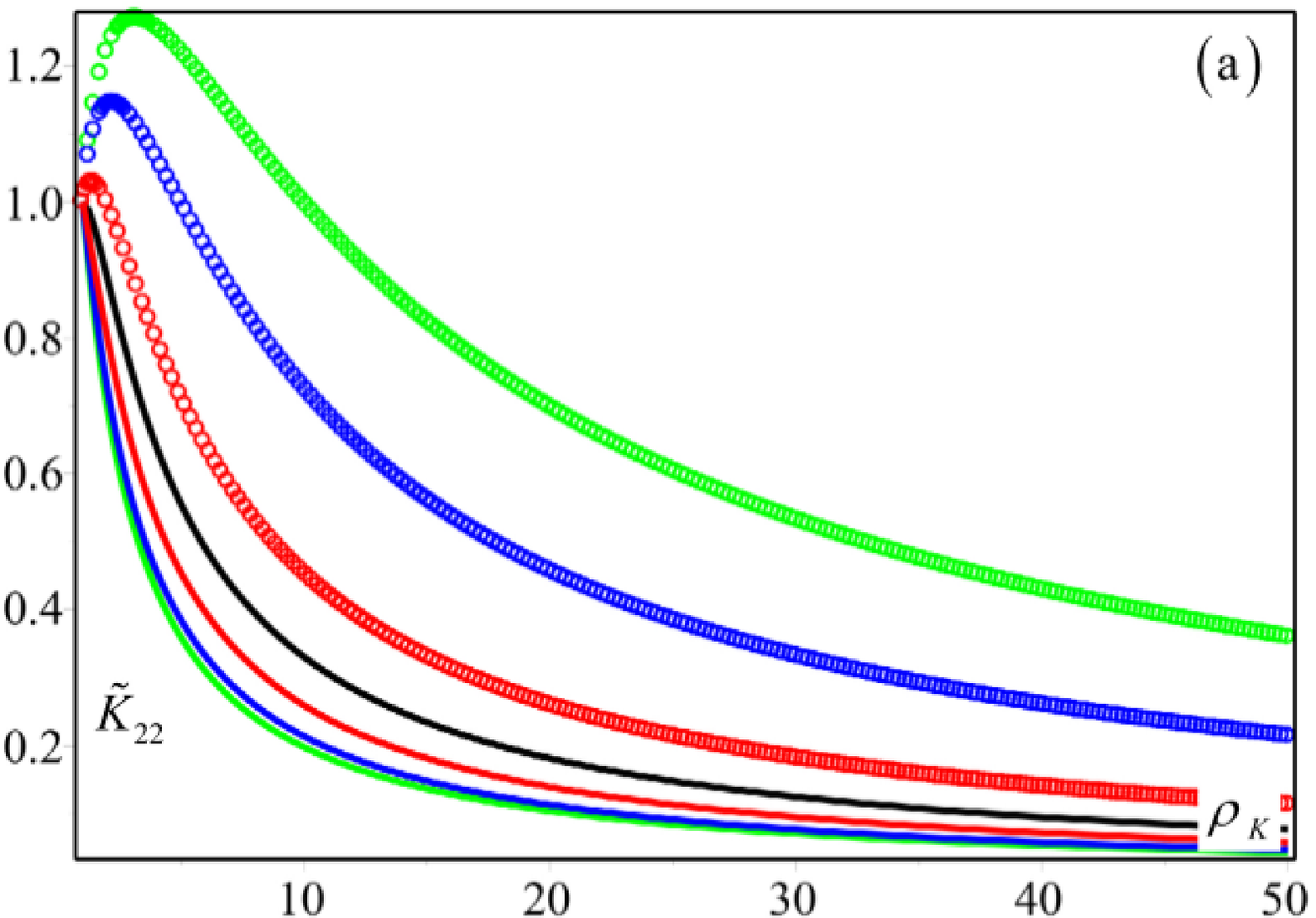}\hspace{5mm}\includegraphics[width=60mm]{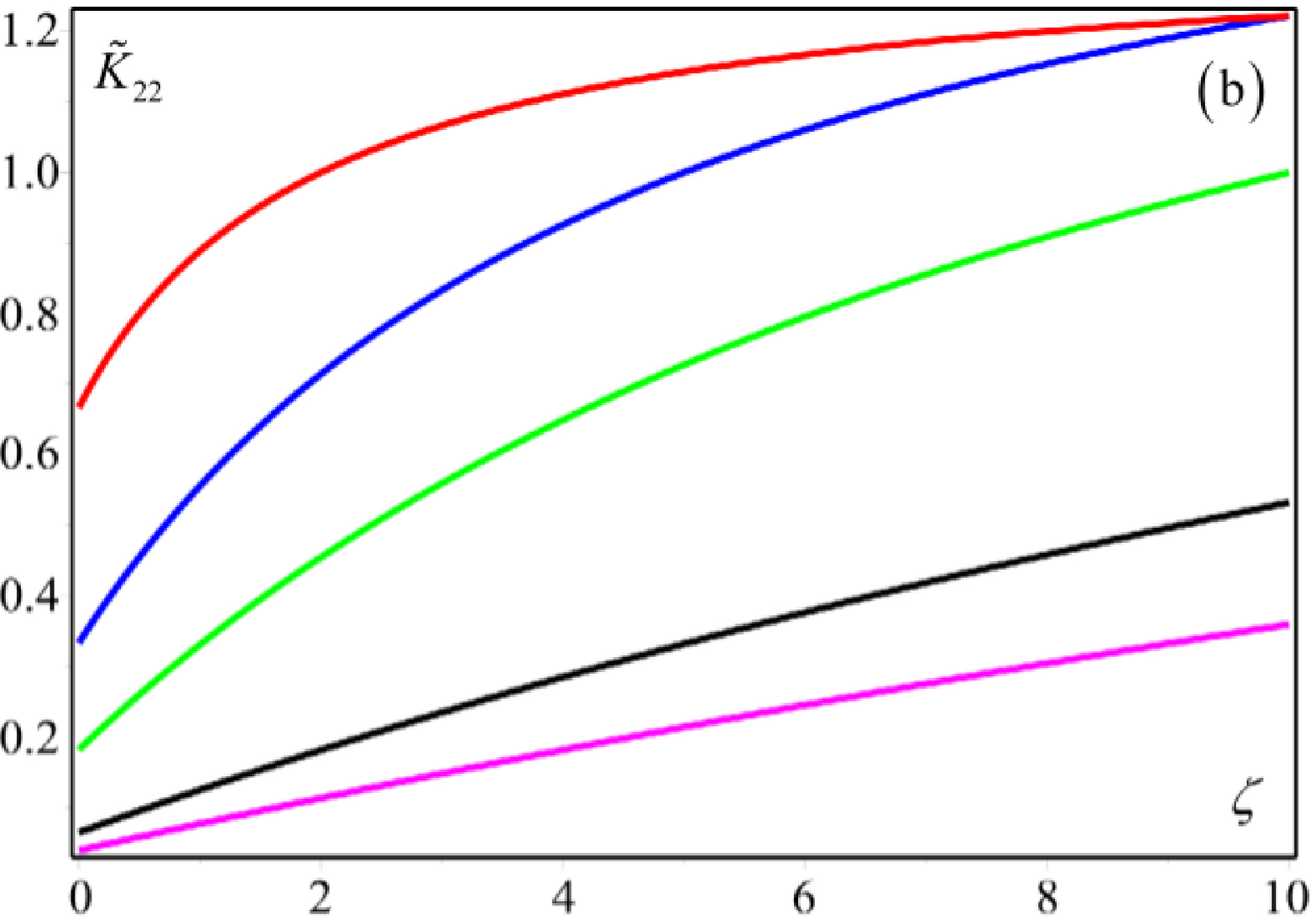}

\protect\protect\caption{Dimensionless component $\tilde{K}_{22}$ vs. the ratio $\rho_{K}$ for different
values of the geometric ratio $\zeta$: $\zeta=1/10$ green line,
$\zeta=1/5$ blue line, $\zeta=1/2$ red line, $\zeta=1/1$ black
line, $\zeta=2$ red points, $\zeta=5$ blue points, $\zeta=10$ green
points. (b) Dimensionless component $\tilde{K}_{22}$ vs. the ratio $\rho_{K}$
for different values of the ratio $\rho_{K}$: $\rho_{K}=2$ red line,
$\rho_{K}=5$ blue line, $\rho_{K}=10$ green line, $\rho_{K}=30$
black line, $\rho_{K}=50$ violet line. }

\label{K22} 
\end{figure}

The variation of the normalized components of the overall heat conduction tensor $\tilde{K}_{11}$ and $\tilde{K}_{22}$ with the non-dimensional
ratio $\rho_{K}$ are shown in Figs. \ref{K11}$/(a)$ and \ref{K22}$/(a)$. Several values of $\zeta$ have been assumed for the computations. It can be observed that for 
$\rho_{K}=1$, we have $\tilde{K}_{iq_{1}}=1$. This is due to the fact that the value $\rho_{K}=1$ corresponds to the case where the heat conduction of the two phases are
identical, and then $K_{iq_{1}}^{a}=K_{iq_{1}}^{b}$. In Figs. \ref{K11}$/(a)$ and \ref{K22}$/(b)$ the same non-dimensional components $\tilde{K}_{11}$ and $\tilde{K}_{22}$ 
are reported as functions of $\zeta$ for different values of $\rho_{K}$. As it is shown by these figures, for $\zeta\rightarrow 0$, 
$\tilde{K}_{11}$ and $\tilde{K}_{22}$ tends to a finite value depending on $\rho_{K}$. This limit correspond to the case of vanishing thickness of the layer $a$, where 
$K_{iq_{1}}=K_{iq_{1}}^{b}$, and then $\tilde{K}_{iq_{1}}=2/(1+\rho_{K})$. In the limit $\zeta\rightarrow +\infty$, 
for which the thickness of the layer $b$ vanishes and $K_{iq_{1}}=K_{iq_{1}}^{a}$, the normalized components of the overall heat conductivity tensor tend to a finite
value given by $\tilde{K}_{iq_{1}}=2\rho_{K}/(1+\rho_{K})$. The non-dimensional components of the overall mass diffusion tensor $\tilde{D}_{11}$ and $\tilde{D}_{22}$ are characterized by properties
similar to those of $\tilde{K}_{11}$ and  $\tilde{K}_{22}$, and can be studied substituting the non-dimensional ratio $\rho_{K}$ with $\rho_{D}$.

\vspace{-1mm}

\clearpage

\subsection{Comparative analysis: homogenized model vs heterogeneous material}

In order to study the capabilities of the proposed homogenization
procedure, the two-dimensional bi-phase orthotropic layered material
shown in Fig.~\ref{fig02} is assumed to be subjected to $\mL$-periodic
harmonic body forces $b_{i}$, directed along the orthotropy direction
$x_{j}$ (see Fig.~\ref{fig02}) and $\mL$-periodic heat and mass
sources $r(x_{j})$ and $s(x_{j})$: 
\begin{equation}
b_{j}(x_{j})=B_{j}e^{i\cfrac{2\pi mx_{j}}{L_{j}}},\quad r(x_{j})=Re^{i\cfrac{2\pi nx_{j}}{L_{j}}},\quad s(x_{j})=Se^{i\cfrac{2\pi px_{j}}{L_{j}}},\label{bsr}
\end{equation}
where: $j=1,2$; $L_{1}=L_{2}=L$; $B_{i},R,S\in\mathbb{C}$; $m,n,p\in\mathbb{Z}$;
and $i^{2}=-1$.

This problem is analyzed by applying the homogenized first-order model
with overall elastic and thermodiffusive constants derived from the
homogenization of the periodic cell through the approach developed
in previous Sections. The obtained results are then compared with
those derived by means of a fully heterogeneous modelling procedure.
Due to the periodicity of the heterogeneous material, body forces,
heat and mass sources considered, only an horizontal (or vertical)
characteristic portion of length $L$ of the heterogeneous model is
analyzed (Fig.~\ref{fig02}b). In order to assess the reliability
of the homogenized model, the macroscopic displacement, temperature
and chemical potential fields are compared to the corresponding fields
in the heterogeneous model by means of the up-scaling relations (\ref{upscaling}).

The overall elastic and thermodiffusive constants involving the fluctuation
functions are obtained in exact analytical forms via expressions (\ref{omC}),
(\ref{omalpha}), (\ref{ombeta}), (\ref{omD}), and (\ref{omK}).
Conversely, the solution of the heterogeneous problem with $\mL-$periodic
harmonic body forces is computed via FE analysis with periodic boundary
conditions on the displacement temperature and chemical potential
fields. For the considered two-dimensional body subject to body forces
along the orthotropy axes, heat and mass sources, the homogenized
field equations (\ref{ovmech})--(\ref{ovdiff})
take the form: 
\begin{equation}
C_{jjjj}\frac{\partial^{2}U_{j}}{\partial x_{j}^{2}}-\alpha_{jj}\frac{\partial\Theta}{\partial x_{j}}-\beta_{jj}\frac{\partial\Upsilon}{\partial x_{j}}+b_{j}=0, \label{ex_mech} \quad 
K_{jj}\frac{\partial^{2}\Theta}{\partial x_{j}^{2}}+r=0, \quad
D_{jj}\frac{\partial^{2}\Upsilon}{\partial x_{j}^{2}}+s=0,
\end{equation}
where $j=1,2$ are not summed indexes. Equations (\ref{ex_mech}) describe an extensional problem
in presence of thermodiffusion. Considering body forces, heat and
mass sources of the form (\ref{bsr}), the macroscopic displacements,
temperature and chemical potential fields are given by 
\begin{equation}
U_{j}(x_{j})=\frac{B_{j}L^{2}}{C_{jjjj}(2\pi m)^{2}}e^{i\frac{2\pi mx_{j}}{L_{j}}}-i\left[\frac{R\alpha_{jj}L^{3}}{C_{jjjj}K_{jj}(2\pi n)^{3}}e^{i\frac{2\pi nx_{j}}{L_{j}}}+\frac{S\beta_{jj}L^{3}}{C_{jjjj}D_{jj}(2\pi p)^{3}}e^{i\frac{2\pi px_{j}}{L_{j}}}\right],\label{UJJ}
\end{equation}
\begin{equation}
\Theta(x_{j})=\frac{RL^{2}}{K_{jj}(2\pi n)^{2}}e^{i\frac{2\pi nx_{j}}{L_{j}}},\quad
\Upsilon(x_{j})=\frac{SL^{2}}{D_{jj}(2\pi p)^{2}}e^{i\frac{2\pi px_{j}}{L_{j}}},\label{UpsilonJ}
\end{equation}
where $j=1,2$ are still not summed indexes. In order to compare the
behavior of the derived analytical solution with the numerical results
provided by the heterogeneous model, only the real part of macroscopic
fields (\ref{UJJ}), (\ref{UpsilonJ}) is accounted.
Moreover, the imaginary part of the amplitudes $B_{j}$, $R$ and
$S$ is assumed to be zero. The real part of expressions (\ref{UJJ}), (\ref{UpsilonJ}) can be written in the non-dimensional
form: 
\begin{equation}
U_{j}^{*}(x_{j})=\cos\left(\frac{2\pi mx_{j}}{L_{j}}\right)+\varXi_{jj}^{\alpha}\frac{m^{2}}{2\pi n^{3}}\sin\left(\frac{2\pi nx_{j}}{L_{j}}\right)+\varXi_{jj}^{\beta}\frac{m^{2}}{2\pi p^{3}}\sin\left(\frac{2\pi px_{j}}{L_{j}}\right),
\label{Uast}
\end{equation}
\begin{equation}
\Theta^{*}(x_{j})=\cos\left(\frac{2\pi nx_{j}}{L_{j}}\right),\quad\Upsilon^{*}(x_{j})=\cos\left(\frac{2\pi px_{j}}{L_{j}}\right),
\label{TUast}
\end{equation}
where $U_{j}^{*}=\frac{U_{j}C_{jjjj}(2\pi m)^{2}}{B_{j}L_{j}^{2}}$,
$\Theta^{*}=\frac{\Theta K_{jj}(2\pi n)^{2}}{RL_{j}^{2}}$, $\Upsilon^{*}=\frac{\Upsilon D_{jj}(2\pi p)^{2}}{SL_{j}^{2}}$;
and $\varXi_{jj}^{\alpha}=\frac{\alpha_{jj}RL_{j}}{K_{jj}B_{j}}$
and $\varXi_{jj}^{\beta}=\frac{\beta_{jj}SL_{j}}{D_{jj}B_{j}}$, $j=1,2$ are still not summed indexes.

\begin{figure}[!htcb]
\centering

\includegraphics[width=60mm]{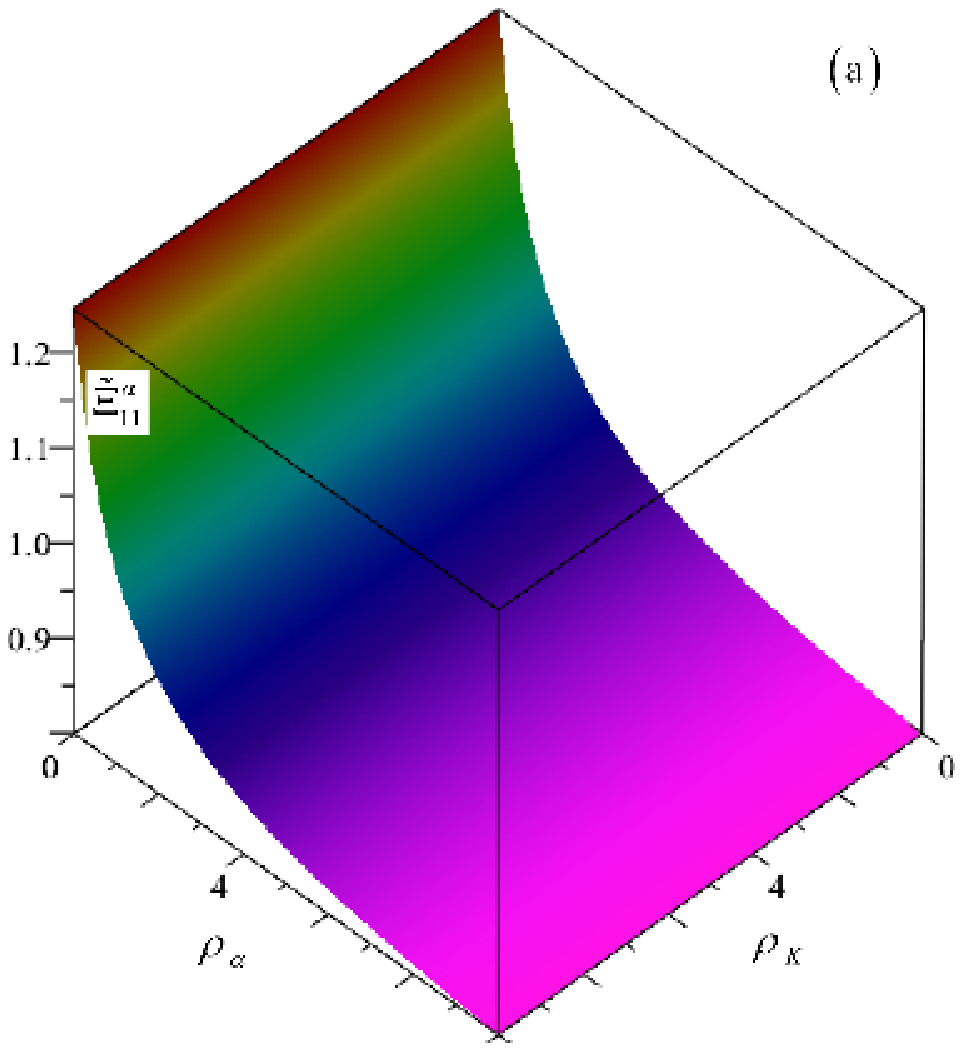}\hspace{10mm}\includegraphics[width=78mm]{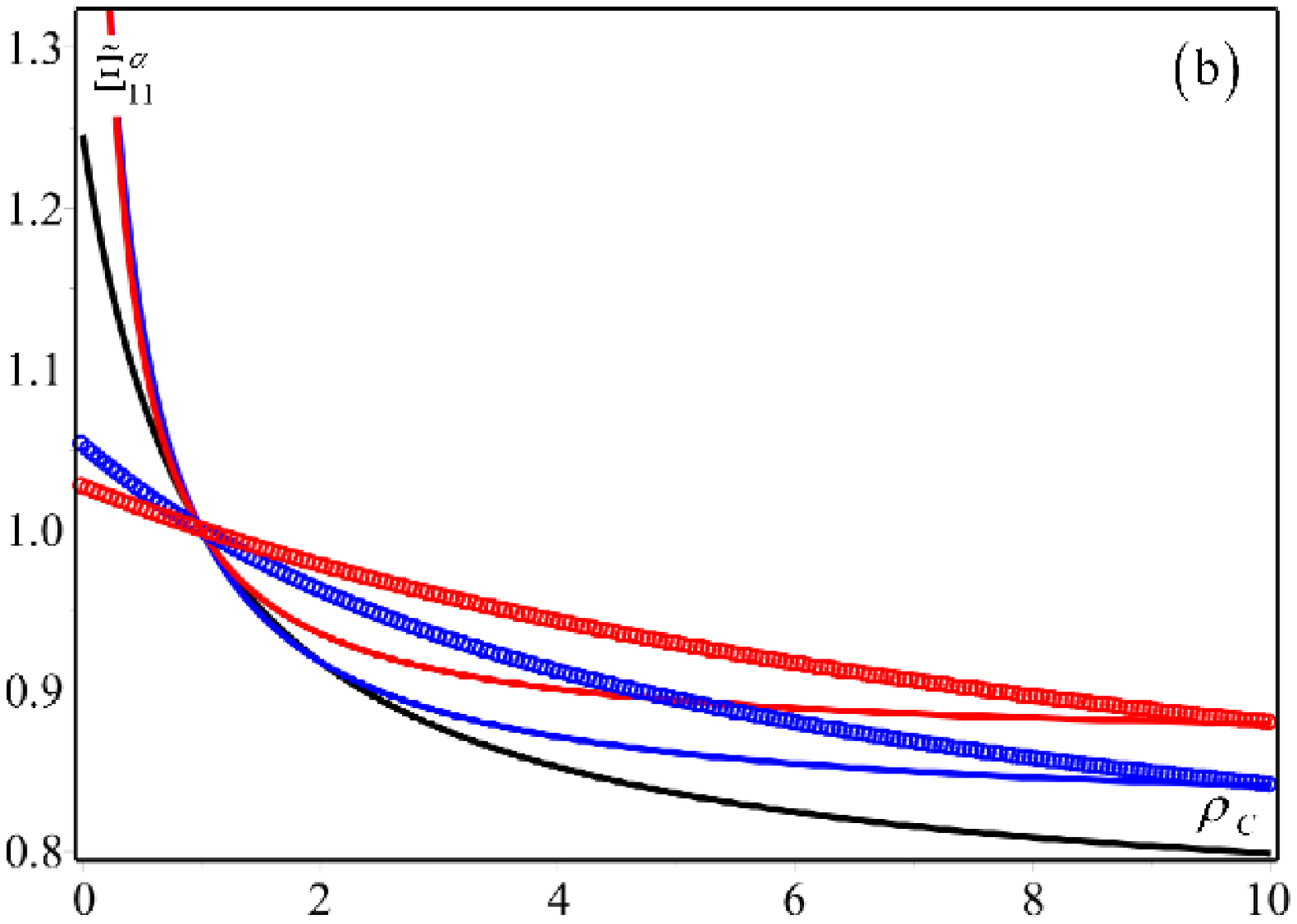}

\protect\protect\caption{(a) Dimensionless amplitude $\tilde{\Xi}{}_{11}^{\alpha}$ vs. the
ratios $\rho_{\alpha}$ and $\rho_{K}$ for $\tilde{\nu}_{a}=\tilde{\nu}_{b}=0.3$,
$\rho_{C}$=10 and $\zeta=1$. (b) Dimensionless amplitude
$\tilde{\Xi}{}_{jj}^{\alpha}$ vs. the ratios $\rho_{C}$ for $\tilde{\nu}_{a}=\tilde{\nu}_{b}=0.3$,
$\rho_{\alpha}=\rho_{K}=10$ and $\zeta=1$ for different
values of the geometric ratio $\zeta$: $\zeta=1/10$ red line, $\zeta=1/5$
blue line, $\zeta=1$ black line, $\zeta=5$ blue line points, $\zeta=10$
red line points. }

\label{Amacro11alpha} 
\end{figure}

\begin{figure}[!htcb]
\centering
\vspace{8mm}
\includegraphics[width=60mm]{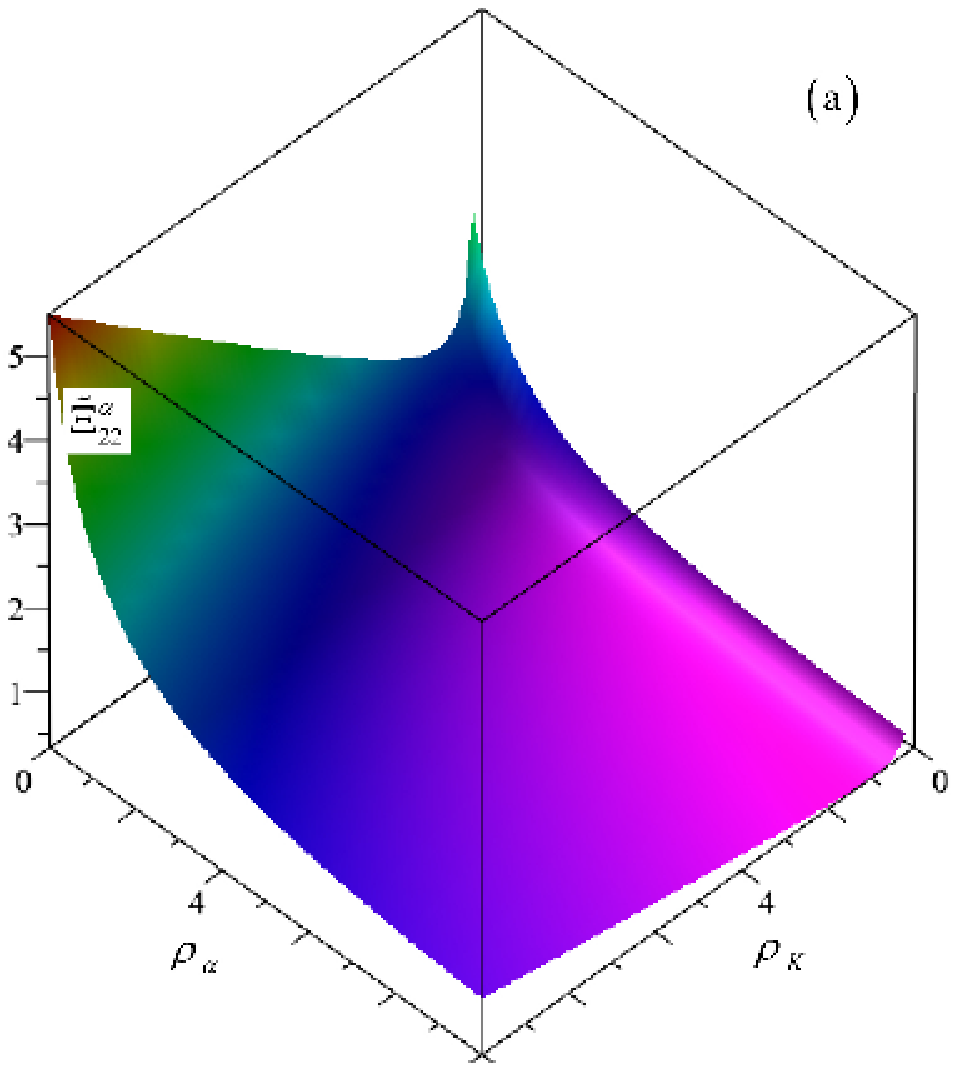}\hspace{10mm}\includegraphics[width=78mm]{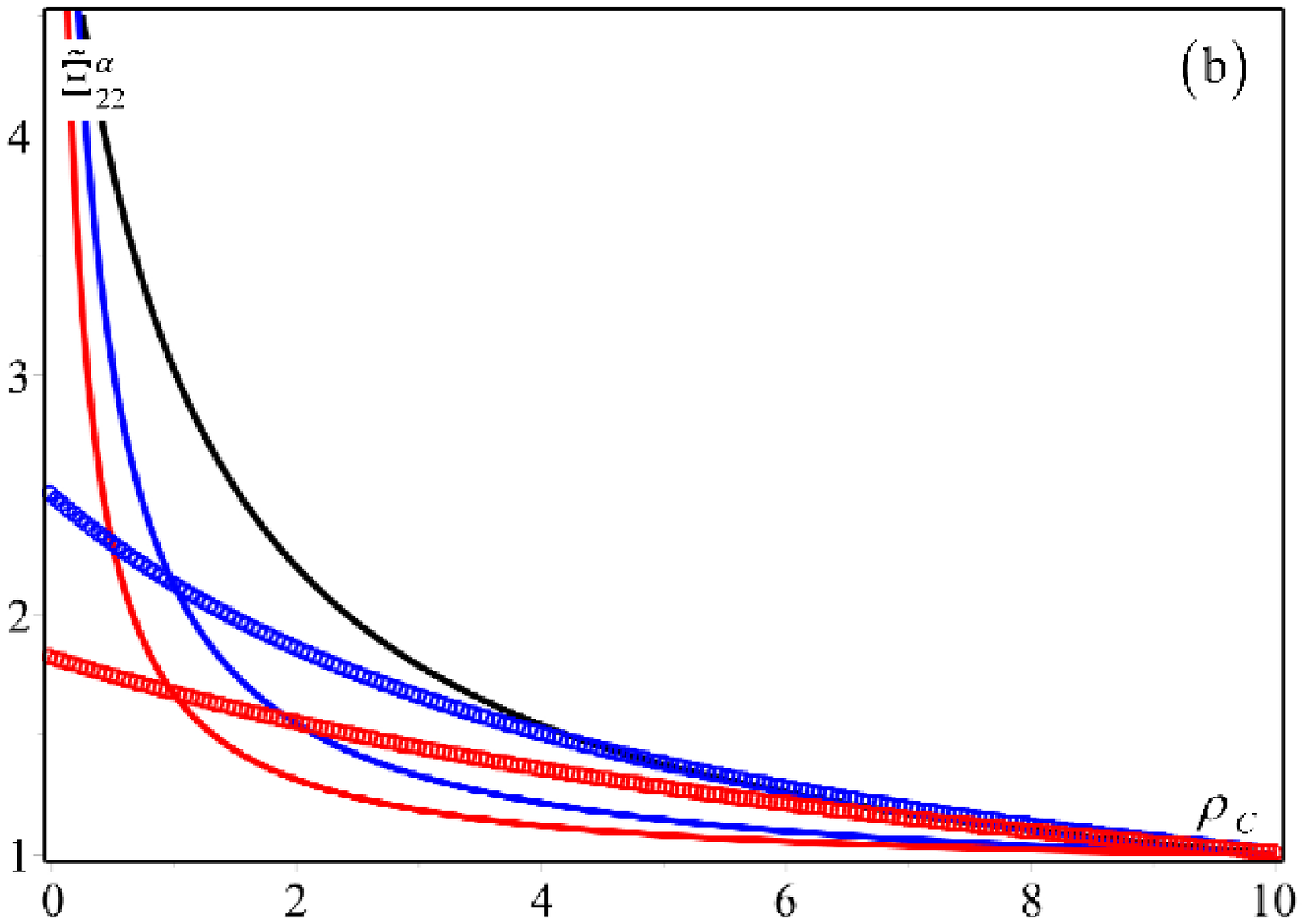}

\protect\protect\caption{(a) Dimensionless amplitude $\tilde{\Xi}{}_{22}^{\alpha}$ vs. the
ratios $\rho_{\alpha}$ and $\rho_{K}$ for $\tilde{\nu}_{a}=\tilde{\nu}_{b}=0.3$,
$\rho_{C}$=10 and $\zeta=1$. (b) Dimensionless amplitude
$\tilde{\Xi}{}_{22}^{\alpha}$ vs. the ratios $\rho_{C}$ for $\tilde{\nu}_{a}=\tilde{\nu}_{b}=0.3$,
$\rho_{\alpha}=\rho_{K}=10$ and $\zeta=1$ for different
values of the geometric ratio $\zeta$: $\zeta=1/10$ red line, $\zeta=1/5$
blue line, $\zeta=1$ black line, $\zeta=5$ blue line points, $\zeta=10$
red line points. }

\label{Amacro22alpha} 
\end{figure}

The amplitude functions $\varXi_{jj}^{\alpha}$ and $\varXi_{jj}^{\beta}$ are associated respectively with the thermal expansion and mass diffusion contribution
to the macroscopic displacement (\ref{Uast}) along the direction $\mathbf{e}_{j}$. In order to study the influence of the geometrical, elastic and thermodiffusive properties of the phases
on $\varXi_{jj}^{\alpha}$ and $\varXi_{jj}^{\beta}$, the following non-dimensional form for these functions is introduced:
\begin{equation}
 \tilde{\varXi}_{jj}^{\alpha}(\rho_{C}, \rho_{\alpha},\rho_{K}, \zeta, \tilde{\nu}_{a}, \tilde{\nu}_{b})
 =\frac{\varXi_{jj}^{\alpha}}{\hat{\varXi}_{jj}^{\alpha}}, \quad \tilde{\varXi}_{jj}^{\beta}(\rho_{C}, \rho_{\beta},\rho_{K}, \zeta, \tilde{\nu}_{a}, \tilde{\nu}_{b})
 =\frac{\varXi_{jj}^{\beta}}{\hat{\varXi}_{jj}^{\beta}}, \quad \mbox{with} \ j=1,2, 
\end{equation}
where $\hat{\varXi}_{jj}^{\alpha}=RL_{j}(\alpha^{a}+\alpha^{b})/B_{j}(K^{a}+K^{b})$ and 
$\hat{\varXi}_{jj}^{\beta}=RL_{j}(\beta^{a}+\beta^{b})/B_{j}(D^{a}+D^{b})$. In the case where the two phases possess the same value of the Poisson's coefficient ($\tilde{\nu}_{a}=\tilde{\nu}_{b}$),
the following property is verified for the normalized amplitude functions:
\begin{equation}
 \tilde{\varXi}_{jj}^{\alpha}(\rho_{C}, \rho_{\alpha},\rho_{K}, \zeta, \tilde{\nu})=\tilde{\varXi}_{jj}^{\alpha}(\rho_{C}^{-1}, \rho_{\alpha}^{-1},\rho_{K}^{-1}, \tilde{\nu}), \quad
 \tilde{\varXi}_{jj}^{\alpha}(\rho_{C}, \rho_{\beta},\rho_{K}, \zeta, \tilde{\nu})=\tilde{\varXi}_{jj}^{\beta}(\rho_{C}^{-1}, \rho_{\alpha}^{-1},\rho_{K}^{-1}, \tilde{\nu}).
\end{equation}
The three-dimensional plot reported in Fig.~\ref{Amacro11alpha}$/(a)$ shows the variation of the normalized amplitude component $\tilde{\varXi}_{11}^{\alpha}$ with $\rho_{\alpha}$
and $\rho_{K}$ for $\tilde{\nu}_{a}=\tilde{\nu}_{b}=0.3$, $\rho_{C}=10$ and $\zeta=1$. In Fig.~\ref{Amacro11alpha}$/(b)$ the same component $\tilde{\varXi}_{11}^{\alpha}$, which represents
the contribution of the thermal expansion to the macroscopic displacement along the direction $\mathbf{e}_{1}$, is plotted as a function of $\rho_{C}$ assuming
$\tilde{\nu}_{a}=\tilde{\nu}_{b}=0.3$, $\rho_{\alpha}=\rho_{K}=10$ and several values of the dimensionless ratio $\zeta$. The variation of the normalized component $\tilde{\varXi}_{22}^{\alpha}$,
corresponding to the contribution of the thermal expansion to the macroscopic displacement along $\mathbf{e}_{2}$ (see Fig.~\ref{fig02}), is reported in  Fig.~\ref{Amacro22alpha}$/(a)$
as a function of $\rho_{\alpha}$ and $\rho_{K}$ for $\tilde{\nu}_{a}=\tilde{\nu}_{b}=0.3$, $\rho_{C}=10$ and $\zeta=1$, and in Fig.~\ref{Amacro22alpha}$/(b)$ as a function of $\rho_{C}$ assuming
$\tilde{\nu}_{a}=\tilde{\nu}_{b}=0.3$, $\rho_{\alpha}=\rho_{K}=10$. Observing the curves reported in the figures, it can be noted that the normalized amplitude $\tilde{\varXi}_{22}^{\alpha}$,
associate to the component of the macroscopic displacement $U^{*}_{2}(x_{2})$ parallel to the stratification direction, is greater than the amplitude $\tilde{\varXi}_{11}^{\alpha}$ which
correspond to the component $U^{*}_{1}(x_{1})$ parallel to the stratification direction. The dimensionless amplitude  $\tilde{\varXi}_{11}^{\beta}$ and $\tilde{\varXi}_{22}^{\beta}$,
associated to the contributions of the mass diffusion respectively to $U^{*}_{1}(x_{1})$ and $U^{*}_{2}(x_{2})$, are characterized by the same properties of $\tilde{\varXi}_{11}^{\alpha}$
and $\tilde{\varXi}_{22}^{\alpha}$, and their behavior can be easily studied substituting the non-dimensional ratios $\rho_{\alpha}$ and $\rho_{K}$ with $\rho_{\beta}$ and $\rho_{D}$.
\\

\newpage

The analytical solution (\ref{Uast}) and (\ref{TUast}), derived by the solution of the homogenized field equations (\ref{ex_mech}) 
is now compared with the results obtained by the finite element analysis of the heterogeneous
problem corresponding to the bi-phase layered material reported in Fig.~\ref{fig02} subject to harmonic body forces,
heat and mass sources. More precisely, finite element analysis of the heterogeneous problem, performed by means of the program COMSOL Multiphysics, provides the local fields $u_{j}$, $\theta$, $\eta$ which
are used together with the \emph{up-scaling} relations (\ref{upscaling}) for obtaining the macro-scopic fields $U_{j}$, $\Theta$ and $\Upsilon$. These macro-scopic quantities are compared with 
the analytical expressions  (\ref{Uast}) and (\ref{TUast}). Plane stress condition has been assumed for both the solution of the homogenized equations and the heterogeneous problem, and 
two isotropic phases with the same value of the Poisson's coefficient $\nu_{a}=\nu_{b}=0.3$ have been considered. 

\begin{figure}[!htcb]
\centering

\includegraphics[width=60mm]{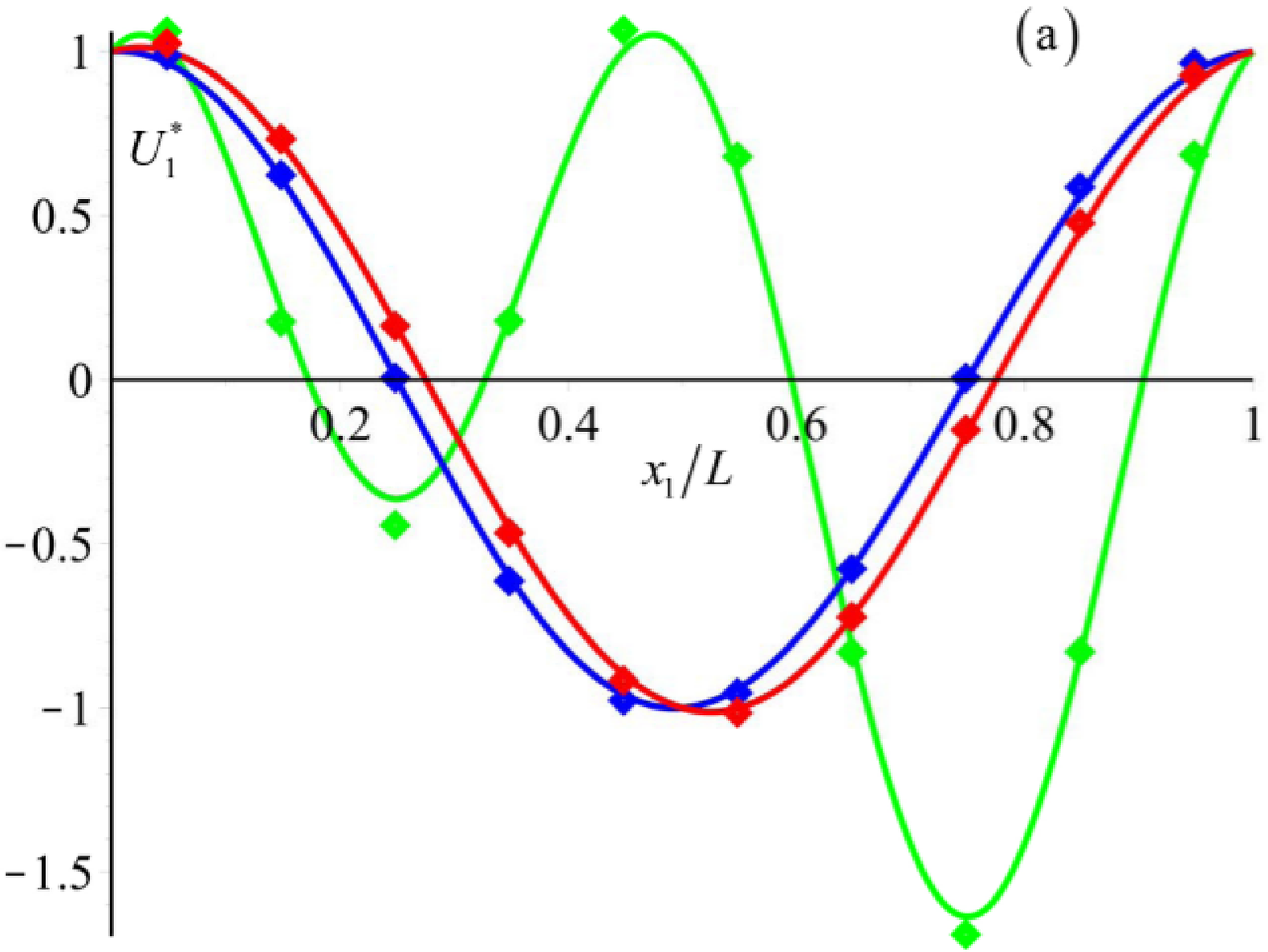}\hspace{10mm}\includegraphics[width=60mm]{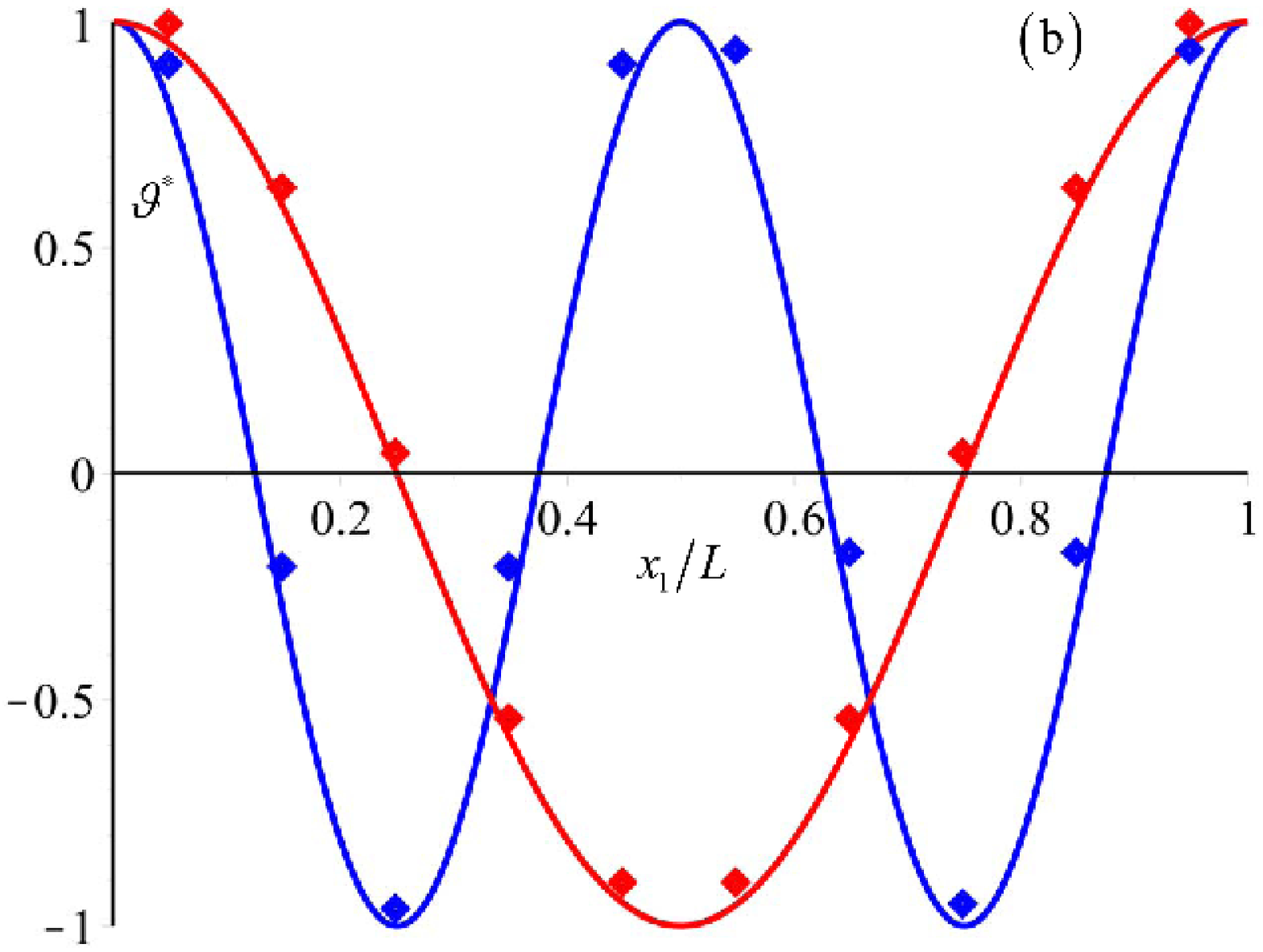}

\protect\protect\caption{Macroscopic displacement component $U_{1}^{*}$ and temperature fields $\Theta^{*}$
due to harmonic body force $b_{1}$ and temperature source $r$. The
heterogeneous model (diamonds) is compared with the homogenized first
order model. (a) Dimensionless macro displacement $U_{1}^{*}$ vs.
the ratio $x_{1}/L$ for $\Xi_{11}^{\alpha}=1$ for different values
of wave number $n$, $m$ ($n=1$, $m=1$ red line; $n=2$, $m=1$
blue line; $n=1$, $m=2$ green line). (b) Macro temperature $\Theta^{*}$
vs. the ratio $x_{1}/L$ for different values of wave number $n$
($n=1$ red line; $n=2$, blue line). }

\label{termoelL} 
\end{figure}

In Fig.~\ref{termoelL}, the macroscopic displacement component
$U_{1}^{*}$ and temperature $\Theta^{*}$ evaluated using analytical expressions (\ref{Uast}) and (\ref{TUast})$_{1}$ are reported as functions of the normalized spatial coordinate $x_{1}/L$ 
(continuous lines in the figure) and compared with the numerical results obtained by the heterogeneous model assuming periodic body forces $b_{1}(x_{1})$ and heat sources $r(x_{1})$ 
and considering the value of the amplitude $\tilde{\varXi}_{11}^{\alpha}=1$ (diamonds in the figure). The following values for the geometrical parameters, the ratios between 
the elastic and of thermodiffusive constants have been assumed: $L/\varepsilon=10$, $\rho_{C}=10$, $\rho_{\alpha}=10$, $\rho_{\beta}=0$ $\rho_{K}=10$, $\rho_{D}=0$, the effects of the
mass diffusion have been neglected in this example. The macroscopic displacement and temperature fields are plotted for the characteristic portion of length $L_{1}=L$,
corresponding to $x_{1}/L=1$ (i. e. for $0\leq x_{1}/L \leq 1$), and several values for the 
wave numbers $m, n \in \mathbb{Z}$ have been considered. Observing the curves, for both the quantities $U_{1}^{*}(x_{1}/L)$ and $\Theta^{*}(x_{1}/L)$ a good agreement 
is detected between the results derived by means of the first order homogenization approach and those obtained by the heterogeneous model.

Results for the macroscopic displacement component $U^{*}_{2}$  and temperature in $\Theta^{*}$ along the characteristic portion of length $L_{2}=L$ 
in direction parallel to $\mathbf{e}_{2}$ (not reported here for conciseness) show a good agreement between the solution obtained by means of
the first-order asymptotic homogenization method and the values obtained by means of finite element analysis of the heterogeneous problem.

\begin{figure}
\centering

\includegraphics[width=60mm]{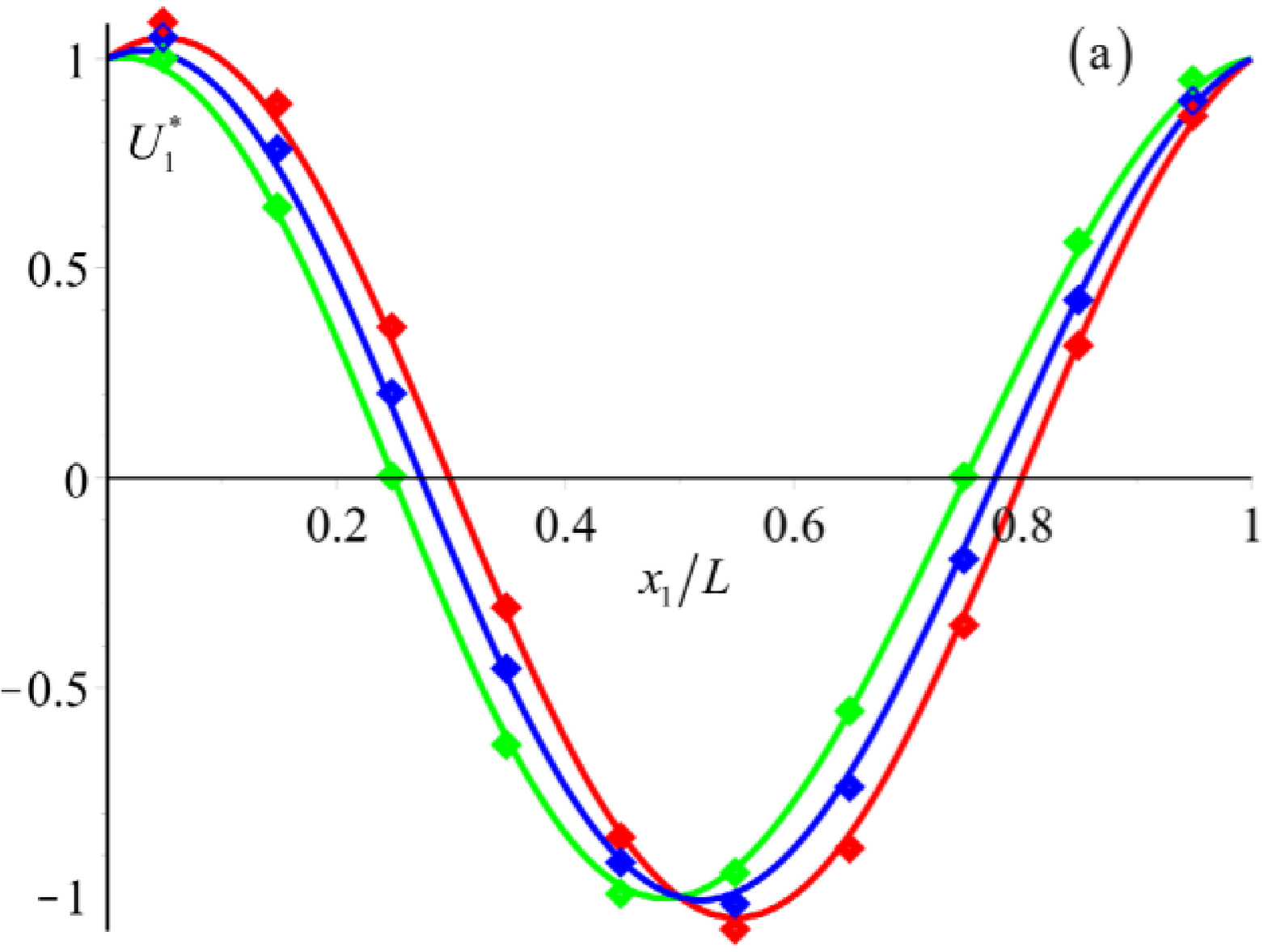}\hspace{10mm}\includegraphics[width=60mm]{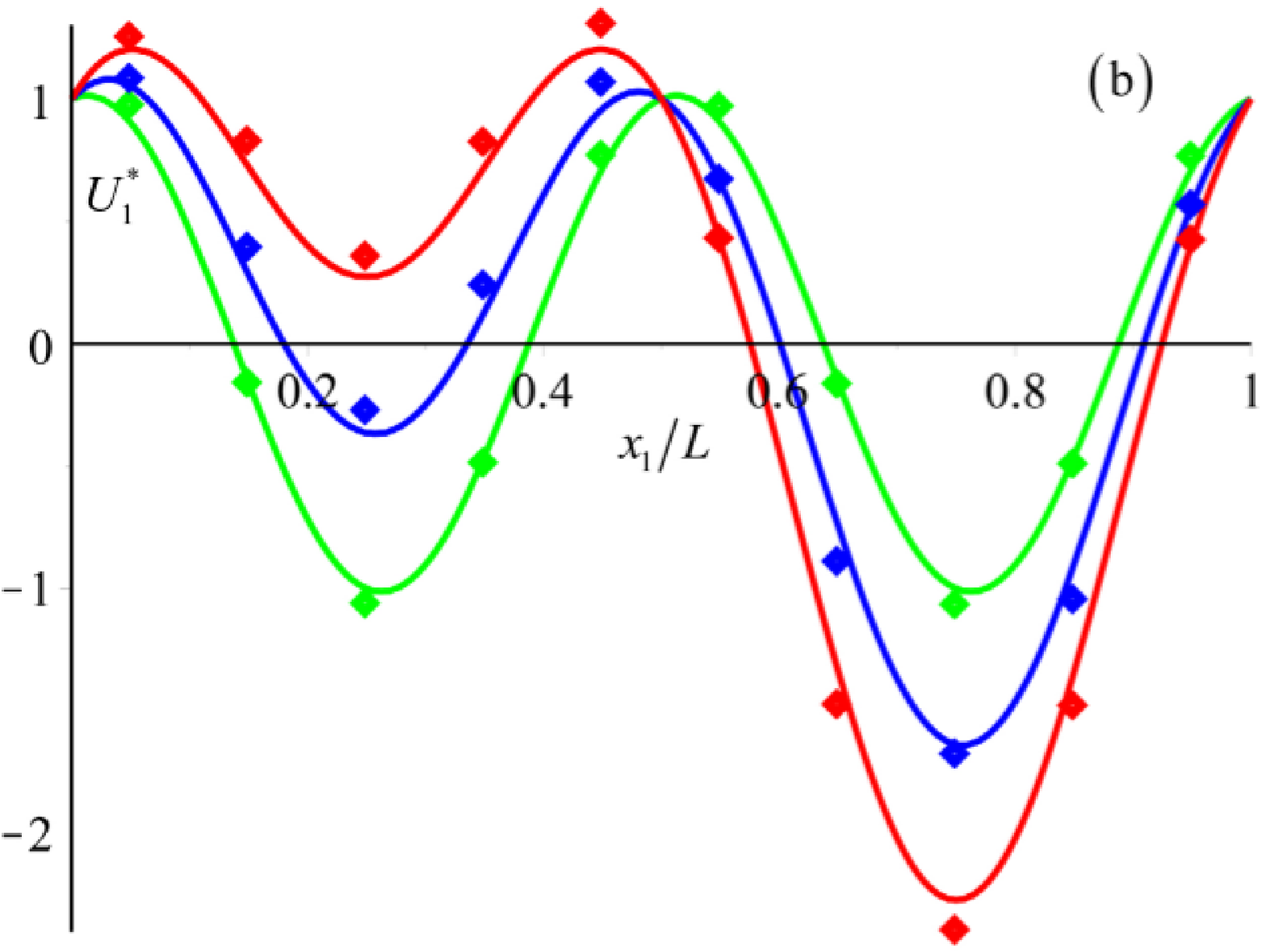}

\includegraphics[width=60mm]{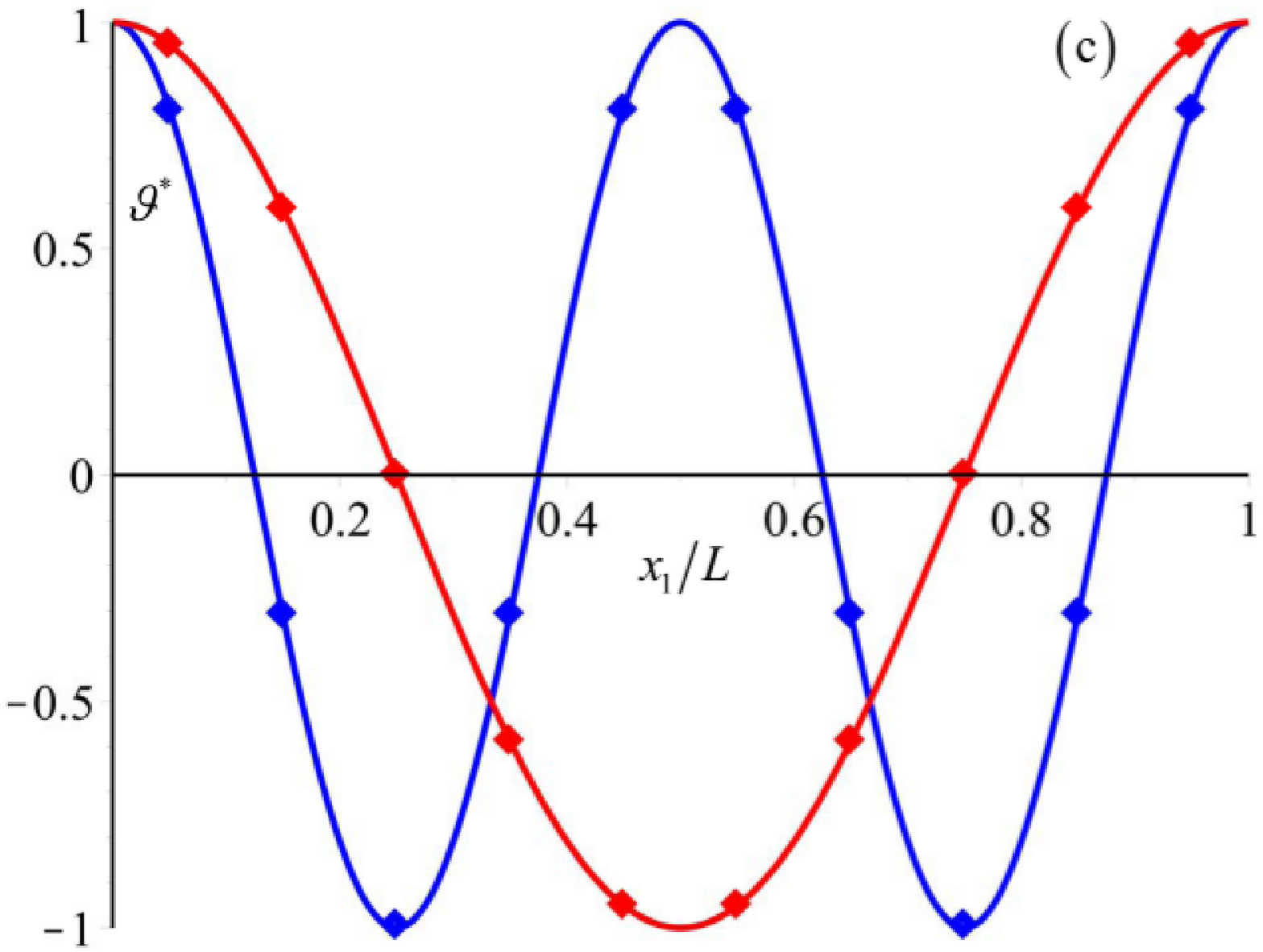}\hspace{10mm}\includegraphics[width=60mm]{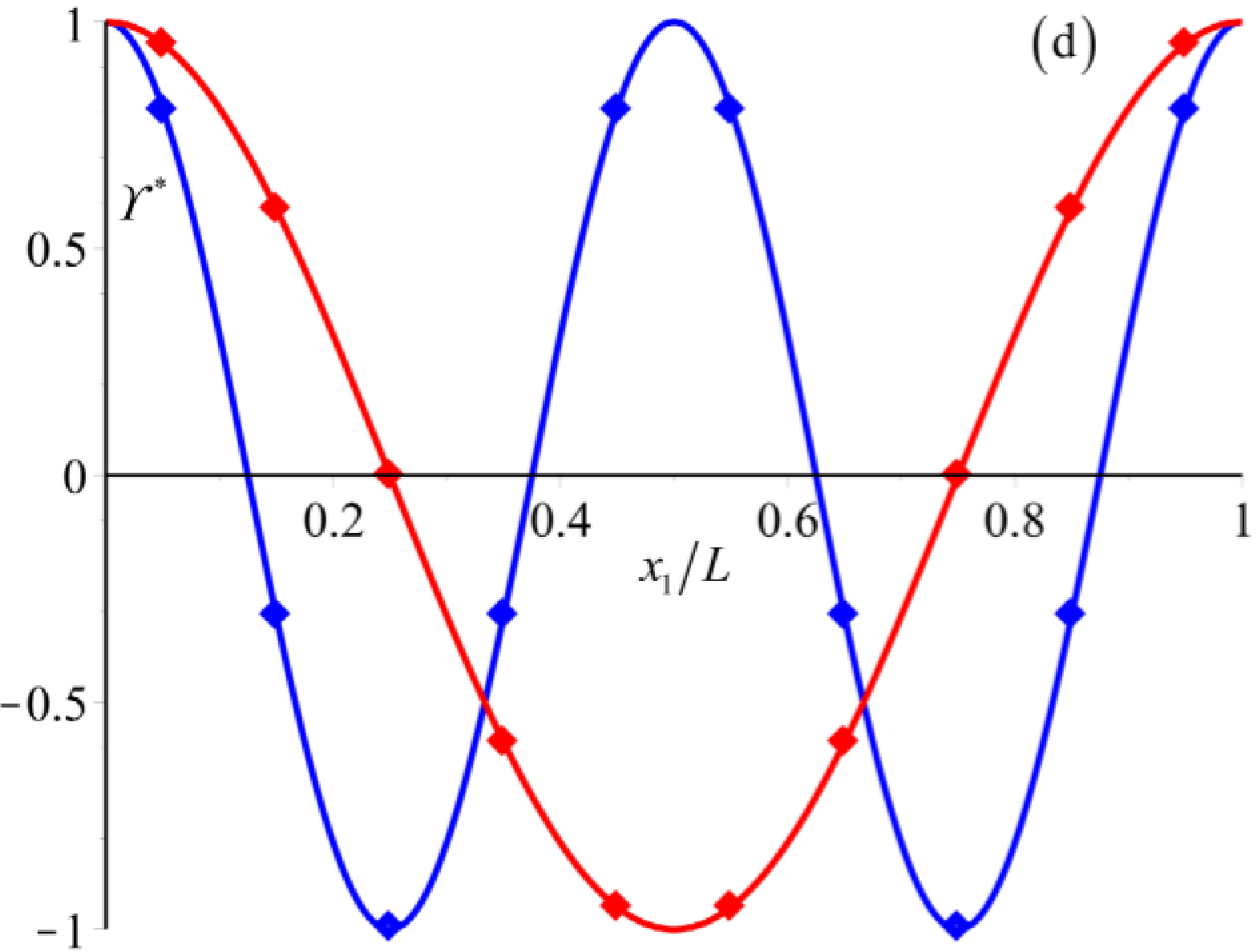}

\protect\protect\caption{Macroscopic displacement component
$U_{1}^{*}$, temperature $\Theta^{*}$, and chemical potential $\varUpsilon^{*}$
due to harmonic body force $b_{1}$ and temperature and mass sources
$r$,$s$ respectively. The heterogeneous model (diamonds) is compared
with the homogenized first order model. Dimensionless macro displacement
$U_{1}^{*}$ vs. the ratio $x_{1}/L$ for $\Xi_{11}^{\alpha}=1$,$\Xi_{11}^{\beta}=1$
and wave number $m=1$ (a), $m=2$ (b) for different values of wave
number $n$, $p$ ($n=1$, $p=1$ red line; $n=2$, $p=1$ blue line;
$n=2$, $p=2$ green line). (c) Macro temperature $\Theta^{*}$ vs.
the ratio $x_{1}/L$ for different values of wave number $n$ ($n=1$
red line; $n=2$, blue line). (d) Macro concentration $\varUpsilon^{*}$
vs. the ratio $x_{1}/L$ for different values of wave number $p$
($p=1$ red line; $p=2$, blue line).}

\label{termoeldiffL} 
\end{figure}

In Fig.~\ref{termoeldiffL}, the variation of the normalized component of the macroscopic displacement $U^{*}_{1}(x_{1}/L)$, temperature $\Theta^{*}(x_{1}/L)$ and
chemical potential $\Upsilon^{*}(x_{1}/L)$ along the characteristic portion of length $L_{1}=L$ is plotted as a function of $x_{1}/L$. Two isotropic phases having the same Poisson's
coefficient $\nu=0.3$ have been assumed, and the same values of the previous example have been assigned to the geometrical, elastic and thermal parameters. The amplitude
of the mass diffusion contribution to the displacement is assumed to be $\tilde{\varXi}_{11}^{\beta}=1$, and $\rho_{K}=10$. Similarly to the previous case, for the finite element analysis 
of the heterogeneous elastic thermodiffusive problem harmonic body forces $b_1(x_1)$, heat and mass sources $r(x_1)$ and $s(x_1)$ have been introduced. The reported curves
show a good agreement between the results obtained by the asymptotic homogenization (continuous lines the figure) and those provided by the heterogeneous elastic thermodiffusive model (diamonds in the figure).
The good agreement between the results coming from the two different approaches can be observed for $U^{*}_{1}(x_{1}/L)$ in Fig.~\ref{termoeldiffL}$/{(a),(b)}$, for $\Theta^{*}(x_{1}/L)$
in Fig.~\ref{termoeldiffL}$/{(b)}$ and for $\Upsilon^{*}(x_{1}/L)$ in Fig.~\ref{termoeldiffL}$/{(d)}$.
Similar results are obtained for the macroscopic displacement $U^{*}_{2}$, temperature $\Theta^{*}$ and
chemical potential $\Upsilon^{*}$ along the characteristic portion of length $L_{2}=L$ in direction parallel to $\mathbf{e}_{2}$


\clearpage

\section{Conclusions}
A general asymptotic homogenization approach for describing the static elastic, thermal and diffusive properties of periodic composite materials in presence
of thermodiffusion is proposed. \emph{Down-scaling} relations associating the displacements, temperature and chemical potential at the micro-scale 
to the corresponding fields at the macro-scale are introduced. 
Perturbation functions are defined for representing the effects of the microstructures on the microscopic displacement, temperature, chemical potential 
and on the coupling effects between these fields. These perturbation functions are obtained through the solution of non-homogeneous
problems on the cell defining periodic boundary conditions and normalization conditions (\emph{up-scaling} relations). 

Averaged field equations of infinite order are derived for the considered class of periodic thermodiffusive materials, 
and an original formal solution is performed by means of power series expansion of the macroscopic displacements, 
temperature and chemical potential fields. Field equation for the homogenized Cauchy thermodiffusive continuum are derived,
and exact expressions for the overall elastic and thermodiffusive constants of this equivalent first order medium are obtained. 

An example of application of the developed general method to the illustrative case of a two-dimensional bi-phase orthotropic layered material is provided. 
The effective elastic and thermodiffusive constants of this particular composite material are determined using the general expressions derived by the asymptotic
homogenization procedure. Analytical expressions for the macroscopic fields derived by the solution of the homogenized equations
corresponding to the first order equivalent continuum. Finite element analysis of the corresponding heterogeneous model is performed assuming 
periodic body forces, heat and mass sources acting on the considered bi-phase layered composite. In order to compare the analytical solution of the homogenized equations with the
numerical results obtained by the heterogeneous model, the microscopic fields computed by finite elements techniques are used to estimate the macroscopic displacements, temperature
and chemical potential fields by means of the \emph{up-scaling} relations defined in the paper. The good agreement detected between the solution derived by the homogenized first order equations 
and the numerical results obtained by the heterogeneous model through the \emph{up-scaling} relations represents an important validation of the accuracy
of the proposed asymptotic homogenization approach.

Thanks to the great versatility of the asymptotic homogenization techniques and to the proposed general rigorous formulation, the method developed in the paper can be adopted
for studying effective elastic and thermodiffusive properties of many composite materials, without any other assumption regarding the geometry of the microstructures 
in addition to the periodicity. In particular, the proposed asymptotic homogenization procedure can have relevant applications in modelling mechanical and thermodiffusive
properties of energy devices with layered configurations, such as lithium ions batteries and solid oxide fuel cells. In standard operative situations, the components of these devices 
are commonly subject to severe thermomechanical and diffusive stress which can cause damages and crack formation compromising their performances. Consequently, evaluating the 
overall elastic and thermodiffusive properties of these battery devices through the asymptotic homogenization approach illustrated in the paper can represent an important issue in order 
to predict damaging phenomena and to improve the efficient design and manufacturing of these systems.

Multi-scale homogenization techniques such as that proposed in the paper provide an accurate description of the macroscopical mechanical and thermodiffusive properties of heterogeneous materials
through the derivation of effective constants of the first order equivalent continuum. Nevertheless, first order homogenization procedures are not enough accurate to model size-effects and
non-local phenomena connected to the microstructural scale length. As a consequence, the developed first order homogenization approach does not provide a precise description of
the behavior of thermodiffusive composite materials in presence of high gradients of stresses, deformations, temperature, chemical potential, heat and mass fluxes, 
as well as of non-local phenomena such as waves dispersion. In order to overcome these limits in the accuracy, non-local higher order homogenization techniques can be used. These methods 
provide constitutive relations of equivalent higher order continuum media including characteristic scale-lengths associate to the microstructural effects. Using the rigorous and general 
approach illustrated in the paper, a better approximation of the elastic and thermodiffusive behavior of composite materials in presence of strong gradients can be obtained through the solution of
higher order cells problems involving the coefficients of the averaged field equations of infinite order reported in Appendix \ref{apphigh}. As it is shown in the same Appendix, 
these equations can be formally solved by means of a double asymptotic expansion performed in terms of the microstructural size.

\section*{Acknowledgments}
AB and LM gratefully acknowledge financial support from the Italian Ministry of Education,
University and Research in the framework of the FIRB project 2010 "Structural mechanics models for renewable energy applications". 
AP would like to acknowledge financial support from the European Union's Seventh Framework Programme FP7/2007-2013/ under REA Grant agreement number PCIG13-GA-2013-618375-MeMic.

\bibliography{SOFC2}
\bibliographystyle{elsarticle-harv} 
\appendix

\section{Higher-order analysis and averaged equations of infinite order}
\label{apphigh}

In this Appendix, explicit expressions for the higher order cells problems associated to the \emph{down-scaling} relations (\ref{Uasym}),
(\ref{Tasym}) and (\ref{Easym}) are reported. Moreover, the averaged field equations of infinite order are derived, and a formal solution is obtained
by means of an asymptotic expansion of the macroscopic fields in terms of the microstructural size.  
\subsection{Higher-order asymptotic analysis and derivation of the corresponding \emph{cell problems}}
\label{appcellproblems}
In order to derive exact expressions for the
fluctuation functions affecting the behavior of the microscopic
fields $u_{k},\theta,\eta$, the \emph{down-scaling} relations (\ref{Uasym}),
(\ref{Tasym}) and (\ref{Easym}) are substituted into the microscopic
field equations (\ref{field1}), (\ref{field2}).
Remembering the property $\frac{\partial}{\partial x_{j}}f(\boldsymbol{x},\boldsymbol{\xi}=\frac{\boldsymbol{x}}{\varepsilon})=\left(\frac{\partial f}{\partial x_{j}}+\frac{1}{\varepsilon}\frac{\partial f}{\partial\xi_{j}}\right)_{\boldsymbol{\xi}=\boldsymbol{x}/\varepsilon}=\left(\frac{\partial f}{\partial x_{j}}+\frac{f_{,j}}{\varepsilon}\right)_{\boldsymbol{\xi}=\boldsymbol{x}/\varepsilon}$,
equation (\ref{field1}) become 
\begin{align}
\varepsilon^{-1} & \biggl\{\left[\left(C_{ijkl}^{\Gve}N_{kpq_{1},l}^{(1)}\right)_{,j}+C_{ijpq_{1},j}^{\Gve}\right]H_{pq_{1}}(\bx)+\left[\left(C_{ijkl}^{\Gve}\tilde{N}_{k,l}^{(1)}\right)_{,j}-\alpha_{ij,j}^{\Gve}\right]\Theta(\boldsymbol{x})\nonumber \\
 & +\left[\left(C_{ijkl}^{\Gve}\hat{N}_{k,l}^{(1)}\right)_{,j}-\beta_{ij,j}^{\Gve}\right]\Upsilon(\boldsymbol{x})\biggr\}\nonumber \\
+\varepsilon^{0} & \biggl\{\left(C_{ijkl}^{\Gve}N_{kpq_{1}q_{2},l}^{(2)}\right)_{,j}+\frac{1}{2}\,\biggl(\left(C_{ijkq_{2}}^{\Gve}N_{kpq_{1}}^{(1)}\right)_{,j}+C_{iq_{2}pq_{1}}^{\Gve}+C_{iq_{2}kl}^{\Gve}N_{kpq_{1},l}^{(1)}\nonumber \\
 & +\left(C_{ijkq_{1}}^{\Gve}N_{kpq_{2}}^{(1)}\right)_{,j}+C_{iq_{1}pq_{2}}^{\Gve}+C_{iq_{1}kl}^{\Gve}N_{kpq_{2},l}^{(1)}\biggr)\kappa_{pq_{1}q_{2}}(\boldsymbol{x})\nonumber \\
 & +\biggl[\left(C_{ijkl}^{\Gve}\tilde{N}_{kq_{1},l}^{(2)}\right)_{,j}+\left(C_{ijkq_{1}}^{\Gve}\tilde{N}_{k}^{(1)}\right)_{,j}+C_{iq_{1}kl}^{\Gve}\tilde{N}_{k,l}^{(1)}-\alpha_{iq_{1}}^{\Gve}-\left(\alpha_{ij}^{\Gve}M_{q_{1}}^{(1)}\right)_{,j}\biggr]\frac{\partial\Theta}{\partial x_{q_{1}}}\nonumber \\
 & +\biggl[\left(C_{ijkl}^{\Gve}\hat{N}_{kq_{1},l}^{(2)}\right)_{,j}+\left(C_{ijkq_{1}}^{\Gve}\hat{N}_{k}^{(1)}\right)_{,j}+C_{iq_{1}kl}^{\Gve}\hat{N}_{k,l}^{(1)}-\beta_{iq_{1}}^{\Gve}-\left(\beta_{ij}^{\Gve}W_{q_{1}}^{(1)}\right)_{,j}\biggr]\frac{\partial\Upsilon}{\partial x_{q_{1}}}\biggl\}\nonumber \\
 & +\cdots\cdots+b_{i}(\bx)=0,\quad\quad i=1,2,\label{eq:asymech2}
\end{align}
where $H_{pq_{1}}=\partial U_{p}/\partial x_{q_{1}}$ are the components
of the macroscopic displacement gradient tensor previously defined,
and $\kappa_{pq_{1}q_{2}}=\partial^{2}U_{p}/\partial x_{q_{1}}\partial x_{q_{2}}$
are the elements of the macroscopic second gradient tensor.
Equations (\ref{field2}) assume the following form
\begin{align}
\varepsilon^{-1} & \biggl[\left(K_{ij}^{\Gve}M_{q_{1},j}^{(1)}\right)_{,i}+K_{iq_{1},i}^{\Gve}\biggr]\frac{\partial\Theta}{\partial x_{q_{1}}}\nonumber \\
+\varepsilon^{0} & \biggl[\left(K_{ij}^{\Gve}M_{q_{1}q_{2},j}^{(2)}\right)_{,i}+\frac{1}{2}\,\biggl(\left(K_{iq_{1}}^{\Gve}M_{q_{2}}^{(1)}\right)_{,i}+K_{q_{2}q_{1}}^{\Gve}+K_{q_{1}j}^{\Gve}M_{q_{2},j}^{(1)}\nonumber \\
 & +\left(K_{iq_{2}}^{\Gve}M_{q_{1}}^{(1)}\right)_{,i}+K_{q_{1}q_{2}}^{\Gve}+K_{q_{2}j}^{\Gve}M_{q_{1},j}^{(1)}\biggl)\biggl]\frac{\partial^{2}\Theta}{\partial x_{q_{1}}\partial x_{q_{2}}}\nonumber \\
 & +\cdots\cdots+r(\bx)=0,\label{eq:asytemp2}
\end{align}
\begin{align}
\varepsilon^{-1} & \biggl[\left(D_{ik}^{\Gve}W_{q_{1},j}^{(1)}\right)_{,i}+D_{iq_{1},i}^{\Gve}\biggr]\frac{\partial\Upsilon}{\partial x_{q_{1}}}\nonumber \\
+\varepsilon^{0} & \biggl[\left(D_{ij}^{\Gve}W_{q_{1}q_{2},j}^{(2)}\right)_{,i}+\frac{1}{2}\,\biggl(\left(D_{iq_{1}}^{\Gve}W_{q_{2}}^{(1)}\right)_{,i}+D_{q_{2}q_{1}}^{\Gve}+D_{q_{1}j}^{\Gve}W_{q_{2},j}^{(1)}\nonumber \\
 & +\left(D_{iq_{2}}^{\Gve}W_{q_{1}}^{(1)}\right)_{,i}+D_{q_{1}q_{2}}^{\Gve}+D_{q_{2}j}^{\Gve}W_{q_{1},j}^{(1)}\biggl)\biggl]\frac{\partial^{2}\Upsilon}{\partial x_{q_{1}}\partial x_{q_{2}}}\nonumber \\
 & +\cdots\cdots+s(\bx)=0.\label{eq:asydiff2}
\end{align}
In order to transform the field equations (\ref{eq:asymech2}), (\ref{eq:asytemp2})
and (\ref{eq:asydiff2}) in a PDEs system with constant coefficients,
in which the unknowns are the macroscopic quantities $U_{k}(\mathbf{x})$, $\Theta(\mathbf{x})$
and $\varUpsilon(\mathbf{x})$, the fluctuation functions have to
satisfy non-homogeneous equations (\emph{cell problems}) reported
below.

At the order $\varepsilon^{-1}$ from the equations (\ref{eq:asymech2}), (\ref{eq:asytemp2}), (\ref{eq:asydiff2}) we derive the \emph{first-order cell problems} reported in Sec.~\ref{cellproblems} in the text of the paper, equations (\ref{cellmech-1}) and (\ref{cellthermodiff-1}).

At the order $\varepsilon^{0}$, equation (\ref{eq:asymech2}) yields the following \emph{second-order cell problems}
\begin{align}
 & \left(C_{ijkl}^{\Gve}N_{kpq_{1}q_{2},l}^{(2)}\right)_{,j}+\frac{1}{2}\,\biggl[\left(C_{ijkq_{2}}^{\Gve}N_{kpq_{1}}^{(1)}\right)_{,j}+C_{iq_{2}pq_{1}}^{\Gve}+C_{iq_{2}kl}^{\Gve}N_{kpq_{1},l}^{(1)}\nonumber \\
 & +\left(C_{ijkq_{1}}^{\Gve}N_{kpq_{2}}^{(1)}\right)_{,j}+C_{iq_{1}pq_{2}}^{\Gve}+C_{iq_{1}kl}^{\Gve}N_{kpq_{2},l}^{(1)}\biggr]=n_{ipq_{1}q_{2}}^{(2)},\nonumber \\
 & \left(C_{ijkl}^{\Gve}\tilde{N}_{kq_{1},l}^{(2)}\right)_{,j}+\left(C_{ijkq_{1}}^{\Gve}\tilde{N}_{k}^{(1)}\right)_{,j}+C_{iq_{1}kl}^{\Gve}\tilde{N}_{k,l}^{(1)}-\alpha_{iq_{1}}^{\Gve}-\left(\alpha_{ij}^{\Gve}M_{q_{1}}^{(1)}\right)_{,j}=\tilde{n}_{iq_{1}}^{(2)},\nonumber \\
 & \left(C_{ijkl}^{\Gve}\hat{N}_{kq_{1},l}^{(2)}\right)_{,j}+\left(C_{ijkq_{1}}^{\Gve}\hat{N}_{k}^{(1)}\right)_{,j}+C_{iq_{1}kl}^{\Gve}\hat{N}_{k,l}^{(1)}-\beta_{iq_{1}}^{\Gve}-\left(\beta_{ij}^{\Gve}W_{q_{1}}^{(1)}\right)_{,j}=\hat{n}_{iq_{1}}^{(2)}\label{cell0mech}
\end{align}
At the same order, from (\ref{eq:asytemp2}), (\ref{eq:asydiff2}) we derive the \emph{second-order thermodiffusive cell problems}:
\begin{align}
 & \left(K_{ij}^{\Gve}M_{q_{1}q_{2},j}^{(2)}\right)_{,i}+\frac{1}{2}\,\biggl[\left(K_{iq_{1}}^{\Gve}M_{q_{2}}^{(1)}\right)_{,i}+K_{q_{2}q_{1}}^{\Gve}+K_{q_{1}j}^{\Gve}M_{q_{2},j}^{(1)}\nonumber \\
 & +\left(K_{iq_{2}}^{\Gve}M_{q_{1}}^{(1)}\right)_{,i}+K_{q_{1}q_{2}}^{\Gve}+K_{q_{2}j}^{\Gve}M_{q_{1},j}^{(1)}\biggl]=m_{q_{1}q_{2}}^{(2)},\label{cell0thermo}
\end{align}
\begin{align}
 & \left(D_{ij}^{\Gve}W_{q_{1}q_{2},j}^{(2)}\right)_{,i}+\frac{1}{2}\,\biggl[\left(D_{iq_{1}}^{\Gve}W_{q_{2}}^{(1)}\right)_{,i}+D_{q_{2}q_{1}}^{\Gve}+D_{q_{1}j}^{\Gve}W_{q_{2},j}^{(1)}\nonumber \\
 & +\left(D_{iq_{2}}^{\Gve}W_{q_{1}}^{(1)}\right)_{,i}+D_{q_{1}q_{2}}^{\Gve}+D_{q_{2}j}^{\Gve}W_{q_{1},j}^{(1)}\biggl]=w_{q_{1}q_{2}}^{(2)}, \label{cell0diff}
\end{align}
where: 
\[
n_{ipq_{1}q_{2}}^{(2)}=\frac{1}{2}\bigg\langle C_{iq_{2}pq_{1}}^{\Gve}+C_{iq_{2}kl}^{\Gve}N_{kpq_{1},l}^{(1)}+C_{iq_{1}pq_{2}}^{\Gve}+C_{iq_{1}kl}^{\Gve}N_{kpq_{2},l}^{(1)}\bigg\rangle,
\]
\[
\tilde{n}_{iq_{1}}^{(2)}=\bigg\langle C_{iq_{1}kl}^{\Gve}\tilde{N}_{k,l}^{(1)}-\alpha_{iq_{1}}^{\Gve}\bigg\rangle,\quad\hat{n}_{iq_{1}}^{(2)}=\bigg\langle C_{iq_{1}kl}^{\Gve}\hat{N}_{k,l}^{(1)}-\beta_{iq_{1}}^{\Gve}\bigg\rangle,
\]
\[
m_{q_{1}q_{2}}^{(2)}=\frac{1}{2}\bigg\langle K_{q_{2}q_{1}}^{\Gve}+K_{q_{1}j}^{\Gve}M_{q_{2},j}^{(1)}+K_{q_{1}q_{2}}^{\Gve}+K_{q_{2}j}^{\Gve}M_{q_{1},j}^{(1)}\bigg\rangle,
\]
\begin{equation}
w_{q_{1}q_{2}}^{(2)}=\frac{1}{2}\bigg\langle D_{q_{2}q_{1}}^{\Gve}+D_{q_{1}j}^{\Gve}W_{q_{2},j}^{(1)}+D_{q_{1}q_{2}}^{\Gve}+D_{q_{2}j}^{\Gve}W_{q_{1},j}^{(1)}\bigg\rangle.\label{nmw1}
\end{equation}
In general, for the $m=1,2,\cdots$, the $m-$order \emph{cell problems} associate to equation (\ref{eq:asymech2}) assume the form: 
\begin{align}
 & \left(C_{ijkl}^{\Gve}N_{kpq_{1}\cdots q_{m},l}^{(m)}\right)_{,j}+\frac{1}{m!}\sum_{\mathcal{P}(q)}\biggl[\left(C_{ijkq_{m}}^{\Gve}N_{kpq_{1}\cdots q_{m-1}}^{(m-1)}\right)_{,j}+C_{iq_{m}kq_{m-1}}^{\Gve}N_{kpq_{1}\cdots q_{m-2}}^{(m-2)}\nonumber \\
 & +C_{iq_{m}kl}^{\Gve}N_{kpq_{1}\cdots q_{m-1},l}^{(m-1)}\biggr]=n_{ipq_{1}\cdots q_{m}}^{(m)},\nonumber \\
 & \left(C_{ijkl}^{\Gve}\tilde{N}_{kq_{1}\cdots q_{m-1},l}^{(m)}\right)_{,j}+\frac{1}{m!}\sum_{\mathcal{P}(q)}\biggl[\left(C_{ijkq_{m-1}}^{\Gve}\tilde{N}_{kq_{1}\cdots q_{m-2}}^{(m-1)}\right)_{,j}+C_{iq_{m-1}kq_{m-2}}^{\Gve}\tilde{N}_{kq_{1}\cdots q_{m-3}}^{(m-2)}\nonumber \\
 & +C_{iq_{m-1}kl}^{\Gve}\tilde{N}_{q_{1}\cdots q_{m-2},l}^{(m-1)}-\alpha_{iq_{m-1}}^{\Gve}M_{q_{1}\cdots q_{m-2}}^{(m-2)}-\left(\alpha_{ij}^{\Gve}M_{q_{1}\cdots q_{m-1}}^{(m-1)}\right)_{,j}\biggr]=\tilde{n}_{iq_{1}\cdots q_{m-1}}^{(m)},\nonumber \\
 & \left(C_{ijkl}^{\Gve}\hat{N}_{kq_{1}\cdots q_{m-1},l}^{(m)}\right)_{,j}+\frac{1}{m!}\sum_{\mathcal{P}(q)}\biggl[\left(C_{ijkq_{m-1}}^{\Gve}\hat{N}_{kq_{1}\cdots q_{m-2}}^{(m-1)}\right)_{,j}+C_{iq_{m-1}kq_{m-2}}^{\Gve}\hat{N}_{kq_{1}\cdots q_{m-3}}^{(m-2)}\nonumber \\
 & +C_{iq_{m-1}kl}^{\Gve}\hat{N}_{q_{1}\cdots q_{m-2},l}^{(m-1)}-\beta_{iq_{m-1}}^{\Gve}W_{q_{1}\cdots q_{m-2}}^{(m-2)}-\left(\beta_{ij}^{\Gve}W_{q_{1}\cdots q_{m-1}}^{(m-1)}\right)_{,j}\biggr]=\hat{n}_{iq_{1}\cdots q_{m-1}}^{(m)},\label{cellmechm-2}
\end{align}
whereas the $m-$order thermodiffusive \emph{cell problems} corresponding to equations
(\ref{eq:asytemp2}) and (\ref{eq:asydiff2}) are: 
\begin{align}
 & \left(K_{ij}^{\Gve}M_{q_{1}\cdots q_{m},j}^{(m)}\right)_{,i}+\frac{1}{m!}\sum_{\mathcal{P}(q)}\biggl[\left(K_{iq_{m}}^{\Gve}M_{q_{1}\cdots q_{m-1}}^{(m-1)}\right)_{,i}\nonumber \\
 & +K_{q_{m}q_{1}}^{\Gve}M_{q_{2}\cdots q_{m-1}}^{(m-2)}+K_{q_{m}j}^{\Gve}M_{q_{1}\cdots q_{m-1},j}^{(m-1)}\biggl]=m_{q_{1}\cdots q_{m}}^{(m)},\label{cellthermom-2}
\end{align}
\begin{align}
 & \left(D_{ij}^{\Gve}W_{q_{1}\cdots q_{m},j}^{(m)}\right)_{,i}+\frac{1}{m!}\sum_{\mathcal{P}(q)}\biggl[\left(D_{iq_{m}}^{\Gve}W_{q_{1}\cdots q_{m-1}}^{(m-1)}\right)_{,i}\nonumber \\
 & +D_{q_{m}q_{1}}^{\Gve}W_{q_{2}\cdots q_{m-1}}^{(m-2)}+D_{q_{m}j}^{\Gve}W_{q_{1}\cdots q_{m-1},j}^{(m-1)}\biggl]=w_{q_{1}\cdots q_{m}}^{(m)},\label{cellmassm-2}
\end{align}
where the symbol $\mathcal{P}(q)$ denotes all possible permutations
of the multi-index $q$, and the constants $n_{ipq_{1}\cdots q_{m}}^{(m)}$,
$\tilde{n}_{iq_{1}\cdots q_{m-1}}^{(m)}$, $\hat{n}_{iq_{1}\cdots q_{m-1}}^{(m)}$,
$m_{q_{1}\cdots q_{m}}^{(m)}$, $w_{q_{1}\cdots q_{m}}^{(m)}$ are
defined as follows: 
\[
n_{ipq_{1}\cdots q_{m}}^{(m)}=\frac{1}{m!}\sum_{\mathcal{P}(q)}\bigg\langle C_{iq_{m}kq_{m-1}}^{\Gve}N_{kpq_{1}\cdots q_{m-2}}^{(m-2)}+C_{iq_{m}kl}^{\Gve}N_{kpq_{1}\cdots q_{m-1},l}^{(m-1)}\bigg\rangle,
\]
\[
\tilde{n}_{iq_{1}\cdots q_{m-1}}^{(m)}=\frac{1}{m!}\sum_{\mathcal{P}(q)}\bigg\langle C_{iq_{m-1}kq_{m-2}}^{\Gve}\tilde{N}_{kq_{1}\cdots q_{m-3}}^{(m-2)}+C_{iq_{m-1}kl}^{\Gve}\tilde{N}_{q_{1}\cdots q_{m-2},l}^{(m-1)}-\alpha_{iq_{m-1}}^{\Gve}M_{q_{1}\cdots q_{m-2}}^{(m-2)}\bigg\rangle,
\]
\[
\hat{n}_{iq_{1}\cdots q_{m-1}}^{(m)}=\frac{1}{m!}\sum_{\mathcal{P}(q)}\bigg\langle C_{iq_{m-1}kq_{m-2}}^{\Gve}\hat{N}_{kq_{1}\cdots q_{m-3}}^{(m-2)}+C_{iq_{m-1}kl}^{\Gve}\hat{N}_{q_{1}\cdots q_{m-2},l}^{(m-1)}-\beta_{iq_{m-1}}^{\Gve}W_{q_{1}\cdots q_{m-2}}^{(m-2)}\bigg\rangle,
\]
\[
m_{q_{1}\cdots q_{m}}^{(m)}=\frac{1}{m!}\sum_{\mathcal{P}(q)}\bigg\langle K_{q_{m}q_{1}}^{\Gve}M_{q_{2}\cdots q_{m-1}}^{(m-2)}+K_{q_{m}j}^{\Gve}M_{q_{1}\cdots q_{m-1},j}^{(m-1)}\bigg\rangle,
\]
\begin{equation}
w_{q_{1}\cdots q_{m}}^{(m)}=\frac{1}{m!}\sum_{\mathcal{P}(q)}\bigg\langle D_{q_{m}q_{1}}^{\Gve}W_{q_{2}\cdots q_{m-1}}^{(m-2)}+D_{q_{m}j}^{\Gve}W_{q_{1}\cdots q_{m-1},j}^{(m-1)}\bigg\rangle.\label{cost_inf}
\end{equation}

The perturbation functions characterizing the \emph{down-scaling}
relations (\ref{Uasym}), (\ref{Tasym}), and (\ref{Easym})
are obtained by the solution of the previously defined cells problems,
derived by imposing the normalization conditions (\ref{renorm-1}).
According to \citet{BakhPan1} and \citet{SmiCher1}, the constants (\ref{nmw1})
and (\ref{cost_inf}) are determined by imposing that the non-homogeneous
terms in equations (\ref{cellmechm-2}), (\ref{cell0mech}), (\ref{cell0thermo}), (\ref{cell0diff}),
(\ref{cellthermom-2}) and (\ref{cellmassm-2}) (associated to the auxiliary body
forces \citep{Bacigalupo2}, heat and mass sources) possess vanishing mean values
over the unit cell $\mQ$. This implies the $\mQ-$periodicity of
the perturbations functions $N_{kpq}^{(m)},\tilde{N}_{kq}^{(m)},\hat{N}_{kq}^{(m)},M_{q}^{(m)},W_{q}^{(m)}$,
and then the continuity and regularity of the microscopic fields (micro-displacements,
micro-temperature and micro-concentration) at the interface between
adjacent cells are guaranteed.

\subsection{Averaged field equation of infinite order and its formal solution}

Using the cell problems (\ref{cellmech-1}),
(\ref{cellthermodiff-1}), (\ref{cellmechm-2}), (\ref{cell0mech}),
(\ref{cell0thermo}), (\ref{cell0diff}), (\ref{cellthermom-2}) and
(\ref{cellmassm-2}) together with the constants definitions (\ref{nmw0}),
(\ref{nmw1}) and (\ref{cost_inf}) into the microscopic field equations
(\ref{eq:asymech}), (\ref{eq:asytemp}) and (\ref{eq:asydiff}),
the averaged equations of infinite order are derived: 
\begin{align}
 & n_{ipq_{1}q_{2}}^{(2)}\frac{\partial^{2}U_{p}}{\partial x_{q_{1}}\partial x_{q_{2}}}+\tilde{n}_{iq_{1}}^{(2)}\frac{\partial\Theta}{\partial x_{q_{1}}}+\hat{n}_{iq_{1}}^{(2)}\frac{\partial\Upsilon}{\partial x_{q_{1}}}+\sum_{n=0}^{+\infty}\varepsilon^{n+1}\sum_{|q|=n+3}n_{ipq}^{(n+3)}\frac{\partial^{n+3}U_{p}}{\partial x_{q}}\nonumber \\
 & +\sum_{n=0}^{+\infty}\varepsilon^{n+1}\sum_{|q|=n+2}\tilde{n}_{iq}^{(n+2)}\frac{\partial^{n+2}\Theta}{\partial x_{q}}+\sum_{n=0}^{+\infty}\varepsilon^{n+1}\sum_{|q|=n+2}\hat{n}_{iq}^{(n+2)}\frac{\partial^{n+2}\Upsilon}{\partial x_{q}}+b_{i}=0\label{mech_inf}
\end{align}
\begin{equation}
m_{q_{1}q_{2}}^{(2)}\frac{\partial^{2}\Theta}{\partial x_{q_{1}}\partial x_{q_{2}}}+\sum_{n=0}^{+\infty}\varepsilon^{n+1}\sum_{|q|=n+3}m_{q}^{(n+3)}\frac{\partial^{n+3}\Theta}{\partial x_{q}}+r=0,\label{thermo_inf}
\end{equation}
\begin{equation}
w_{q_{1}q_{2}}^{(2)}\frac{\partial^{2}\Upsilon}{\partial x_{q_{1}}\partial x_{q_{2}}}+\sum_{n=0}^{+\infty}\varepsilon^{n+1}\sum_{|q|=n+3}w_{q}^{(n+3)}\frac{\partial^{n+3}\Upsilon}{\partial x_{q}}+s=0,\label{diff_inf}
\end{equation}
where $q$ is a multi-index, $\partial^{n+j}(\cdot)/\partial x_{q}=\partial^{n+j}(\cdot)/\partial x_{q_{1}}\cdots\partial x_{q_{n+j}}$
with $j\in\mathbb{N}$, $n_{ipq}^{(n+3)}=n_{ipq_{1}\cdots q_{n+3}}$,
$\tilde{n}_{iq}^{(n+2)}=\tilde{n}_{iq_{1}\cdots q_{n+2}}$, $\hat{n}_{iq}^{(n+2)}=\hat{n}_{iq_{1}\cdots q_{n+2}}$,
$m_{q}^{(n+3)}=m_{q_{1}\cdots q_{n+3}}^{(n+3)}$ and $w_{q}^{(n+3)}=w_{q_{1}\cdots q_{n+3}}^{(n+3)}$.

A formal solution of the averaged field equations of infinite order
(\ref{mech_inf}), (\ref{thermo_inf}) and (\ref{diff_inf}) is obtained
by means of an asymptotic expansion of the macroscopic fields $U_{i},\Theta$
and $\Upsilon$ in terms of the microstructural size $\varepsilon$,
i.e. 
\begin{equation}
U_{i}(\mathbf{x})=\sum_{j=0}^{+\infty}\varepsilon^{m}U_{i}^{(m)}(\mathbf{x}),\quad\Theta(\mathbf{x})=\sum_{j=0}^{+\infty}\varepsilon^{m}\Theta^{(m)}(\mathbf{x}),\quad\Upsilon(\mathbf{x})=\sum_{j=0}^{+\infty}\varepsilon^{m}\Upsilon^{(m)}(\mathbf{x}).\label{asym2}
\end{equation}

By substituting the series (\ref{asym2}) into (\ref{mech_inf}),
(\ref{thermo_inf}) and (\ref{diff_inf}), a sequence of equations
for determining the terms of the asymptotic expansion $U_{i}^{(m)}$,
$\Theta^{(m)}$ and $\Upsilon^{(m)}$ is obtained. At the order $\varepsilon^{0}$,
from the equation (\ref{mech_inf}) we derive: 
\begin{equation}
n_{ipq_{1}q_{2}}^{(2)}\frac{\partial^{2}U_{p}^{(0)}}{\partial x_{q_{1}}\partial x_{q_{2}}}+\tilde{n}_{iq_{1}}^{(2)}\frac{\partial\Theta^{(0)}}{\partial x_{q_{1}}}+\hat{n}_{iq_{1}}^{(2)}\frac{\partial\Upsilon^{(0)}}{\partial x_{q_{1}}}+b_{i}=0.\label{inf_asymech}
\end{equation}
whereas thermodiffusion equations (\ref{thermo_inf}) and (\ref{diff_inf})
yield respectively 
\begin{equation}
m_{q_{1}q_{2}}^{(2)}\frac{\partial^{2}\Theta^{(0)}}{\partial x_{q_{1}}\partial x_{q_{2}}}+r=0, \quad
w_{q_{1}q_{2}}^{(2)}\frac{\partial^{2}\Upsilon^{(0)}}{\partial x_{q_{1}}\partial x_{q_{2}}}+s=0.\label{inf_diff0}
\end{equation}
At the generic order $m$ from (\ref{mech_inf}) we obtain 
\begin{align}
 & n_{ipq_{1}q_{2}}^{(2)}\frac{\partial^{(2)}U_{p}^{(m)}}{\partial x_{q_{1}}\partial x_{q_{2}}}+\tilde{n}_{iq_{1}}^{(2)}\frac{\partial\Theta^{(m)}}{\partial x_{q_{1}}}+\hat{n}_{iq_{1}}^{(2)}\frac{\partial\Upsilon^{(m)}}{\partial x_{q_{1}}}+\sum_{r=3}^{m+2}\sum_{|q|=r}n_{ipq}^{(r)}\frac{\partial^{r}U_{p}^{(m+2-r)}}{\partial x_{q}}+\nonumber \\
 & \sum_{r=3}^{m+2}\sum_{|q|=r-1}\tilde{n}_{iq}^{(r)}\frac{\partial^{r-1}\Theta^{(m+2-r)}}{\partial x_{q}}+\sum_{r=3}^{m+2}\sum_{|q|=r-1}\hat{n}_{iq}^{(r)}\frac{\partial^{r-1}\Upsilon^{(m+2-r)}}{\partial x_{q}}=0,\label{asym_infinity}
\end{align}
and (\ref{thermo_inf}) and (\ref{diff_inf}) are given by 
\begin{equation}
m_{q_{1}q_{2}}^{(2)}\frac{\partial^{2}\Theta^{(m)}}{\partial x_{q_{1}}x_{q_{2}}}+\sum_{p=3}^{m+2}\sum_{|h|=p}m_{h}^{(p)}\frac{\partial^{p}\Theta^{(m+2-p)}}{\partial x_{h}}=0,
\end{equation}
\begin{equation}
w_{q_{1}q_{2}}^{(2)}\frac{\partial^{2}\Upsilon^{(m)}}{\partial x_{q_{1}}x_{q_{2}}}+\sum_{p=3}^{m+2}\sum_{|h|=p}w_{h}^{(p)}\frac{\partial^{p}\Upsilon^{(m+2-p)}}{\partial x_{h}}=0,\label{inf_asydiff}
\end{equation}
where $h$ and $q$ are multi-indexes. The solution of equations (\ref{inf_asymech})-(\ref{inf_asydiff})
requires that the following normalization conditions are satisfied:
\begin{equation}
\frac{1}{\delta{L}^{2}}\int_{\mathcal{L}}U_{p}^{(m)}(\mathbf{x})d\mathbf{x}=0,\quad\frac{1}{\delta{L}^{2}}\int_{\mathcal{L}}\Theta^{(m)}(\mathbf{x})d\mathbf{x}=0,\quad\frac{1}{\delta{L}^{2}}\int_{\mathcal{L}}\Upsilon^{(m)}(\mathbf{x})d\mathbf{x}=0,
\end{equation}
where the $\mathcal{L}-$periodic domain is the same defined in previous
Section as $\mathcal{L}=[0,L]\times[0,\delta L]$.

The averaged field equation (\ref{mech_inf}), (\ref{thermo_inf})
and (\ref{diff_inf}) (or alternatively the sequence of PDEs (\ref{inf_asymech})-(\ref{inf_asydiff})),
obtained by means of the proposed rigorous asymptotic procedure, are
used in Sec.~\ref{seccauchy} of the text of the paper for deriving the field equation of the first order (Cauchy)
homogeneous continuum equivalent to the considered periodic thermodiffusive
material. 

The approximation of the average field equations (\ref{mech_inf})-(\ref{diff_inf})
yielded by solution of homogenized differential problems of generic
order $m$ (\ref{asym_infinity}) is more accurate with respect to
that obtained by the assumption (\ref{approx}). This implies also
a more precise approximation of the solution of the microscopic field
equation (\ref{field1})-(\ref{field2}) by means of the down-scaling
relation (\ref{eq:asymech}), (\ref{eq:asytemp}) and (\ref{eq:asydiff})
involving the macroscopic field (\ref{asym2}). As it is explained
for periodic elastic composites in \citet{PeerFleck1} and \citet{Bacigalupo4}, the truncation of
the average equations of infinite order (\ref{mech_inf})-(\ref{diff_inf})
at a generic order $m$ with the aim to derive higher order field
equations for generalized thermodiffusive continua may lead to problems
in which the symmetries of the higher order elastic and thermodiffusive
constants is not guaranteed. Moreover a loss of ellipticity of the
governing equations can be observed. 
Asymptotic-variational homogenization techniques similar to those
illustrated in \citet{SmiCher1} and \citet{Bacigalupo4} represent an appropriate and powerful tool in order to avoid these problems. The generalization of
these methods to the case of elastic materials in presence of thermodiffusion
is still missing in literature.

\section{Symmetry and positive definiteness of elastic and thermodiffusive tensors}
\label{appsym}

In this Appendix, the symmetry properties of the tensors of components
$n_{ipq_{1}q_{2}}^{(2)}$, $m_{q_{1}q_{2}}^{(2)}$, $w_{q_{1}q_{2}}^{(2)}$,
and the ellipticity of the field equations (\ref{inf_asymech}) and (\ref{inf_diff0}) are demonstrated.

\subsection{Symmetry and positive definiteness of tensor of components $n_{ipq_{1}q_{2}}^{(2)}$
(vs. $C_{iq_{2}pq_{1}}$)}

Let us consider the cell problem (\ref{cellmech-1})$_{1}$, remembering
that $n_{ipq_{1}}^{(1)}=0$, it becomes 
\begin{equation}
\left(C_{ijkl}^{m}N_{kpq_{1},l}^{(1)}\right)_{,j}+C_{ijpq_{1},j}^{m}=0,\label{cellmech-test}
\end{equation}
where $C_{ijkl}^{m}$ are $\mQ-$periodic functions. The weak form
of equation (\ref{cellmech-test}), using $N_{riq_{2}}^{(1)}$ as
$\mQ-$periodic test function, is given by 
\begin{equation}
\left\langle \left(C_{ijkl}^{m}N_{kpq_{1},l}^{(1)}+C_{ijpq_{1}}^{m}\right)_{,j}N_{riq_{2}}^{(1)}\right\rangle =0,\label{cellmech-weak}
\end{equation}
applying the divergence theorem to (\ref{cellmech-weak}), and remembering
that for the $\mQ-$periodicity of $C_{ijkl}^{m}$ and $N_{riq_{2}}^{(1)}$
the path integrals evaluated on the boundary of the unit cell $\mQ$
vanish, we obtain: 
\begin{equation}
\left\langle \left(C_{ijkl}^{m}N_{kpq_{1},l}^{(1)}+C_{ijpq_{1}}^{m}\right)N_{riq_{2},j}^{(1)}\right\rangle =0.\label{cellmech-int}
\end{equation}
Using the result (\ref{cellmech-int}), expression (\ref{nmw1})$_{1}$
can be written in the equivalent form: 
\begin{align}
n_{ipq_{1}q_{2}}^{(2)} & =\frac{1}{2}\bigg\langle\left(C_{iq_{2}pq_{1}}^{\Gve}+C_{iq_{2}kl}^{\Gve}N_{kpq_{1},l}^{(1)}\right)+\left(C_{iq_{1}pq_{2}}^{\Gve}+C_{iq_{1}kl}^{\Gve}N_{kpq_{2},l}^{(1)}\right)\bigg\rangle\nonumber \\
 & =\frac{1}{2}\left\langle C_{iq_{2}pq_{1}}^{\Gve}+C_{iq_{2}kl}^{\Gve}N_{kpq_{1},l}^{(1)}+\left(C_{ijkl}^{m}N_{kpq_{1},l}^{(1)}+C_{ijpq_{1}}^{m}\right)N_{riq_{2},j}^{(1)}+\right.\nonumber \\
 & \left.C_{iq_{1}pq_{2}}^{\Gve}+C_{iq_{1}kl}^{\Gve}N_{kpq_{2},l}^{(1)}+\left(C_{rjkl}^{m}N_{kpq_{2},l}^{(1)}+C_{rjpq_{2}}^{m}\right)N_{riq_{1},j}^{(1)}\right\rangle \nonumber \\
 & =\frac{1}{2}\left[\frac{1}{4}\left\langle C_{rjkl}^{m}\left(N_{riq_{2},j}^{(1)}+\delta_{ri}\delta_{jq_{2}}+N_{rq_{2}i,j}^{(1)}+\delta_{rq_{2}}\delta_{ij}\right)\cdot\right.\right.\nonumber \\
 & \left.\left(N_{kpq_{1},l}^{(1)}+\delta_{kp}\delta_{lq_{1}}+N_{kq_{1}p,l}^{(1)}+\delta_{kq_{1}}\delta_{lp}\right)\right\rangle +\nonumber \\
 & \left.\frac{1}{4}\left\langle C_{rjkl}^{m}\left(N_{riq_{1},j}^{(1)}+\delta_{ri}\delta_{jq_{1}}+N_{rq_{1}i,j}^{(1)}+\delta_{rq_{1}}\delta_{ij}\right)\cdot\right.\right.\nonumber \\
 & \left.\left.\left(N_{kpq_{2},l}^{(1)}+\delta_{kp}\delta_{q_{2}l}+N_{kq_{2}p,l}^{(1)}+\delta_{kq_{2}}\delta_{lp}\right)\right\rangle \right],
\end{align}
as a consequence, we can observe that: 
\begin{equation}
n_{ipq_{1}q_{2}}^{(2)}=\frac{1}{2}(C_{iq_{2}pq_{1}}+C_{iq_{1}pq_{2}}),
\end{equation}
where the components $C_{iq_{2}pq_{1}}$ of the overall elastic tensor
take the form: 
\begin{equation}
C_{iq_{2}pq_{1}}=\frac{1}{4}\left\langle C_{rjkl}^{m}\left(N_{riq_{2},j}^{(1)}+\delta_{ri}\delta_{jq_{2}}+N_{rq_{2}i,j}^{(1)}+\delta_{rq_{2}}\delta_{ij}\right)\cdot\left(N_{kpq_{1},l}^{(1)}+\delta_{kp}\delta_{lq_{1}}+N_{kq_{1}p,l}^{(1)}+\delta_{kq_{1}}\delta_{lp}\right)\right\rangle ,\label{ovC}
\end{equation}
Observing expression (\ref{ovC}), it is easy to deduce that the tensor
of components $C_{iq_{2}pq_{1}}$ is symmetric and positive definite.

\subsection{Symmetry and positive definiteness of tensors of components $m_{q_{1}q_{2}}^{(2)}$
and $w_{q_{1}q_{2}}^{(2)}$ (vs. $K_{q_{1}q_{2}}$ and $D_{q_{1}q_{2}}$)}

Remembering that $m_{q_{1}}^{(1)}=0$, the cell problems (\ref{cellthermodiff-1})$_{1}$,
possesses the form 
\begin{equation}
\left(K_{ij}^{m}M_{q_{1},j}^{(1)}\right)_{,i}+K_{iq_{1},i}^{m}=0\label{celltemp-test}
\end{equation}
where $K_{ij}^{m}$ are $\mQ-$periodic functions. The weak form of
equation (\ref{celltemp-test}), using $M_{q_{2}}^{(1)}$ as $\mQ-$periodic
test function, is given by 
\begin{equation}
\left\langle \left(K_{ij}^{m}M_{q_{1},j}^{(1)}+K_{iq_{1}}^{m}\right)_{,i}M_{q_{2}}^{(1)}\right\rangle =0,\label{celltemp-weak}
\end{equation}
applying the divergence theorem to (\ref{celltemp-weak}), and remembering
that for the $\mQ-$periodicity of $K_{ij}^{m}$ and $M_{q_{2}}^{(1)}$
the path integrals evaluated on the boundary of the unit cell $\mQ$
vanish, we obtain: 
\begin{equation}
\left\langle \left(K_{ij}^{m}M_{q_{1},j}^{(1)}+K_{iq_{1}}^{m}\right)M_{q_{2},i}^{(1)}\right\rangle =0.\label{celltemp-int}
\end{equation}
Using the result (\ref{celltemp-int}), expression (\ref{nmw1})$_{(4)}$
can be written in the equivalent form: 
\begin{align}
m_{q_{1}q_{2}}^{(2)} & =\frac{1}{2}\bigg\langle(K_{q_{2}q_{1}}^{m}+K_{q_{1}j}^{m}M_{q_{2},j}^{(1)})+(K_{q_{1}q_{2}}^{m}+K_{q_{2}j}^{m}M_{q_{1},j}^{(1)})\bigg\rangle\nonumber \\
 & =\frac{1}{2}\left\langle K_{q_{2}q_{1}}^{m}+K_{q_{1}j}^{m}M_{q_{2},j}^{(1)}+\left(K_{ij}^{m}M_{q_{1},j}^{(1)}+K_{iq_{1}}^{m}\right)M_{q_{2},i}^{(1)}+\right.\nonumber \\
 & \left.K_{q_{1}q_{2}}^{m}+K_{q_{2}j}^{m}M_{q_{1},j}^{(1)}+\left(K_{ij}^{m}M_{q_{2},j}^{(1)}+K_{iq_{2}}^{m}\right)M_{q_{1},i}^{(1)}\right\rangle \nonumber \\
 & =\frac{1}{2}\left[\left\langle K_{ij}^{m}(M_{q_{2},i}^{(1)}+\delta_{iq_{2}})(M_{q_{1},j}^{(1)}+\delta_{q_{1}j})\right\rangle +\right.\nonumber \\
 & \left.\left\langle K_{ji}^{m}(M_{q_{1},i}^{(1)}+\delta_{iq_{1}})(M_{q_{2},j}^{(1)}+\delta_{q_{2}j})\right\rangle \right]\nonumber \\
 & =\left\langle K_{ij}^{m}(M_{q_{2},i}^{(1)}+\delta_{iq_{2}})(M_{q_{1},j}^{(1)}+\delta_{q_{1}j})\right\rangle 
\end{align}
as a consequence, we can observe that $m_{q_{1}q_{2}}^{(2)}=K_{q_{1}q_{2}}$,
i.e. 
\begin{equation}
K_{q_{1}q_{2}}=\left\langle K_{ij}^{m}(M_{q_{2},i}^{(1)}+\delta_{iq_{2}})(M_{q_{1},j}^{(1)}+\delta_{q_{1}j})\right\rangle .\label{ovK}
\end{equation}
Observing expression (\ref{ovK}), it is easy to deduce that the tensor
of components $K_{q_{1}q_{2}}$ is symmetric and positive definite.
Since the the equations of heat and mass diffusion possess an identical
form, the components of the tensors $K_{q_{1}q_{2}}$ and $D_{q_{1}q_{2}}$
have the same properties, and then the results obtained for the components
of the overall heat conduction tensor can be extended to the case
of the overall mass diffusion tensor of components $D_{q_{1}q_{2}}$.
These components are given by the following expression: 
\begin{equation}
D_{q_{1}q_{2}}=\left\langle D_{ij}^{m}(W_{q_{2},i}^{(1)}+\delta_{iq_{2}})(W_{q_{1},j}^{(1)}+\delta_{q_{1}j})\right\rangle .\label{ovD}
\end{equation}

\section{Overall elastic and thermodiffusive constants for bi-phase isotropic layered materials}
\label{appover}

In this Appendix the explicit expressions for the overall elastic
and thermodiffusive constant of a bi-phase layered material with isotropic
phases are reported. The components of the overall elastic tensor
take the form: 
\[
C_{1111}=\frac{-\zeta^{2}\tilde{E}_{a}\tilde{E}_{b}+\zeta[(\tilde{E}_{a}\tilde{\nu}_{b})^{2}-2\tilde{E}_{a}\tilde{\nu}_{a}\tilde{E}_{b}\tilde{\nu}_{b}
+(\tilde{E}_{b}\tilde{\nu}_{a})^{2}-(\tilde{E}_{a})^{2}-(\tilde{E}_{b})^{2}]-\tilde{E}_{a}\tilde{E}_{b}}{(\zeta+1)[\zeta(\tilde{E}_{b}(\tilde{\nu}_{a})^{2}
-\tilde{E}_{b})+\tilde{E}_{a}(\tilde{\nu}_{b})^{2}-\tilde{E}_{a}]};
\]
\[
C_{2222}=-\frac{(\zeta+1)\tilde{E}_{a}\tilde{E}_{b}}{\zeta(\tilde{E}_{b}(\tilde{\nu}_{a})^{2}-\tilde{E}_{b})+\tilde{E}_{a}(\tilde{\nu}_{b})^{2}-
\tilde{E}_{a}};
\]
\[
C_{1212}=\frac{(\zeta+1)\tilde{E}_{a}\tilde{E}_{b}}{2[\tilde{E}_{a}+\tilde{E}_{a}\tilde{\nu}_{b}+\zeta(\tilde{E}_{b}\tilde{\nu}_{a}+\tilde{E}_{b})]};
\]
\begin{equation}
C_{1122}=-\frac{\tilde{E}_{a}\tilde{E}_{b}(\tilde{\nu}_{b}+\zeta\tilde{\nu}_{a})}{\zeta(\tilde{E}_{b}(\tilde{\nu}_{a})^{2}-\tilde{E}_{b})+
\tilde{E}_{a}(\tilde{\nu}_{b})^{2}-\tilde{E}_{a}}.
\end{equation}
The components of the overall thermal dilatation tensor and diffusive
expansion tensor are respectively given by 
\[
\alpha_{11}=\frac{A_{11}\alpha^{a}-B_{11}\alpha^{a}}{\Delta_{11}};
\]
\begin{equation}
\alpha_{22}=\frac{\zeta(\tilde{E}_{b}(\tilde{\nu}_{a})^{2}-\tilde{E}_{b})\alpha^{a}+\tilde{E}_{a}((\tilde{\nu}_{b})^{2}-1)\alpha^{b}}
{\zeta\tilde{E}_{b}((\tilde{\nu}_{a})^{2}-1)+\tilde{E}_{a}((\tilde{\nu}_{b})^{2}-1)};
\end{equation}
\[
\beta_{11}=\frac{A_{11}\beta^{a}-B_{11}\beta^{a}}{\Delta_{11}};
\]
\begin{equation}
\beta_{22}=\frac{\zeta(\tilde{E}_{b}(\tilde{\nu}_{a})^{2}-\tilde{E}_{b})\beta^{a}+\tilde{E}_{a}((\tilde{\nu}_{b})^{2}-1)\beta^{b}}{\zeta\tilde{E}_{b}
((\tilde{\nu}_{a})^{2}-1)+\tilde{E}_{a}((\tilde{\nu}_{b})^{2}-1)};
\end{equation}
where: 
\[
A_{11}=\zeta^{2}[\tilde{E}_{b}(\tilde{\nu}_{a})^{2}-\tilde{E}_{b}]+\zeta[\tilde{E}_{a}(\tilde{\nu}_{b})^{2}-\tilde{E}_{b}\tilde{\nu}_{b}+\tilde{E}_{b}
\tilde{\nu}_{b}(\tilde{\nu}_{a})^{2}+\tilde{E}_{a}\tilde{\nu}_{a}-\tilde{E}_{a}\tilde{\nu}_{a}(\tilde{\nu}_{b})^{2}-\tilde{E}_{a}];
\]
\[
B_{11}=\zeta[\tilde{E}_{a}\tilde{\nu}_{a}(\tilde{\nu}_{b})^{2}-\tilde{E}_{b}+\tilde{E}_{b}(\tilde{\nu}_{a})^{2}+\tilde{E}_{b}\tilde{\nu}_{b}
-\tilde{E}_{b}\tilde{\nu}_{b}(\tilde{\nu}_{a})^{2}-\tilde{E}_{a}\tilde{\nu}_{a}]+\tilde{E}_{a}(\tilde{\nu}_{b})^{2}-\tilde{E}_{a};
\]
\begin{equation}
\Delta_{11}=(\zeta+1)[\zeta\tilde{E}_{b}((\tilde{\nu}_{a})^{2}-1)+\tilde{E}_{a}((\tilde{\nu}_{b})^{2}-1)].
\end{equation}
Finally, the components of the overall heat conduction and mass diffusion
tensors become 
\begin{equation}
K_{11}=\frac{K^{b}-\zeta K^{a}}{\zeta+1},\quad K_{22}=\frac{(\zeta+1)K^{a}K^{b}}{K^{a}+\zeta K^{b}};
\end{equation}
\begin{equation}
D_{11}=\frac{D^{b}-\zeta D^{a}}{\zeta+1},\quad D_{22}=\frac{(\zeta+1)D^{a}D^{b}}{D^{a}+\zeta D^{b}}.
\end{equation}

\section{Down-scaling relations vs cells problems}
\label{appdown}

In this Appendix, we provide more details regarding the structure of the down-scaling relations \eq{Uasym}, \eq{Tasym} and \eq{Easym} and of the related cells problems. Following the 
approaches proposed by \citet{Bensou1, BakhPan1, Allaire1, Boutin1, Meguid1} and \citet{Boutin2}, the microscopic fields can be represented through an asymptotic expansion in the general form:
\begin{equation}
u_{h}\left(\bx,\frac{\bx}{\varepsilon}\right)=\sum_{l=1}^{+\infty}\varepsilon^{l}u_{h}^{(l)}\left(\bx,\frac{\bx}{\varepsilon}\right)=u_{h}^{(0)}\left(\bx,\frac{\bx}{\varepsilon}\right)
+\varepsilon u_{h}^{(1)}\left(\bx,\frac{\bx}{\varepsilon}\right)+\varepsilon^{2}u_{h}^{(2)}\left(\bx,\frac{\bx}{\varepsilon}\right)+\cdots, 
\label{Uasymapp}
\end{equation}
\begin{equation}
\theta\left(\bx,\frac{\bx}{\varepsilon}\right)=\sum_{l=1}^{+\infty}\varepsilon^{l}\theta^{(l)}\left(\bx,\frac{\bx}{\varepsilon}\right)=\theta^{(0)}\left(\bx,\frac{\bx}{\varepsilon}\right)
+\varepsilon \theta^{(1)}\left(\bx,\frac{\bx}{\varepsilon}\right)+\varepsilon^{2}\theta^{(2)}\left(\bx,\frac{\bx}{\varepsilon}\right)+\cdots, 
\label{Tasymapp}
\end{equation}
\begin{equation}
\eta\left(\bx,\frac{\bx}{\varepsilon}\right)=\sum_{l=1}^{+\infty}\varepsilon^{l}\eta^{(l)}\left(\bx,\frac{\bx}{\varepsilon}\right)=\eta^{(0)}\left(\bx,\frac{\bx}{\varepsilon}\right)
+\varepsilon \eta^{(1)}\left(\bx,\frac{\bx}{\varepsilon}\right)+\varepsilon^{2}\eta^{(2)}\left(\bx,\frac{\bx}{\varepsilon}\right)+\cdots.
\label{Easymapp}
\end{equation}
Substituting expressions \eq{Uasymapp}, \eq{Tasymapp} and \eq{Easymapp} into the microscopic field equations \eq{field1}, \eq{field2}, and remembering the property
$\frac{\partial}{\partial x_{j}}f(\boldsymbol{x},\boldsymbol{\xi}=\frac{\boldsymbol{x}}{\varepsilon})=\left(\frac{\partial f}{\partial x_{j}}+\frac{1}{\varepsilon}
\frac{\partial f}{\partial\xi_{j}}\right)_{\boldsymbol{\xi}=\boldsymbol{x}/\varepsilon}=\left(\frac{\partial f}{\partial x_{j}}+\frac{f_{,j}}{\varepsilon}\right)_{\boldsymbol{\xi}
=\boldsymbol{x}/\varepsilon}$, we obtain
\begin{align}
& \varepsilon^{-2}\left(C_{ijhl}^{\Gve}u_{h,l}^{(0)}\right)_{,j}+\varepsilon^{-1}\Biggl\{\left[C_{ijhl}^{\Gve}\left(\frac{\partial u_{h}^{(0)}}{\partial x_{l}}+u_{h,l}^{(1)}\right)\right]_{,j}
+\left(C_{ijhl}^{\Gve}u_{h,l}^{(0)}\right)_{,j}-\left(\alpha_{ij}^{\Gve}\theta^{(0)}\right)_{,j}-\left(\beta_{ij}^{\Gve}\eta^{(0)}\right)_{,j}\Biggl\}\nonumber \\
+ &  \varepsilon^{0}\Biggl\{\left[C_{ijhl}^{\Gve}\left(\frac{\partial u_{h}^{(1)}}{\partial x_{l}}+u_{h,l}^{(2)}\right)\right]_{,j}+\frac{\partial}{\partial x_{j}}
\left[C_{ijhl}^{\Gve}\left(\frac{\partial u_{h}^{(0)}}{\partial x_{l}}+u_{h,l}^{(1)}\right)\right]-\left(\alpha_{ij}^{\Gve}\theta^{(1)}\right)_{,j}
-\frac{\partial}{\partial x_{j}}\left(\alpha_{ij}^{\Gve}\theta^{(0)}\right)\nonumber \\
- & \left(\beta_{ij}^{\Gve}\eta^{(1)}\right)_{,j}-\frac{\partial}{\partial x_{j}}\left(\beta_{ij}^{\Gve}\eta^{(0)}\right)\Biggl\}+ \varepsilon\Biggl\{\left[C_{ijhl}^{\Gve}
\left(\frac{\partial u_{h}^{(2)}}{\partial x_{l}}+u_{h,l}^{(3)}\right)\right]_{,j}+\frac{\partial}{\partial x_{j}}
\left[C_{ijhl}^{\Gve}\left(\frac{\partial u_{h}^{(1)}}{\partial x_{l}}+u_{h,l}^{(2)}\right)\right]\nonumber \\
- &\left(\alpha_{ij}^{\Gve}\theta^{(2)}\right)_{,j}-\frac{\partial}{\partial x_{j}}\left(\alpha_{ij}^{\Gve}\theta^{(1)}\right)-
\left(\beta_{ij}^{\Gve}\eta^{(2)}\right)_{,j}-\frac{\partial}{\partial x_{j}}\left(\beta_{ij}^{\Gve}\eta^{(1)}\right)\Biggl\}
+\cdots\cdots+b_{i}=0,\quad\quad i=1,2,\label{eq:asymechapp}
\end{align}
\begin{align}
& \varepsilon^{-2}\left(K_{ij}^{\Gve}\theta_{,j}^{(0)}\right)_{,i}+\varepsilon^{-1}\biggl\{\left[K_{ij}^{\Gve}\left(\frac{\partial\theta^{(0)}}{\partial x_{j}}+\theta^{(1)}_{,j}\right)\right]_{,i}
+\frac{\partial}{\partial x_{i}}\left(K_{ij}^{\Gve}\theta_{,j}^{(0)}\right)\biggr\}\nonumber \\
+ & \varepsilon^{0}\biggl\{\left[K_{ij}^{\Gve}\left(\frac{\partial\theta^{(1)}}{\partial x_{j}}+\theta^{(2)}_{,j}\right)\right]_{,i}
+\frac{\partial}{\partial x_{i}}\left[K_{ij}^{\Gve}\left(\frac{\partial\theta^{(0)}}{\partial x_{j}}+\theta^{(1)}_{,j}\right)\right]\biggr\}\nonumber \\
+ &  \varepsilon\biggl\{\left[K_{ij}^{\Gve}\left(\frac{\partial\theta^{(2)}}{\partial x_{j}}+\theta^{(3)}_{,j}\right)\right]_{,i}
+\frac{\partial}{\partial x_{i}}\left[K_{ij}^{\Gve}\left(\frac{\partial\theta^{(1)}}{\partial x_{j}}+\theta^{(2)}_{,j}\right)\right]\biggr\}+\cdots\cdots+r=0,\label{eq:asytempapp}
\end{align}
\begin{align}
& \varepsilon^{-2}\left(D_{ij}^{\Gve}\eta_{,j}^{(0)}\right)_{,i}+\varepsilon^{-1}\biggl\{\left[D_{ij}^{\Gve}\left(\frac{\partial\eta^{(0)}}{\partial x_{j}}+\eta^{(1)}_{,j}\right)\right]_{,i}
+\frac{\partial}{\partial x_{i}}\left(D_{ij}^{\Gve}\eta_{,j}^{(0)}\right)\biggr\}\nonumber \\
+ & \varepsilon^{0}\biggl\{\left[D_{ij}^{\Gve}\left(\frac{\partial\eta^{(1)}}{\partial x_{j}}+\eta^{(2)}_{,j}\right)\right]_{,i}
+\frac{\partial}{\partial x_{i}}\left[D_{ij}^{\Gve}\left(\frac{\partial\eta^{(0)}}{\partial x_{j}}+\eta^{(1)}_{,j}\right)\right]\biggr\}\nonumber \\
+ &  \varepsilon\biggl\{\left[D_{ij}^{\Gve}\left(\frac{\partial\eta^{(2)}}{\partial x_{j}}+\eta^{(3)}_{,j}\right)\right]_{,i}
+\frac{\partial}{\partial x_{i}}\left[D_{ij}^{\Gve}\left(\frac{\partial\eta^{(1)}}{\partial x_{j}}+\eta^{(2)}_{,j}\right)\right]\biggr\}+\cdots\cdots+s=0. \label{eq:asydiffapp}
\end{align}

At the order $\varepsilon^{-2}$ from equation \eq{eq:asymechapp} we derive:
\begin{equation}
\left(C_{ijhl}^{\Gve}u_{h,l}^{(0)}\right)_{,j}=f^{(0)}_i(\bx),
\label{u0}
\end{equation}
whereas from heat conduction and mass diffusion equations \eq{eq:asytempapp} and \eq{eq:asydiffapp} we get respectively:
\begin{equation}
\left(K_{ij}^{\Gve}\theta_{,j}^{(0)}\right)_{,i}=g^{(0)}(\bx), \quad
\left(D_{ij}^{\Gve}\eta_{,j}^{(0)}\right)_{,i}=h^{(0)}(\bx).
\label{c0}
\end{equation}
The interface conditions \eq{intc1}-\eq{intc3}, expressed with respect to $\BGx$, become:
\begin{equation}
\left.\jump{0.1}{u_h^{(0)}}\right|_{\BGx\in\Sigma_1}=0, \qquad 
\left.\bjump{0.3}{C_{ijhl}^{\Gve}u_{h,l}^{(0)}n_{j}}\right|_{\BGx\in\Sigma_1}=0, 
\label{iu0}
\end{equation}
\begin{equation}
\left.\jump{0.1}{\theta^{(0)}}\right|_{\BGx\in\Sigma_1}=0, \qquad 
\left.\bjump{0.3}{K_{ij}^{\Gve}\theta^{(0)}_{,j}n_{i}}\right|_{\BGx\in\Sigma_1}=0, 
\label{it0}
\end{equation}
\begin{equation}
\left.\jump{0.1}{\eta^{(0)}}\right|_{\BGx\in\Sigma_1}=0, \qquad 
\left.\bjump{0.3}{D_{ij}^{\Gve}\eta^{(0)}_{,j}n_{i}}\right|_{\BGx\in\Sigma_1}=0,
\label{ic0}
\end{equation}
where $\Sigma_1$ is the representation of the interface $\Sigma$ bewteen two different phases of the material in the non-dimensional space of the variable $\BGx$.

At the order $\varepsilon^{-1}$ equation \eq{eq:asymechapp} yields
\begin{equation}
 \left[C_{ijhl}^{\Gve}\left(\frac{\partial u_{h}^{(0)}}{\partial x_{l}}+u_{h,l}^{(1)}\right)\right]_{,j}
+\frac{\partial}{\partial x_{j}}\left(C_{ijhl}^{\Gve}u_{h,l}^{(0)}\right)_{,j}-\left(\alpha_{ij}^{\Gve}\theta^{(0)}\right)_{,j}-\left(\beta_{ij}^{\Gve}\eta^{(0)}\right)_{,j}=f^{(1)}_i(\bx),
\label{u1}
\end{equation}
at the same order, from equations \eq{eq:asytempapp} and \eq{eq:asydiffapp} we obtain:
\begin{equation}
 \left[K_{ij}^{\Gve}\left(\frac{\partial\theta^{(0)}}{\partial x_{j}}+\theta^{(1)}_{,j}\right)\right]_{,i}
+\frac{\partial}{\partial x_{i}}\left(K_{ij}^{\Gve}\theta_{,j}^{(0)}\right)=g^{(1)}(\bx),
\label{t1}
\end{equation}
\begin{equation}
 \left[D_{ij}^{\Gve}\left(\frac{\partial\eta^{(0)}}{\partial x_{j}}+\eta^{(1)}_{,j}\right)\right]_{,i}
+\frac{\partial}{\partial x_{i}}\left(D_{ij}^{\Gve}\eta_{,j}^{(0)}\right)=h^{(1)}(\bx),
\label{c1}
\end{equation}
and the interface conditions are given by
\begin{equation}
  \left.\jump{0.1}{u_h^{(1)}}\right|_{\BGx\in\Sigma_1}=0; \qquad 
 \left.\bjump{0.3}{\left(C_{ijhl}^{\Gve}\left(\frac{\partial u_{h}^{(0)}}{\partial x_{l}}+u_{h,l}^{(1)}\right)-\alpha_{ij}^{\Gve}\theta^{(0)}-\beta_{ij}^{\Gve}\eta^{(0)}\right)n_{j}}\right|_{\BGx\in\Sigma_1}=0,
\label{iu1}
 \end{equation}
\begin{equation}
 \left.\jump{0.1}{\theta^{(1)}}\right|_{\BGx\in\Sigma_1}=0; \qquad 
 \left.\bjump{0.3}{K_{ij}^{\Gve}\left(\theta^{(1)}_{,j}+\frac{\partial\theta^{(0)}}{\partial x_{j}}\right)n_{i}}\right|_{\BGx\in\Sigma_1}=0,
\label{it1}
 \end{equation}
\begin{equation}
 \left.\jump{0.1}{\eta^{(1)}}\right|_{\BGx\in\Sigma_1}=0; \qquad 
 \left.\bjump{0.3}{D_{ij}^{\Gve}\left(\eta^{(1)}_{,j}+\frac{\partial\eta^{(0)}}{\partial x_{j}}\right)n_{i}}\right|_{\BGx\in\Sigma_1}=0.
\label{ic1}
 \end{equation}

At the order $\varepsilon^{0}$, the cells problems associate to equation \eq{eq:asymechapp} assume the form:
\begin{align}
& \left[C_{ijhl}^{\Gve}\left(\frac{\partial u_{h}^{(1)}}{\partial x_{l}}+u_{h,l}^{(2)}\right)\right]_{,j}+\frac{\partial}{\partial x_{j}}
\left[C_{ijhl}^{\Gve}\left(\frac{\partial u_{h}^{(0)}}{\partial x_{l}}+u_{h,l}^{(1)}\right)\right]\nonumber\\
- & \left(\alpha_{ij}^{\Gve}\theta^{(1)}\right)_{,j}
-\frac{\partial}{\partial x_{j}}\left(\alpha_{ij}^{\Gve}\theta^{(0)}\right)-\left(\beta_{ij}^{\Gve}\eta^{(1)}\right)_{,j}-\frac{\partial}{\partial x_{j}}
\left(\beta_{ij}^{\Gve}\eta^{(0)}\right)=f^{(2)}_i(\bx),
\label{u2}
\end{align}
whereas the cells problems correspoding to equations \eq{eq:asytempapp} and \eq{eq:asydiffapp} are:
\begin{equation}
 \left[K_{ij}^{\Gve}\left(\frac{\partial\theta^{(1)}}{\partial x_{j}}+\theta^{(2)}_{,j}\right)\right]_{,i}
+\frac{\partial}{\partial x_{i}}\left[K_{ij}^{\Gve}\left(\frac{\partial\theta^{(0)}}{\partial x_{j}}+\theta^{(1)}_{,j}\right)\right]=g^{(2)}(\bx),
\label{t2}
\end{equation}
\begin{equation}
 \left[D_{ij}^{\Gve}\left(\frac{\partial\eta^{(1)}}{\partial x_{j}}+\eta^{(2)}_{,j}\right)\right]_{,i}
+\frac{\partial}{\partial x_{i}}\left[D_{ij}^{\Gve}\left(\frac{\partial\eta^{(0)}}{\partial x_{j}}+\eta^{(1)}_{,j}\right)\right]=h^{(2)}(\bx),
\label{c2}
\end{equation}
and the interface conditions assume the form:
\begin{equation}
  \left.\jump{0.1}{u_h^{(2)}}\right|_{\BGx\in\Sigma_1}=0, \qquad 
 \left.\bjump{0.3}{\left(C_{ijhl}^{\Gve}\left(\frac{\partial u_{h}^{(1)}}{\partial x_{l}}+u_{h,l}^{(2)}\right)-\alpha_{ij}^{\Gve}\theta^{(1)}-\beta_{ij}^{\Gve}\eta^{(1)}\right)n_{j}}\right|_{\BGx\in\Sigma_1}=0,
\label{iu2}
 \end{equation}
\begin{equation}
 \left.\jump{0.1}{\theta^{(2)}}\right|_{\BGx\in\Sigma_1}=0, \qquad 
 \left.\bjump{0.3}{K_{ij}^{\Gve}\left(\theta^{(2)}_{,j}+\frac{\partial\theta^{(1)}}{\partial x_{j}}\right)n_{i}}\right|_{\BGx\in\Sigma_1}=0,
\label{it2}
 \end{equation}
\begin{equation}
 \left.\jump{0.1}{\eta^{(2)}}\right|_{\BGx\in\Sigma_1}=0, \qquad 
 \left.\bjump{0.3}{D_{ij}^{\Gve}\left(\eta^{(2)}_{,j}+\frac{\partial\eta^{(1)}}{\partial x_{j}}\right)n_{i}}\right|_{\BGx\in\Sigma_1}=0.
\label{ic2}
 \end{equation}

At the order $\varepsilon^{-2}$, the solvibility conditions in the class of the functions $\mQ-$periodic with respect to the fast variable $\BGx$ 
implies that $f^{(0)}_i(\bx)=g^{(0)}(\bx)=h^{(0)}(\bx)=0$, then the cell problems
\eq{u0}-\eq{c0} become:
\begin{equation}
 \left(C_{ijhl}^{\Gve}u_{h,l}^{(0)}\right)_{,j}=0, \quad
 \left(K_{ij}^{\Gve}\theta_{,j}^{(0)}\right)_{,i}=0,\quad
 \left(D_{ij}^{\Gve}\eta_{,j}^{(0)}\right)_{,i}=0,
 \label{cell0app}
\end{equation}
as a consequence, the solution of problems \eq{cell0app} does not depend by the fast variable $\BGx$ and then $u_{h}^{(0)}(\bx, \BGx)=U_{h}(\bx), \theta^{(0)}(\bx, \BGx)=\Theta(\bx)$ and $\eta^{(0)}(\bx, \BGx)=\Upsilon(\bx)$.

At the order $\varepsilon^{-1}$, the solvability conditions in the class of the functions $\mQ-$periodic with respect to the fast variable $\BGx$ together with
the interface conditions \eq{iu1}-\eq{ic1} yield to
\begin{equation}
 \langle C_{ijhl,j}^{\Gve}\rangle\frac{\partial U_{h}}{\partial x_{l}}
-\langle\alpha_{ij,j}^{\Gve}\rangle\Theta^{(1)}
-\langle\beta_{ij,j}^{\Gve}\rangle\Upsilon=f^{(1)}_i(\bx),
\label{cell1uapp}
\end{equation}
\begin{equation}
 \langle K_{ij,i}^{\Gve}\rangle\frac{\partial\Theta}{\partial x_{j}}=g^{(1)}(\bx), \qquad  \langle D_{ij,i}^{\Gve}\rangle\frac{\partial\Upsilon}{\partial x_{j}}=h^{(1)}(\bx).
\label{cell1tcapp}
 \end{equation}
For the $\mQ-$periodicity of the functions $C_{ijhl}^{\Gve}, \alpha_{ij}^{\Gve}, \beta_{ij}^{\Gve}, K_{ij}^{\Gve}$ and $D_{ij}^{\Gve}$, we have $f^{(1)}_i(\bx)=g^{(1)}(\bx)=h^{(1)}(\bx)=0$,
and then at this order the solution of the fields equations assumes the form:
\begin{equation}
 u_{h}^{(1)}(\bx, \BGx)=N_{hpq_{1}}^{(1)}(\BGx)\frac{\partial U_{p}}{\partial x_{q_{1}}}+\tilde{N}_{h}^{(1)}(\BGx)\Theta(\bx)+\hat{N}_{h}^{(1)}(\BGx)\Upsilon(\bx),
\label{u1app}
 \end{equation}
\begin{equation}
 \theta^{(1)}(\bx, \BGx)=M_{q_{1}}^{(1)}(\BGx)\frac{\partial\Theta}{\partial x_{q_{1}}}, \qquad \eta^{(1)}(\bx, \BGx)=W_{q_{1}}^{(1)}(\BGx)\frac{\partial\Upsilon}{\partial x_{q_{1}}},
\label{t1c1app}
 \end{equation}
where $N_{hpq_{1}}^{(1)}, \tilde{N}_{h}^{(1)}, \hat{N}_{h}^{(1)}, M_{q_{1}}^{(1)}$ and $W_{q_{1}}^{(1)}$ are the same fluctuations functions introduced in Section \ref{multiscale}.
Substituting expressions \eq{u1app}-\eq{t1c1app} into the cell problems \eq{cell1uapp}-\eq{cell1tcapp} and considering the interface conditions  \eq{iu1}-\eq{ic1}, we derive:
\begin{equation}
 \left(C_{ijhl}^{\Gve}N_{hpq_{1},l}^{(1)}\right)_{,j}+C_{ijpq_{1},j}^{\Gve}=0, \quad \left(C_{ijhl}^{\Gve}\tilde{N}_{h,l}^{(1)}\right)_{,j}-\alpha_{ij,j}^{\Gve}=0, \quad
 \left(C_{ijhl}^{\Gve}\hat{N}_{h,l}^{(1)}\right)_{,j}-\beta_{ij,j}^{\Gve}=0, 
\label{c1uflut}
 \end{equation}
\begin{equation}
 \left(K_{ij}^{\Gve}M_{q_{1},j}^{(1)}\right)_{,i}+K_{iq_{1},i}^{\Gve}=0, \qquad \left(D_{ij}^{\Gve}W_{q_{1},j}^{(1)}\right)_{,i}+W_{iq_{1},i}^{\Gve}=0, 
\label{c1tcflut}
 \end{equation}
and then the interface conditions \eq{iu1}-\eq{ic1} become;
\begin{equation}
  \left.\jump{0.1}{N_{hpq_{1}}^{(1)}}\right|_{\BGx\in\Sigma_1}=0, \qquad 
 \left.\bjump{0.3}{C_{ijhl}^{\Gve}\left(N_{hpq_{1},l}^{(1)}+\delta_{hp}\delta_{lq_{1}}\right)n_{j}}\right|_{\BGx\in\Sigma_1}=0,
\label{ic1uflut}
 \end{equation}
 \begin{equation}
  \left.\jump{0.1}{\tilde{N}_{h}^{(1)}}\right|_{\BGx\in\Sigma_1}=0, \qquad 
 \left.\bjump{0.3}{\left(C_{ijhl}^{\Gve}\tilde{N}_{h,l}^{(1)}-\alpha_{ij}^{\Gve}\right)n_{j}}\right|_{\BGx\in\Sigma_1}=0,
 \end{equation}
 \begin{equation}
  \left.\jump{0.1}{\hat{N}_{h}^{(1)}}\right|_{\BGx\in\Sigma_1}=0, \qquad 
 \left.\bjump{0.3}{\left(C_{ijhl}^{\Gve}\hat{N}_{h,l}^{(1)}-\beta_{ij}^{\Gve}\right)n_{j}}\right|_{\BGx\in\Sigma_1}=0,
 \end{equation}
\begin{equation}
 \left.\jump{0.1}{M_{q_{1}}^{(1)}}\right|_{\BGx\in\Sigma_1}=0, \qquad 
 \left.\bjump{0.3}{K_{ij}^{\Gve}\left(M_{q_{1},j}^{(1)}+\delta_{q_{1}j}\right)n_{i}}\right|_{\BGx\in\Sigma_1}=0,
 \end{equation}
\begin{equation}
 \left.\jump{0.1}{W_{q_{1}}^{(1)}}\right|_{\BGx\in\Sigma_1}=0, \qquad 
 \left.\bjump{0.3}{D_{ij}^{\Gve}\left(W_{q_{1},j}^{(1)}+\delta_{q_{1}j}\right)n_{i}}\right|_{\BGx\in\Sigma_1}=0.
 \label{ic1ctflut}
 \end{equation}
The solution of the cell problems \eq{c1uflut}-\eq{c1tcflut} taking into account the interface conditions \eq{ic1uflut}-\eq{ic1ctflut} provides the $\mQ-$periodic perturbation functions 
$N_{hpq_{1}}^{(1)}, \tilde{N}_{h}^{(1)}, \hat{N}_{h}^{(1)}, M_{q_{1}}^{(1)}$ and $W_{q_{1}}^{(1)}$.

Taking into account the solvability conditions in the class of the functions $\mQ-$periodic  with respect to the fast variable $\BGx$ and the interface conditions 
\eq{iu2}-\eq{ic2}, the cell problems \eq{u2}-\eq{c2} associate to the 
$\varepsilon^{0}$ become:
\begin{equation}
 \langle C_{iq_{1}pl}^{\Gve}+C_{ilhj}^{\Gve}N_{hpq_{1},j}^{(1)}\rangle\frac{\partial^2 U_{p}}{\partial x_{q_{1}}\partial x_{l}}
-\langle C_{ilhj}^{\Gve}\tilde{N}_{h,j}^{(1)}-\alpha_{il}^{\Gve}\rangle\frac{\partial \Theta}{\partial x_{l}}
-\langle C_{ilhj}^{\Gve}\hat{N}_{h,j}^{(1)}-\beta_{il}^{\Gve}\rangle\frac{\partial \Upsilon}{\partial x_{l}}=f^{(2)}_i(\bx),
\label{cell2uapp}
\end{equation}
\begin{equation}
 \left\langle K_{ij}^{\Gve}\left(M_{q_{1},j}^{(1)}+\delta_{jq_{1}}\right)\right\rangle\frac{\partial^2\Theta}{\partial x_{i}\partial x_{q_{1}}}=g^{(2)}(\bx), \qquad  
 \left\langle D_{ij}^{\Gve}\left(W_{q_{1},j}^{(1)}+\delta_{jq_{1}}\right)\right\rangle\frac{\partial^2\Upsilon}{\partial x_{i}\partial x_{q_{1}}}=h^{(2)}(\bx).
\label{cell2tcapp}
 \end{equation}
these cell problems possess a solution satisfying the conditions \eq{iu2}-\eq{ic2} in the form:
\begin{equation}
 u_{h}^{(2)}(\bx, \BGx)=N_{hpq_{1}q_{2}}^{(2)}\frac{\partial^2 U_{p}}{\partial x_{q_{1}}\partial x_{q_{2}}}+\tilde{N}_{hq_{1}}^{(2)}\frac{\partial \Theta}{\partial x_{q_{1}}}
 +\hat{N}_{hq_{1}}^{(2)}\frac{\partial \Upsilon}{\partial x_{q_{1}}},
\end{equation}
\begin{equation}
 \theta^{(2)}(\bx, \BGx)=M_{q_{1}q_{2}}^{(2)}\frac{\partial^2\Theta}{\partial x_{q_{2}}\partial x_{q_{1}}}, \qquad 
 \eta^{(2)}(\bx, \BGx)=W_{q_{1}q_{2}}^{(2)}\frac{\partial^2\Upsilon}{\partial x_{q_{2}}\partial x_{q_{1}}},
\end{equation}
where $N_{hpq_{1}q_{2}}^{(2)}, \tilde{N}_{hq_{1}}^{(2)} \hat{N}_{hq_{1}}^{(2)}, M_{q_{1}q_{2}}^{(2)}$ and $W_{q_{1}q_{2}}^{(2)}$ are second order fluctuations functions 
just introduced in Section \ref{multiscale}. As a consequence, from the cell problems \eq{cell2uapp}-\eq{cell2tcapp} we derive:
\begin{equation}
 \left(C_{ijhl}^{\Gve}N_{hpq_{1}q_{2},l}^{(2)}\right)_{,j}+\left(C_{ijhq_{2}}^{\Gve}N_{hpq_{1}}^{(1)}\right)_{,j}+C_{iq_{1}pq_{2}}^{\Gve}+C_{iq_{2}hj}^{\Gve}N_{hpq_{1},j}^{(1)}=
 \langle C_{iq_{1}pq_{2}}^{\Gve}+C_{iq_{2}hj}^{\Gve}N_{hpq_{1},j}^{(1)}\rangle,
\label{c2uflut}
 \end{equation}
\begin{equation}
 \left(C_{ijhl}^{\Gve}\tilde{N}_{hq_{1},l}^{(2)}\right)_{,j}+\left(C_{ijhq_1}^{\Gve}\tilde{N}_{h}^{(1)}\right)_{,j}+C_{iq_1kj}^{\Gve}\tilde{N}_{h,j}^{(1)}
 -\left(\alpha_{ij}^{\Gve}M_{q_{1}}^{(1)}\right)_{,j}-\alpha_{iq_{1}}^{\Gve}
 =\langle C_{iq_{1}hj}^{\Gve}\tilde{N}_{h,j}^{(1)}-\alpha_{iq_{1}}^{\Gve}\rangle,
\end{equation}
\begin{equation}
\left(C_{ijhl}^{\Gve}\hat{N}_{hq_{1},l}^{(2)}\right)_{,j}+\left(C_{ijhq_1}^{\Gve}\hat{N}_{h}^{(1)}\right)_{,j}+C_{iq_1kj}^{\Gve}\hat{N}_{h,j}^{(1)}
 -\left(\beta_{ij}^{\Gve}W_{q_{1}}^{(1)}\right)_{,j}-\beta_{iq_{1}}^{\Gve}=\left\langle C_{iq_{1}hj}^{\Gve}\hat{N}_{h,j}^{(1)}-\beta_{iq_{1}}^{\Gve}\right\rangle,
\end{equation}
\begin{equation}
 \left(K_{ij}^{\Gve}M_{q_{1}q_{2},j}^{(2)}\right)_{,i}+\left(K_{iq_{2}}^{\Gve}M_{q_{1}}^{(1)}\right)_{,i}+K_{iq_{2}}^{\Gve}\left(M_{q_{1},j}^{(1)}+\delta_{jq_{1}}\right)
 =\left\langle K_{q_{2}j}^{\Gve}\left(M_{q_{1},j}^{(1)}+\delta_{jq_{1}}\right)\right\rangle,
\end{equation}
\begin{equation}
 \left(D_{ij}^{\Gve}W_{q_{1}q_{2},j}^{(2)}\right)_{,i}+\left(D_{iq_{2}}^{\Gve}W_{q_{1}}^{(1)}\right)_{,i}+D_{iq_{2}}^{\Gve}\left(W_{q_{1},j}^{(1)}+\delta_{jq_{1}}\right)
 =\left\langle D_{q_{2}j}^{\Gve}\left(W_{q_{1},j}^{(1)}+\delta_{jq_{1}}\right)\right\rangle,
\label{c2ctflut}
 \end{equation}
and then the interface conditions \eq{iu2}-\eq{ic2} become;
\begin{equation}
  \left.\jump{0.1}{N_{hpq_{1}q_{2}}^{(2)}}\right|_{\BGx\in\Sigma_1}=0, \qquad 
 \left.\bjump{0.3}{\left(C_{ijhl}^{\Gve}N_{hpq_{1}q_{2},l}^{(2)}+C_{ijhq_{2}}^{\Gve}N_{hpq_{1}}^{(1)}\right)n_{j}}\right|_{\BGx\in\Sigma_1}=0,
\label{ic2uflut}
 \end{equation}
 \begin{equation}
  \left.\jump{0.1}{\tilde{N}_{hq_{1}}^{(2)}}\right|_{\BGx\in\Sigma_1}=0, \qquad 
 \left.\bjump{0.3}{\left(C_{ijhl}^{\Gve}\tilde{N}_{hq_{1},l}^{(2)}+C_{ijhq_{1}}^{\Gve}\tilde{N}_{h}^{(1)}-\alpha_{ij}^{\Gve}M_{q_{1}}^{(1)}\right)n_{j}}\right|_{\BGx\in\Sigma_1}=0,
 \end{equation}
 \begin{equation}
  \left.\jump{0.1}{\hat{N}_{hq_{1}}^{(2)}}\right|_{\BGx\in\Sigma_1}=0, \qquad 
  \left.\bjump{0.3}{\left(C_{ijhl}^{\Gve}\hat{N}_{hq_{1},l}^{(2)}+C_{ijhq_{1}}^{\Gve}\hat{N}_{h}^{(1)}-\beta_{ij}^{\Gve}W_{q_{1}}^{(1)}\right)n_{j}}\right|_{\BGx\in\Sigma_1}=0,
 \end{equation}
\begin{equation}
 \left.\jump{0.1}{M_{q_{1}q_{2}}^{(2)}}\right|_{\BGx\in\Sigma_1}=0, \qquad 
 \left.\bjump{0.3}{K_{ij}^{\Gve}\left(M_{q_{1}q_{2},j}^{(2)}+M_{q_{1}}^{(1)}\delta_{jq_{2}}\right)n_{i}}\right|_{\BGx\in\Sigma_1}=0,
 \end{equation}
\begin{equation}
 \left.\jump{0.1}{W_{q_{1}q_{2}}^{(2)}}\right|_{\BGx\in\Sigma_1}=0, \qquad 
 \left.\bjump{0.3}{S_{ij}^{\Gve}\left(W_{q_{1}q_{2},j}^{(2)}+W_{q_{1}}^{(1)}\delta_{jq_{2}}\right)n_{i}}\right|_{\BGx\in\Sigma_1}=0.
 \label{ic2ctflut}
 \end{equation}
The solution of the cell problems \eq{c2uflut}-\eq{c2ctflut} taking into account the interface conditions \eq{ic2uflut}-\eq{ic2ctflut} provides the $\mQ-$periodic perturbation functions 
$N_{hpq_{1}q_{2}}^{(2)}, \tilde{N}_{hq_{1}}^{(2)}, \hat{N}_{hq_{1}}^{(2)}, M_{q_{1}q_{2}}^{(2)}$ and $W_{q_{1}q_{2}}^{(2)}$.

The general procedure here reported can be applied to higher order cell problems for deriving the averaged field equations of infinite order, which assume the form:
\begin{align}
& \langle C_{iq_{1}pl}^{\Gve}+C_{ilhj}^{\Gve}N_{hpq_{1},j}^{(1)}\rangle\frac{\partial^2 U_{p}}{\partial x_{q_{1}}\partial x_{l}}
-\langle C_{ilhj}^{\Gve}\tilde{N}_{h,j}^{(1)}-\alpha_{il}^{\Gve}\rangle\frac{\partial \Theta}{\partial x_{l}}
-\langle C_{ilhj}^{\Gve}\hat{N}_{h,j}^{(1)}-\beta_{il}^{\Gve}\rangle\frac{\partial \Upsilon}{\partial x_{l}}\nonumber\\
+&\langle C_{iq_{3}hq_{2}}^{\Gve}N_{hpq_{1}}^{(1)}+C_{iq_{3}hl}^{\Gve}N_{hpq_{1}q_{2},l}^{(2)}\rangle
\frac{\partial^3 U_{p}}{\partial x_{q_{1}}\partial x_{q_{2}}\partial x_{q_{3}}}
+\langle C_{iq_{1}hq_{2}}^{\Gve}\tilde{N}_{h}^{(1)}+C_{iq_{2}hl}^{\Gve}\tilde{N}_{hq_{1},l}^{(2)}-
\alpha_{iq_{2}}^{\Gve}M^{(1)}_{q_{1}}\rangle\frac{\partial^2 \Theta}{\partial x_{q_{1}}\partial x_{q_{2}}}\nonumber\\
+&\langle C_{iq_{1}hq_{2}}^{\Gve}\hat{N}_{h}^{(1)}+C_{iq_{2}hl}^{\Gve}\hat{N}_{hq_{1},l}^{(2)}-
\beta_{iq_{2}}^{\Gve}W^{(1)}_{q_{1}}\rangle\frac{\partial^2 \Upsilon}{\partial x_{q_{1}}\partial x_{q_{2}}}+\cdots\cdots+b_{i}(\bx)=0,
\label{av1app}
\end{align}
\begin{align}
 &\left\langle K_{q_{2}j}^{\Gve}\left(M_{q_{1},j}^{(1)}+\delta_{jq_{1}}\right)\right\rangle\frac{\partial^2\Theta}{\partial x_{q_{2}}\partial x_{q_{1}}}+
 \left\langle K_{q_{3}q_{2}}^{\Gve}M_{q_{2}}^{(1)}+K_{q_{3}j}^{\Gve}M_{q_{1}q_{2},j}^{(2)}\right\rangle
 \frac{\partial^3\Theta}{\partial x_{q_{1}}\partial x_{q_{2}}\partial x_{q_{3}}}\nonumber\\
 +&\cdots\cdots+r(\bx)=0,
\label{av2app}
 \end{align}
\begin{align}
 &\left\langle D_{q_{2}j}^{\Gve}\left(W_{q_{1},j}^{(1)}+\delta_{jq_{1}}\right)\right\rangle\frac{\partial^2\Upsilon}{\partial x_{q_{2}}\partial x_{q_{1}}}+
 \left\langle D_{q_{3}q_{2}}^{\Gve}W_{q_{2}}^{(1)}+D_{q_{3}j}^{\Gve}W_{q_{1}q_{2},j}^{(2)}\right\rangle
 \frac{\partial^3\Upsilon}{\partial x_{q_{1}}\partial x_{q_{2}}\partial x_{q_{3}}}\nonumber\\
 +&\cdots\cdots+s(\bx)=0.
 \label{av3app}
\end{align}
Note that applying the permutation of the saturated indexes, \eq{av1app}, \eq{av2app} and \eq{av3app} 
become identical to the averaged field equations derived in Appendix \ref{apphigh}. The structure of the down-scaling relations
is defined by the solutions of the various cells problems associate to the different orders of the asymptotic expansion. These 
down-scaling relations assume the form:
\begin{align}
u_{h}\left(\bx,\frac{\bx}{\varepsilon}\right)
 & =\left[U_{h}(\bx)+\varepsilon \left(N_{hpq_{1}}^{(1)}(\BGx)\frac{\partial U_{p}(\bx)}{\partial x_{q_{1}}}+\tilde{N}_{h}^{(1)}(\BGx)\Theta(\bx)
 +\hat{N}_{h}^{(1)}(\BGx)\Upsilon(\bx)\right)+\right.\nonumber \\
 & \left.\;\;\;\;+\varepsilon^{2}\left(N_{hpq_{1}q_{2}}^{(2)}(\BGx)\frac{\partial^{2}U_{p}(\bx)}{\partial x_{q_{1}}\partial x_{q_{2}}}+
 \tilde{N}_{hq_{1}}^{(2)}(\BGx)\frac{\partial\Theta(\bx)}{\partial x_{q_{1}}}
 +\hat{N}_{hq_{1}}^{(2)}(\BGx)\frac{\partial\Upsilon(\bx)}{\partial x_{q_{1}}}\right)+\cdots\right]_{\BGx=\bx/\varepsilon},
\label{Uasymappfin}
\end{align}
\begin{align}
\theta\left(\bx,\frac{\bx}{\varepsilon}\right) 
 & =\left[\Theta(\bx)+\varepsilon M_{q_{1}}^{(1)}(\BGx)\frac{\partial\Theta(\bx)}{\partial x_{q_{1}}}
 +\varepsilon^{2}M_{q_{1}q_{2}}^{(2)}(\BGx)\frac{\partial^{2}\Theta(\bx)}{\partial x_{q_{1}}\partial x_{q_{2}}}+\cdots\right]_{\BGx=\bx/\varepsilon}
 \label{Tasymappfin}
 \end{align}
\begin{align}
\eta\left(\bx,\frac{\bx}{\varepsilon}\right) 
 & =\left[\Upsilon(\bx)+\varepsilon W_{q_{1}}^{(1)}(\BGx)\frac{\partial\Upsilon(\bx)}{\partial x_{q_{1}}}
 +\varepsilon^{2}W_{q_{1}q_{2}}^{(2)}(\BGx)\frac{\partial^{2}\Upsilon(\bx)}{\partial x_{q_{1}}\partial x_{q_{2}}}+\cdots\right]_{\BGx=\bx/\varepsilon}.
\label{Easymappfin}
\end{align}
The relations \eq{Uasymappfin}, \eq{Tasymappfin} and \eq{Easymappfin} are identical to expressions \eq{Uasym}, \eq{Tasym} and \eq{Easym}
introduced in Section \ref{multiscale} and used for developing the homogenization method illustrated in the paper.
\end{document}